# Unified Information Dynamic Analysis of Quantum Decision-Making and Search Algorithms: Computational Intelligence Measure

Ulyanov Sergey V.*, Ghisi F. †, Kurawaki I. ** and Ulyanov Viktor S.††

*Institute of System Analysis and Management, Dubna State University
* Meshcheryakov Laboratory of Information Technologies, Joint Institute for Nuclear Research (JINR)
† Polo Didattico e di Ricerca di Crema - Milan University, Via Bramante, 65 - 26013 CREMA(CR) - Italy
**Yamaha Motor Co. Ltd., Automotive operations Dpt.
††Department of Information Technologies, Moscow State University of Geodesy and Cartography (MIIGAiK)
*Email: srg.v.ulyanov@gmail.com
†Email: ghisif@tin.com
** Email: kurawakii@yamaha-motor.co.jp
†† Email: ulyanovik@gmail.com

**Abstract**

There are important algorithms built upon a mixture of basic techniques described; for example, the Fast Fourier Transform (FFT) employs both Divide-and-Conquer and Transform-and-Conquer techniques. In this article, the evolution of a quantum algorithm (QA) is examined from an information theory viewpoint. The complex vector entering the quantum algorithmic gate - QAG is considered as an information source both from the classical and the quantum level. The analysis of the classical and quantum information flow in Deutsch-Jozsa, Shor and Grover algorithms is used. It is shown that QAG, based on superposition of states, quantum entanglement and interference, when acting on the input vector, stores information into the system state, minimizing the gap between classical Shannon entropy and quantum von Neumann entropy. Minimizing of the gap between Shannon and von Neumann entropies is considered as a termination criterion of QA computational intelligence measure.

## 1. Information Analysis Axioms of Quantum Algorithm Dynamic Evolution.

Let us discuss the main properties of classical and quantum information that in dynamic analysis of quantum algorithms are used. Additional necessary detail description of general properties of information amounts in *Appendix 1* to this article is given.

Any computation (both classical and quantum) is formally identical to a communication in time. By considering quantum computation as a communication process, it is possible to relate its efficiency to its classical communication capacity. At time $t = 0$ , the programmer $(\underline{M})$ sets the computer to accomplish any one of several possible tasks. Each of these tasks can be regarded as embodying a different message. Another programmer $(\underline{C})$ can obtain this message by looking at the output of the computer when the computation is finished at time $t = T$ . Computation based on quantum principles allows for more efficient algorithms for solving certain problems than algorithms based on pure classical principles [1].

*Remark.* The sender conveys the maximum information when all the message states have equal *a priori* probability (which also maximizes the channel capacity). In that case the mutual information (channel capacity) at the end of the computation is $\log N$ .

The communication capacity gives an index of efficiency of a quantum computation [1]:
A necessary target of a quantum computation is to achieve the maximum possible communication capacity consistent with given initial states of the quantum computing.

Let us consider any peculiarities of information axioms and information capability of quantum computing as the dynamic evolution of QAs.

If one breaks down the general unitary transformation $U_i$ of a QA into a number of successive unitary blocks, then the maximum capacity may be achieved only after the number of applications of the blocks. In each of the smaller unitary blocks, the mutual information between the $\underline{M}$ and the $\underline{C}$ registers (i.e., the communication capacity) increases by a certain amount [1]. When its total value reaches the maximum possible value consistent with a given initial state of the quantum computing, the computation is regarded as being complete (see, in details [2,3]).

The classical capacity of a quantum communication channel is connected with the efficiency of quantum computing using *entropic* arguments [1-9]. This formalism allows us to derive lower bounds on the computational complexity of QA's in the most general context.

The following qualitative axiomatic descriptions of dynamic evolution of information flow in a QA are provided:

| N | **Axiomatic Rules** |
|---|---|
| 1 | The information amount (information content) of a successful result increases while the QA is in execution |
| 2 | 2.1. The quantity of information becomes the fitness function for the recognition of successful results on "intelligent states" and introduces the measures of accuracy and reliability (robustness) for successful results. |
| | 2.2. In this case the principle of Minimum of Classical / Quantum Entropy (MCQE) corresponds to recognition of successful results on "intelligent states" of the QA computation |
| 3 | If the classical entropy of the output vector is small, then the degree of order for this output state relatively larger, and the output of measurement process on "intelligent states" of a QA gives the necessary information to solve the initial problem with success |

*Remark.* These three information axioms mean that the algorithm can automatically guarantee convergence of information amount to a desired precision with a minimum decision-making risk. This is used to provide robust and stable results for fault-tolerant quantum computation. Main information measures in classical and quantum domains are shown in *Tables 1 and 2*. Physical meaning of "intelligent state" definition in *Appendix* is given.

*Remark.* Five main information-based approaches in optimal design of QA's computation can be used: 1) the maximum entropy (*ME*) principle; 2) minimum Fisher information (*MFI*) principle; 3) principle of extreme physical information (*EPI*); 4) principle of maximum of mutual information (*MMI*) between computational and measurement dynamic evolution (computational and memory registers) of *QA*'s; and 5) principle of maximal intelligence of *QA*'s based on minimum gap of difference between classical and quantum entropies in intelligent states of successful results. The first three principles (*ME*, *MFI*, and *EPI*) are based on the physical laws and may be derived through variation on appropriate Lagrangian's, and includes in last two principles according to relations between the classical and quantum entropies, and mutual information amount. The main properties of quantum information and entropy measures, and interrelations between information amounts (classical and quantum measures) are described in [1-20].

*Remark.* The *EPI* principle differs significantly from the *EM* approach or the *MFI* approach: (1) In its aims (establishing on ontology in *EPI* and eliciting the laws of physics from a consideration of the flow of information in the measurement process, *vs* subjectively estimating the laws in *ME* or *MFI*); (2) in its reason for extremization (conservation of information in *EPI*, *vs* arbitrary, subjective and sometimes inappropriate choice of «maximum smoothness» in *ME* and *MFI*); (3) how «constraints» are chosen (via the invariance of information to a symmetry operation principle in EPI *vs* arbitrary subjective choice in *ME* or *MFI*); and (4) in its solutions (to a differential equation in EPI and MFI, *vs* a solution, always in the form of an exponential of a function, in ME). Only *EPI* applies broadly to all of physics principles [5,6].

| Title | Classical $(\mathrm{Cl})$ | Quantum $(\mathrm{Q})$ |
|---|---|---|
| *Fisher* | $F^{Cl}(x)=\int \frac{1}{p(\xi\|x)}\left(\frac{\partial p(\xi\|x)}{\partial x}\right)^2 d\xi$ | $F^{Q}(x)=\int \frac{Tr\left[\left(\hat{E}(\xi)\frac{\partial \rho(x)}{\partial x}\right)\right]^2}{Tr\left(\hat{E}(\xi)\rho(x)\right)} d\xi$ |
| *Boltzman – Shannon* $\Downarrow$ *von Neumann* | $S^{Sh}=-\sum_i p_i \ln p_i$ | $S^{vN}=-Tr(\rho \ln \rho)$ |
| *Relative Information – Kullback – Leibler* | $I^{Cl}(p:q)=-\sum_i p_i \ln \frac{q_i}{p_i}$ | $I^{Q}(\rho:\sigma)=-Tr\left(\rho \ln \frac{\rho}{\sigma}\right)$ |
| *Renyi* | $S_q^{(Cl)R}=\frac{1}{1-q}\ln\left(\sum_{i=1}^{W} p_i^q\right)$ | $S_q^{(Q)R}=\frac{\ln\left[Tr\left(\rho^q\right)\right]}{1-q}$ |
| *Havrda & Charvat – Daròczy (Tsallis)* | $S_q^{(Cl)T}=\frac{\left(1-\sum_{i=1}^{W} p_i^q\right)}{1-q}$ | $S_q^{(Q)T}=\frac{\left(1-Tr\left(\rho^q\right)\right)}{1-q}$ |

*Table 1: Typical Measures of Information Amount.*

| |
|---|
| 1. *Shannon and von Neumann Entropy Relation* $S^{vN}\le S^{Sh}$ For Diagonal Density Matrix: $\rho=\rho_{ii}$ and $S^{vN}=S^{Sh}$ |
| 2. *Tsallis q-Entropy and Shannon Entropy Relation* $\lim_{q\to 1} S_q^{(Cl)T}=\lim_{q\to 1} S_q^{(Cl)R}=-\sum_{i=1}^{W} p_i \ln p_i$ |
| 3. *Tsallis q-Entropy and Renyi Entropy Relation* $S_q^{(Cl)R}=\frac{\ln\left[1+(1-q)S_q^{(Cl)T}\right]}{1-q}$ |

*Table 2: Relations between Different Typical Measures of Information Amounts.*

## 2. Information Intelligent Measure of QA's (principle 5).

The information QA intelligent measure as $I_T(\|\psi\rangle)$ of the state $\|\psi\rangle$ with respect to the qubits in $T$ and to the basis $B=\{\|i_1\rangle\otimes\cdots\otimes\|i_n\rangle\}$ is [10]

$$\mathcal{J}_T\left(|\psi\rangle\right) = 1 - \frac{S_T^{Sh}\left(|\psi\rangle\right) - S_T^{VN}\left(|\psi\rangle\right)}{|T|} \quad \textbf{(1)}$$

The measure (1) is minimal (i.e., 0) when $S_T^{Sh}\left(|\psi\rangle\right) = |T|$ and $S_T^{VN}\left(|\psi\rangle\right) = 0$, it is maximal (i.e., 1) when $S_T^{Sh}\left(|\psi\rangle\right) = S_T^{VN}\left(|\psi\rangle\right)$.

The intelligence of the QA state is maximal if the gap between the Shannon and the von Neumann entropy for the chosen result qubit is minimal. Information QA-intelligent measure (1) and interrelations between information measures in *Table 2* are used together with the step-by-step natural majorization principle for solution of QA-stopping problem.

Due to the presence of quantum entropy, QA cannot obviate Bellman's «*the curse of dimensionality*» encountered in solving many complex numerical and optimization problems. And finally, the stringent condition that quantum computers have to be interaction-free, leave them with little versatility and practical utility. It has been seen that large entanglement of the quantum register is a necessary condition for exponential speed-up in quantum computation. It is one of reasons to why the quantum paradigm is not so easy to extend to all the classical computational algorithms and also explain the failure of programmability and scalability in quantum speed-up.

*Remark*. To be concrete, a quantum registers such that the maximum Schmidt number (see below) of any bipartition is bounded at most by a polynomial in the size of the system can be simulated efficiently by classical means. The universality study of scaling of entanglement in Shor's factoring algorithm and in adiabatic QAs across a quantum phase transition for both the *NP*-complete Exact Cover problem (as a particular case of the 3-SAT problem) as well as the Grover's problem shows as following: (i) analytical result for Shor's QA's is a linear scaling of the entropy in terms of the number of qubits, therefore difficult the possibility of an efficient classical simulation protocol; (ii) a similar result is obtained numerically for the quantum adiabatic evolution Exact Cover algorithm, which also universality of the quantum phase transition the system evolves nearby; and (iii) entanglement in Grover's adiabatic QSA remains a bounded quantity even at the critical point. For these cases a classification of scaling of entanglement appears as a natural grading of the computational complexity of simulating quantum phase transitions.

## 3. Information Analysis and Measures of the Quantum Decision Making Algorithm Computational Intelligence.

Most of the applications of quantum information theory have been developed in the domain of quantum communications systems, in particular in quantum source coding, quantum data compressing and quantum error-correcting codes. In parallel, QA's have been studied as computational processes, concentrating attention on their dynamics and ignoring the information aspects involved in quantum computation. In the following section, the application of tools and techniques from quantum information theory in the domain QA's synthesis and simulation is described. For this purpose, the analysis of the classical and quantum information flow in Deutsch-Jozsa algorithm is used. It is shown that the quantum algorithmic gate (QAG) *G*, based on superposition of states, quantum entanglement and interference, when acting on the input vector, stores information into the system state, minimizing the gap between classical Shannon entropy and quantum von Neumann entropy.

This principle is fairly general, resulting in both a methodology to design a QAG and a technique to simulate (efficiently) its behavior on a classical computer.

The following disclosure uses classical and quantum correlations to describe QA computation. Classical correlations play a more prominent role than quantum correlations in the speed-up of certain

QAs.

### 3.1. Information Analysis of Deutsch Algorithm.

The advantage of quantum computing lies in the exploitation of the phenomenon of superposition. The great importance of the quantum theory of computation lies in the fact that it reveals the fundamental connections between the laws of physics and the nature of computation. There is a great simplification in understanding quantum computation: a quantum computer is formally equivalent to a multi-particle *Mach-Zender*-like interferometer. Deutsch's QA is a simple example that illustrates the advantages of quantum computation.

Deutsch's QA as discussed in [2,3] is based on the assumption that a binary function of a binary variable $f:\{0,1\}\to\{0,1\}$ is given. Thus, $f(0)$ can be either 0 or 1, and $f(1)$ likewise can be either 0 or 1, giving altogether four possibilities. The problem posed by Deutsch's QA is to determine whether the function is constant [i.e., $f(0)=f(1)$], or varying [i.e., $f(0)\neq f(1)$].

Deutsch poses the following task: by computing $f$ only once, determine whether it is constant or balanced. This kind of problem is generally referred to as a promise algorithm, because one property out of a certain number of properties is initially promised to hold, and the task is to determine computationally which one holds (see, *Fig. 1*).

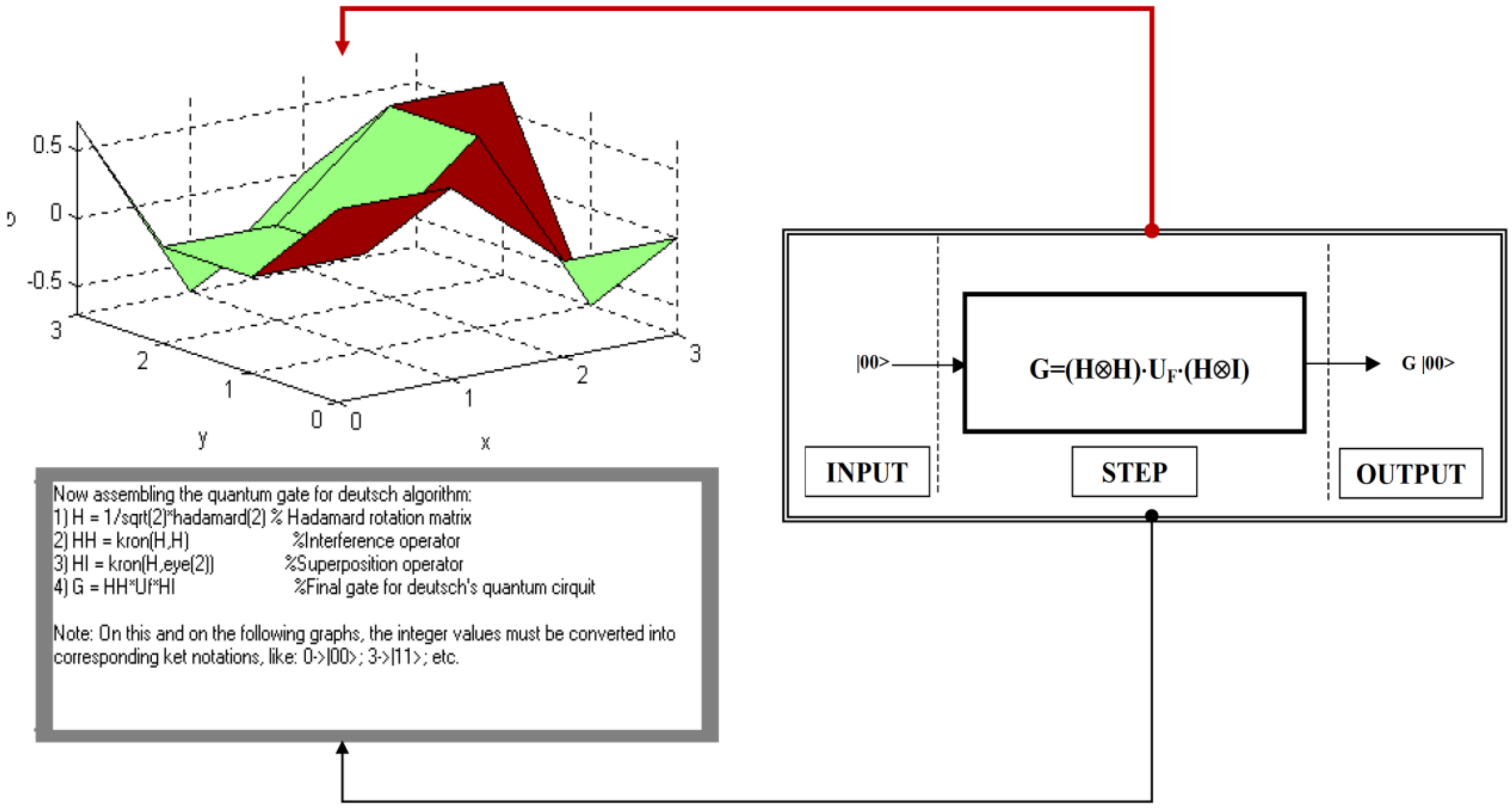

*Figure 1: Deutsch's quantum algorithm simulation - Quantum general gate assembling.*

Classically, finding out in one step whether a function is constant or balanced is clearly impossible. One would need to compute $f(0)$ and then compute $f(1)$ in order to compare them. There is no way out of this double evaluation. Quantum mechanically, however, there is a simple method for performing this task by computing $f$ only once. Two qubits are needed for the computation. In reality only one qubit is really needed, but the second qubit is there to implement the necessary transformation. Imagine that the first qubit is the input to the quantum computing whose internal Hardware (HW) part is represented by the second qubit.

The computational process itself will implement the following transformation on the two qubits (this is performed quantum mechanically, i.e., not using «classical» devices such as beam-splitters): $|x\rangle|y\rangle\to|x\rangle|y\oplus f(x)\rangle$, where $x$ is the input and $y$ is the HW. Note that this transformation is reversible and thus there is a unitary transformation to implement it (in the basic principle). The function $f$ has been used only once. The trick is to prepare the input in such a state to make use of quantum superposition.

The solution beings with the input $|x\rangle|y\rangle = (|0\rangle + |1\rangle)(|0\rangle - |1\rangle)$, where $|x\rangle$ is the actual input and $|y\rangle$ is part of the computing HW. Thus, before the transformation is implemented, the state of the computing is an equal superposition of all four basis states, which are obtained by simply expanding the state $|x\rangle|y\rangle$ as $|x\rangle|y\rangle = |\psi_{in}\rangle = |00\rangle - |01\rangle + |10\rangle - |11\rangle$.

Note that there are negative phase factors before the second and fourth terms.

When this state now undergoes the transformation in the abovementioned way $|x\rangle|y\rangle \to |x\rangle|y \oplus f(x)\rangle$, following output state is produced:

| | | |
|---|---|---|
| $\lvert\psi_{out}\rangle$ | $=$ | $\lvert 0 f(0)\rangle - \lvert 0 \bar{f}(0)\rangle + \lvert 1 f(1)\rangle - \lvert 1 \bar{f}(1)\rangle$ |
| | $=$ | $\lvert 0\rangle\left[\lvert f(0)\rangle - \lvert \bar{f}(0)\rangle\right] + \lvert 1\rangle\left[\lvert f(1)\rangle - \lvert \bar{f}(1)\rangle\right]$ |

,

where the over-bar indicates the opposite of that value, so that, for example, $\bar{0} = 1$.

The power of quantum computing is realized in that each of the components in the superposition of $|\psi_{in}\rangle$ underwent the same evolution «simultaneously» leading to the powerful «quantum parallelism». This feature is true for quantum computation in general.

The possibilities are:

| | |
|---|---|
| (1) | *If $f$ is constant then* |
| $\lvert\psi_{out}\rangle =$ | $(\lvert 0\rangle + \lvert 1\rangle)\left[\lvert f(0)\rangle - \lvert \bar{f}(0)\rangle\right]$ |
| (2) | *if $f$ is balanced then* |
| $\lvert\psi_{out}\rangle =$ | $(\lvert 0\rangle - \lvert 1\rangle)\left[\lvert f(0)\rangle - \lvert \bar{f}(0)\rangle\right]$ |

Note that the output qubit (in this case the first qubit) emerges in two different orthogonal states, depending on the type of function $f$. These two states can be distinguished with probability 1 of efficiency. A Hadamard transformation performed on this qubit leads to the state $|0\rangle$ if the function is constant and to the state $|1\rangle$ if the state function is balanced.

Now a single projective measurement in $\{|0\rangle, |1\rangle\}$ basis determines the type of the function.

Therefore, unlike their classical counterparts, quantum computing can solve Deutsch's problem.

The input could also be of the form $(|0\rangle - |1\rangle)(|0\rangle - |1\rangle) \equiv |-\rangle|-\rangle$. A constant function would then lead to the state $|-\rangle|-\rangle$ and a balanced function would lead to $|+\rangle|-\rangle$. So the $|+\rangle$ and $|-\rangle$ are equally good as input states of the first qubit and both lead to quantum speed-up. Their equal mixture, on the other hand, is not. This means that the output would be an equal mixture $\left[(|+\rangle\langle+|) + (|-\rangle\langle-|)\right]$ no matter whether $f(0) = f(1)$ or $f(0) \neq f(1)$, i.e., the two possibilities would be indistinguishable.

Thus, for the QA to work well, one needs the first register to be highly correlated to the two different types of functions. If the output state of the first qubit is $\rho_1$ then function is balanced. The efficiency of Deutsch's algorithm depends on distinguishing the two states $\rho_1$ and $\rho_2$. This is given by the Holevo bound, $I_{acc} = S(\rho) - \frac{1}{2}\left[S(\rho_1) + S(\rho_2)\right]$, where $\rho = \frac{1}{2}(\rho_1 + \rho_2)$. Therefore, if $\rho_1 = \rho_2$, then $I_{acc} = 0$ and the QA has no speed-up over the classical one. At the other extreme, if $\rho_1$ and $\rho_2$ are pure and orthogonal, then $I_{acc} = 1$ and the computation gives the right result in one step [7].

In between these two extremes lie all other computations with varying degrees of efficiency as quantified by the Holevo-bound. These are purely classical correlations and there is no entanglement between the first and the second qubit. In fact, the Holevo-bound is the same as the formula for classical correlations. The key to understanding the efficiency of Deutsch's algorithm is, therefore, through the mixedness of the first register.

If the initial state has the entropy of $S_0$, then the final Holevo-bound is $\Delta S = S(\rho) - S_0$.

So, the more mixed the first qubit, the less efficient the computation.

Note that the quantum mutual information between the first qubits is zero throughout the entire computation (so there are neither classical nor quantum correlations between them).

### 3.2. Information Analysis of QAG Dynamics and Intelligent Output States: Deutsch-Jozsa Algorithm

Deutsch-Jozsa algorithm's dynamics of quantum computation states are analyzed from classical and quantum information theory standpoint. Shannon entropy is interpreted as the degree of information accessibility through the measurement, and von Neumann entropy is employed to measure quantum correlation information of entanglement. A maximally intelligent state is defined as a QA successful computation output state with minimum gap between classical and quantum entropy. The Walsh-Hadamard transform creates maximally intelligent states for Deutsch-Jozsa's problem, since it annihilates the qubit gap between classical and quantum entropy for every state.

***3.2.1 Information Analysis of QAG Dynamics and Intelligent Output States: Deutsch - Jozsa Algorithm.*** In the Deutsch-Jozsa algorithm, an integer number $n > 0$ and a truth-function $f:\{0,1\}^n \to \{0,1\}$ are given such that $f$ is either constant or balanced (where $f$ is constant if it computes the same output for every input, it is balanced if it takes values 0 and 1 on $2^{n-1}$ input strings each.) The problem is to decide whether $f$ is constant or balanced. Function $f$ is encoded into a unitary operator $U_F$ corresponding to a squared matrix of order $2^{n+1}$, where $F:\{0,1\}^n \times \{0,1\} \to \{0,1\}^n \times \{0,1\}$ is an injective function such that:

$$F(x,y) = \left(x, y \oplus f(x)\right) \tag{2a}$$

($\oplus$ is the XOR operator) and $U_F$ is such that:

$$[U_F]_{i,j} = \delta_{i,1+\left[F\left([j-1]\right)_{(2)}\right]_{(10)}} \tag{2b}$$

($[r](b)$ is the basis $b$ representation of number $r$, $\delta_{i,j}$ is the Kronecker delta); $H$ denotes the unitary Walsh-Hadamard transform $H = \frac{1}{\sqrt{2}}\begin{pmatrix} 1 & 1 \\ 1 & -1 \end{pmatrix}$ and ${}^nH$ the $n-$power of matrix $H$ through tensor product. Operator $U_F$ is embedded into a more general unitary operator (like as on *Fig. 1*) called quantum algorithmic gate (QAG) as $G$

$$G = \left({}^nH \otimes I\right) \cdot U_F \cdot {}^{n+1}H \tag{3}$$

($I$ denote the identity matrix of order 2), which is applied to the input vector of dimension $2^{n+1}$. In this context, ${}^{n+1}H$ plays the role of the superposition operator (*Sup*), $U_F$ stands for the entanglement operator (*Ent*) and finally ${}^{n}H \otimes I$ is the interference operator (*Int*). The corresponding computation is described by the following steps:

| *Step* | *Computation Algorithm* | *Formula* |
|---|---|---|
| *Step*0 | $\lvert input\rangle = \lvert 0\rangle \otimes \lvert 0\rangle \otimes \cdots \otimes \lvert 0\rangle \otimes \lvert 1\rangle$ | $(4(a))$ |
| *Step*1 | $\lvert \psi_1\rangle = H_{n+1}\lvert input\rangle = \frac{1}{\sqrt{2^n}} \sum_{i_1,\ldots,i_n} \lvert i_1 \ldots i_n\rangle \otimes \frac{(\lvert 0\rangle - \lvert 1\rangle)}{\sqrt{2}}$ | $(4(b))$ |
| *Step*2 | $\lvert \psi_2\rangle = U_F\lvert \psi_1\rangle = \frac{1}{\sqrt{2^n}} \sum_{i_1,\ldots,i_n} \left(-1^{f(i_1,\ldots,i_n)}\right)\lvert i_1 \ldots i_n\rangle \otimes \frac{(\lvert 0\rangle - \lvert 1\rangle)}{\sqrt{2}}$ | $(4(c))$ |
| *Step*3 | $\lvert output\rangle = (H_n \otimes I)\lvert \psi_2\rangle = \sum_{j_1,\ldots,j_n} a_{j_1\ldots j_n} \lvert j_1 \ldots j_n\rangle \otimes \frac{(\lvert 0\rangle - \lvert 1\rangle)}{\sqrt{2}}$ | $(4(d))$ |

With

$$a_{j_1\cdots j_n} = \frac{1}{2^n} \sum_{i_1,\ldots,i_n} (-1)^{f(i_1,\ldots,i_n)} (-1)^{(i_1,\ldots,i_n)\cdot(j_1,\ldots,j_n)} \quad ,$$

where $(i_1,\ldots,i_n)\cdot(j_1,\ldots,j_n)$ denotes $(i_1 \wedge j_1)\oplus\ldots\oplus(i_n \wedge j_n)$, being $i_1,\ldots,i_n, j_1,\ldots,j_n \in \{0,1\}$. If $f$ is constant then $a_0 = 1$ and $a_j$ is null for all $j \neq 0$. If it is balanced function, then $a_0 = 0$. Therefore, if after performing measurement on vector $\lvert output\rangle$ a basis vector is obtained in the form

$$\lvert 0\rangle \otimes \lvert 0\rangle \otimes \cdots \otimes \lvert 0\rangle \otimes \underbrace{\lvert Measurement\ \ basis\rangle}_{\lvert 0\rangle}$$

or

$$\lvert 0\rangle \otimes \lvert 0\rangle \otimes \cdots \otimes \lvert 0\rangle \otimes \underbrace{\lvert Measurement\ \ basis\rangle}_{\lvert 1\rangle}$$

the $f$ is constant. Otherwise, it is balanced. For example, let $n = 3$ and $f_1, f_2$ be defined as in *Table 3*.

| $x \in \{0,1\}^3$ | $f_1(x)$ | $f_2(x)$ |
|---|---|---|
| 000 | 0 | 1 |
| 001 | 0 | 0 |
| 010 | 0 | 1 |
| 011 | 0 | 1 |
| 100 | 0 | 0 |
| 101 | 0 | 0 |
| 110 | 0 | 0 |
| 111 | 0 | 1 |

*Table 3: Example of constant and balanced functions.*

Consider two cases of quantum entanglement operators as

$$U_{F_1} = I_4 \ (Case\ 1) \quad \textbf{(5)}$$

and $U_{F_2}$ is written as a block matrix with $C = \begin{pmatrix} 0 & 1 \\ 1 & 0 \end{pmatrix}$ as follows:

$$U_{F_2} = \begin{pmatrix} C & 0 & 0 & 0 & 0 & 0 & 0 & 0 \\ 0 & I & 0 & 0 & 0 & 0 & 0 & 0 \\ 0 & 0 & C & 0 & 0 & 0 & 0 & 0 \\ 0 & 0 & 0 & C & 0 & 0 & 0 & 0 \\ 0 & 0 & 0 & 0 & I & 0 & 0 & 0 \\ 0 & 0 & 0 & 0 & 0 & I & 0 & 0 \\ 0 & 0 & 0 & 0 & 0 & 0 & I & 0 \\ 0 & 0 & 0 & 0 & 0 & 0 & 0 & C \end{pmatrix} \quad (Case\ 2) \qquad \textbf{(6)}$$

The computation involved by these two operators is resumed in *Table 4*.

| *Step* | *State*(*Case* 1) | *State*(*Case* 2) |
|---|---|---|
| *Input* | $\|000\rangle \otimes \|1\rangle$ | $\|000\rangle \otimes \|1\rangle$ |
| *Step* 1 | $\boxed{\left(\frac{\|0\rangle+\|1\rangle}{\sqrt{2}}\right) \otimes \left(\frac{\|0\rangle+\|1\rangle}{\sqrt{2}}\right) \otimes \left(\frac{\|0\rangle+\|1\rangle}{\sqrt{2}}\right)} \otimes \underbrace{\left(\frac{\|0\rangle-\|1\rangle}{\sqrt{2}}\right)}_{Ancilla\ qubit}$ | $\boxed{\left(\frac{\|0\rangle+\|1\rangle}{\sqrt{2}}\right) \otimes \left(\frac{\|0\rangle+\|1\rangle}{\sqrt{2}}\right) \otimes \left(\frac{\|0\rangle+\|1\rangle}{\sqrt{2}}\right)} \otimes \underbrace{\left(\frac{\|0\rangle-\|1\rangle}{\sqrt{2}}\right)}_{Ancilla\ qubit}$ |
| *Step* 2 | $\boxed{\left(\frac{\|0\rangle+\|1\rangle}{\sqrt{2}}\right) \otimes \left(\frac{\|0\rangle+\|1\rangle}{\sqrt{2}}\right) \otimes \left(\frac{\|0\rangle+\|1\rangle}{\sqrt{2}}\right)} \otimes \underbrace{\left(\frac{\|0\rangle-\|1\rangle}{\sqrt{2}}\right)}_{Ancilla\ qubit}$ | $\boxed{\frac{-\|000\rangle+\|001\rangle-\|010\rangle-\|011\rangle+\|100\rangle+\|101\rangle+\|110\rangle-\|111\rangle}{\sqrt{2^3}}} \otimes \underbrace{\left(\frac{\|0\rangle-\|1\rangle}{\sqrt{2}}\right)}_{Ancilla\ qubit}$ |
| *Step* 3 | $\boxed{\|000\rangle} \otimes \underbrace{\left(\frac{\|0\rangle-\|1\rangle}{\sqrt{2}}\right)}_{Ancilla\ qubit}$ | $\boxed{\frac{\|010\rangle-\|011\rangle-\|100\rangle-\|101\rangle}{\sqrt{2^3}}} \otimes \underbrace{\left(\frac{\|0\rangle-\|1\rangle}{\sqrt{2}}\right)}_{Ancilla\ qubit}$ |

*Table 4: Deutsch-Jozsa QAG state dynamic.*

***3.2.2 Shannon and von Neumann entropy.*** A vector in a Hilbert space of dimension $2^k$ acts as a classical information source if the measurement with respect to a given orthonormal basis is performed. The possible outputs are the $2^k$ basis vectors, each one with probability given by the squared modulus of its probability amplitude. More in general, given a vector:

$$|\varphi\rangle = \sum_{i_1,\ldots,i_n \in \{0,1\}} a_{i_1,\ldots,i_n} |i_1\rangle \otimes \ldots \otimes |i_n\rangle \qquad \textbf{(7)}$$

in a Hilbert space of dimension $2^n$.

Let $T = \{j_1, \ldots, j_k\} \subseteq \{1, \ldots, n\}$ and $\{1, \ldots, n\} - T = \{l_1, \ldots, l_{n-k}\}$.

Define

$$|\psi\rangle\langle\psi|_T = \sum_{\substack{t_{j_1},\ldots,t_{j_k} \\ i_{j_1},\ldots,i_{j_k}}} b_{t_{j_1},\ldots,t_{j_k}}^{t_{j_1},\ldots,t_{j_k}} |i_{j_1},\ldots,i_{j_k}\rangle\langle t_{j_1},\ldots,t_{j_k}|, \qquad \textbf{(8)}$$

where

$$b_{t_{j_1},\ldots,t_{j_k}}^{t_{j_1},\ldots,t_{j_k}} = \sum_{i_{l_1}=t_{l_1} \ldots i_{l_{n-k}}=t_{l_{n-k}}} a_{i_1,\ldots,i_n} a^*_{t_1,\ldots,t_n} \qquad \textbf{(9)}$$

Choosing $T$ means selecting a subspace of the Hilbert space of $|\varphi\rangle$ in Eq. (7). If $T=\{j\}$, this subspace has dimension 2 and it is the subspace of the qubit *j*. Similarly, if $T=\{j_1,\dots j_k\}$, the subspace of qubits is $j_1,\dots,j_k$. The density operator $|\psi\rangle\langle\psi|_T$ describes the projection of the density matrix corresponding to $|\psi\rangle$ on this subspace.

Define the Shannon entropy of $T$ in $|\psi\rangle$ with respect to the basis $\mathcal{B}=\{|i_1\rangle\otimes\dots\otimes|i_n\rangle\}$ as

$$E_T^{Sh}\left(|\psi\rangle\right)=-\sum_{i=1}^{2^k}\left[|\psi\rangle\langle\psi|_T\right]_{i,i}\log_2\left[|\psi\rangle\langle\psi|_T\right]_{i,i} \tag{10}$$

The Shannon entropy of $T$ expresses the mean information gained by measuring the projection of $|\psi\rangle$ with respect to the projections of the vectors in $\mathcal{B}$ on the subspace of the qubits in $T$. The Shannon entropy can be interpreted as the degree of disorder involved by vector $|\psi\rangle$ when the qubits in $T$ are measured. Vector $|\psi\rangle$ does not act only as a classical information source. On the contrary, it stores also information in a non-local correlation that is through entanglement. In order to measure the quantity of entanglement of a set $T=\{j_1,\dots,j_k\}$ of qubits in $|\psi\rangle$ the Von Neumann entropy of $T$ in $|\psi\rangle$ as following

$$E_T^{vN}=-Tr\left(|\psi\rangle\langle\psi|_T\log_2|\psi\rangle\langle\psi|_T\right) \tag{11}$$

is used.

The von Neumann entropy of the qubits in $T$ is interpreted as the measure of the degree of entanglement of these qubits with the rest of the system.

***3.2.3 Information analysis of Deutsch-Jozsa QAG.*** Before ${}^{n+1}H$ is applied, the input vector defined in Eq. (7) is such that for every qubit $j$ as follows:

$$E_{\{j\}}^{Sh}\left(|input\rangle\right)=E_{\{j\}}^{vN}\left(|input\rangle\right)=0 \tag{12}$$

Eq. (12) is easily proved by observing that

$$|input\rangle\langle input|_{\{j\}}=|0\rangle\langle 0| \tag{13}$$

for the first $n$ qubits and

$$|input\rangle\langle input|_{\{n+1\}}=|1\rangle\langle 1| \tag{14}$$

Since $\log_2 1=0$ and both $\log_2|0\rangle\langle 0|$ and $\log_2|1\rangle\langle 1|$ correspond to the null squared matrix of order 2, the values for Shannon and von Neumann entropy are 0 for every qubit from Eq. (10) and Eq. (11).

When ${}^{n+1}H$ is applied [*Step 1*, Eq. (4 (b))], every qubit undergoes a unitary change of basis through the operator $H$. This means

$$|\psi_1\rangle\langle\psi_1|_{\{j\}}=H^{-1}\left(|input\rangle\langle input|_{\{j\}}\right)H \tag{15}$$

Eq. (15) can be rewritten as

$$\forall j\in\{1,\dots,n\}:|\psi_1\rangle\langle\psi_1|_{\{j\}}=\frac{1}{2}\begin{pmatrix}1&1\\1&1\end{pmatrix} \tag{16}$$

and

$$|\psi_1\rangle\langle\psi_1|_{\{n+1\}}=\frac{1}{2}\begin{pmatrix}1&-1\\-1&1\end{pmatrix}, \tag{17}$$

since it is known that von Neumann entropy is left unchanged by a unitary change of basis, then

$$E^{Sh}_{\{j\}}\left(|\psi_1\rangle\right)=1,\quad E^{vN}_{\{j\}}\left(|\psi_1\rangle\right)=0 \tag{18}$$

for all qubits $j$.

The application of the operator $U_F$ [*Step 2*, Eq. (4 (c))] leaves the situation unchanged for qubit $n+1$, whereas it entangles the first qubits:

$$\forall j\in\{1,\dots,n\}:|\psi_2\rangle\langle\psi_2|_{\{j\}}=\frac{1}{2}\begin{pmatrix}1 & \alpha_j\\ \alpha_j & 1\end{pmatrix} \tag{19}$$

where

$$\alpha_j=\sum_{i_1,\dots,i_{j-1},i_{j+1},\dots,i_n}\frac{(-1)^{\sum_{i_j} f(i_1,\dots,i_n)}}{2^{n-1}}. \tag{20}$$

From the structure of the matrix in Eq. (19) the elements on the main diagonal are the same as in Eq. (16). This means the Shannon entropy has not changed. Moreover,

$$H^{-1}\left(|\psi_2\rangle\langle\psi_2|_{\{j\}}\right)H=\frac{1}{2}\begin{pmatrix}1+\alpha_j & 0\\ 0 & 1-\alpha_j\end{pmatrix}. \tag{21}$$

Since von Neumann entropy is left unchanged by a unitary change, for the first $n$ qubits

$$E^{Sh}_{\{j\}}\left(|\psi_2\rangle\right)=1 \tag{22}$$

and

$$E^{vN}_{\{j\}}\left(|\psi_2\rangle\right)=\frac{1+\alpha_j}{2}\log_2\frac{2}{1+\alpha_j}+\frac{1-\alpha_j}{2}\log_2\frac{2}{1-\alpha_j}. \tag{23}$$

Finally, when ${}^nH\otimes I$ is applied [*Step 3*, Eq. (4(d)], qubit $n+1$ is left unchanged again, whereas all other qubits undergo a unitary change of basis again through operator $H$. From Eq. (21), the Shannon and Von Neumann entropy are calculated as:

$$E^{Sh}_{\{j\}}\left(|output\rangle\right)=E^{vN}_{\{j\}}\left(|output\rangle\right)=\frac{1+\alpha_j}{2}\log_2\frac{2}{1+\alpha_j}+\frac{1-\alpha_j}{2}\log_2\frac{2}{1-\alpha_j}. \tag{24}$$

Since in general (see, *Table 2*)

$$E^{Sh}_{\{j\}}\left(|output\rangle\right)\geq E^{vN}_{\{j\}}\left(|output\rangle\right), \tag{25}$$

the action of $H_n\otimes I$ is to preserve the von Neumann entropy and to reduce the Shannon entropy of the first $n$ qubits as much as possible. The two operators represented in *Table 4* produce the information flow shown in *Table 5*.

| Step | Case 1 | | Case 2 | |
|---|---|---|---|---|
| | $E^{Sh}_{\{j\}}(\lvert\psi\rangle)$ | $E^{vN}_{\{j\}}(\lvert\psi\rangle)$ | $E^{Sh}_{\{j\}}(\lvert\psi\rangle)$ | $E^{vN}_{\{j\}}(\lvert\psi\rangle)$ |
| Input | 0, 0, 0 | 0, 0, 0 | 0, 0, 0 | 0, 0, 0 |
| Step 1 | 1, 1, 1 | 0, 0, 0 | 1, 1, 1 | 0, 0, 0 |
| Step 2 | 1, 1, 1 | 0, 0, 0 | 1, 1, 1 | 1, 1, 1 |
| Step 3 | 0, 0, 0 | 0, 0, 0 | 1, 1, 1 | 1, 1, 1 |

*Table 5: Deutsch-Jozsa QAG information flow* $(1\leq j\leq 3)$.

In Case 2, the von Neumann entropy is maximal for every qubits and, therefore, it is not possible for the interference operator to reduce the Shannon entropy.

From this analysis, the following conclusions can be drawn:

| | |
|---|---|
| 1 | When the QA computation starts, the Shannon entropy coincides with the Von Neumann entropy, but they are both null. |
| 2 | The superposition operator increases the Shannon entropy of each qubit to its maximum, but leaves the von Neumann entropy unchanged. |
| 3 | The entanglement operator increases the von Neumann entropy of each qubit according to the property of the function $f$ , but leaves the Shannon entropy unchanged. |
| 4 | The interference operator does not change the value of the von Neumann entropy introduced by the entanglement operator, but decreases the value of the Shannon entropy to its minimum, that is, to the value of the von Neumann entropy itself. |

***3.2.4 The information intelligent measure.*** The measure $\Im_T(|\psi\rangle)$ of the state $|\psi\rangle$ with respect to the qubits in T and to the basis $\mathcal{B} = \{|i_1\rangle \otimes \ldots \otimes |i_n\rangle\}$ is:

$$\Im_T(|\psi\rangle) = 1 - \frac{S_T^{Sh}(|\psi\rangle) - S_T^{VN}(|\psi\rangle)}{|T|}. \quad \textbf{(26)}$$

The intelligence of the QA state is maximal if the gap between the Shannon and the von Neumann entropy in Eq. (26) for the chosen resultant qubit is minimal. Information QA-intelligent measure $\Im_T(|\psi\rangle)$ and interrelations between information measures $S_T^{Sh}(|\psi\rangle) \geq S_T^{VN}(|\psi\rangle)$ are used together with entropic relations of the step-by-step natural majorization principle for solution of the QA-termination problem. From Eq. (26) it can be seen that for pure states

$$\max \Im_T(|\psi\rangle) \mapsto 1 - \min\left(\frac{S_T^{Sh}(|\psi\rangle) - S_T^{VN}(|\psi\rangle)}{|T|}\right) \mapsto \min S_T^{Sh}(|\psi\rangle), \quad S_T^{VN}(|\psi\rangle) = 0, \quad \textbf{(27)}$$

From Eq. (27) the principle of Shannon entropy minimum is described as follows. Calculation of the Shannon entropy is applied to the quantum state after the interference operation.

The minimum of Shannon entropy in Eq. (27) corresponds to the state when there are few state vectors with high probability (states with minimum uncertainty are intelligent states, see *Appendix*). Selection of an appropriate termination condition is important since QAs are periodical.

$$H = -\sum_{i=0}^{2^{m+n}} p_i \log p_i . \quad \textbf{(28)}$$

The rules of quantum measurement ensure that only one state will be detected each time.

If the algorithm works perfectly, the marked state orbital is revealed with unit efficiency, and the entropy drops to zero. Otherwise, unmarked orbitals may occasionally be detected by mistake. The entropy reduction can be calculated from the probability distribution, using Eq. (28). The minimum Shannon entropy criteria is used for successful termination of Grover's QSA and realized in this case in digital circuit implementation.

***3.2.5 Design of intelligent quantum state.*** Let us consider one of possible approach to optimal criteria choice of priority state extraction from designed superposition of possible coding states (for example, coefficient gains of fuzzy PID-controller). For this goal we apply the definition of "*intelligent quantum state*" as the state with minimal uncertainty (minimum in Heisenberg inequality uncertainty). This definition is correlated with solutions of quantum wave equations (Schrödinger-like etc.) when wave function of quantum system is described coherent state and uncertainty elation have global minimum (see, in details *Appendix 2*). Definition and computation of intelligent state in QA are described in [9] using the definition of von Neumann entropy and Shannon information entropy in this quantum state. According to the definition [9] "*intelligent quantum state*" is the minimum of difference between Shannon information entropy and physical entropy von Neumann on this quantum state:

$$\mathbb{I}\left(\left|\text{Quantum state}\right\rangle\right)=\min\left(H^{Sh}-S^{vN}\right), \qquad \textbf{(29)}$$

where $H^{Sh}$ and $S^{vN}$ are Shannon and von Neumann entropies, correspondingly.

According to quantum information law (see, in details [2,5-20]) we have the following inequality:

$$H^{Sh}\geq S^{vN}\ \text{, i.e., } \mathbb{I}\left(\left|\text{Quantum state}\right\rangle\right)\geq 0 .$$

*Remark.* In quantum mechanics according to Born's rule probability $p$ of quantum system state equal to amplitude probability $\psi$ in degree 2: $p=|\psi|^2$ that strong discussed last time in physics. From quantum information viewpoint pure quantum state have zero value of von Neumann entropy. Thus "*intelligent quantum state*" (29) in QA corresponds to minimum of Shannon information entropy of quantum state. This minimum is achieved for maximum probability state (according to definition of Shannon information entropy $H^{Sh}=-\sum_i p_i \ln p_i$ , i.e., global minimum observed for maximum probability $p_i$). While $p=|\psi|^2$, i.e., amplitude probability in degree 2, thus principle of amplitude probability maximum at correlated state can be used as priority selection criteria of "intelligent" correlated (coherent) state in superposition of possible candidates. *Figure 2* shows the selection of the intelligent state by maximum probability amplitude.

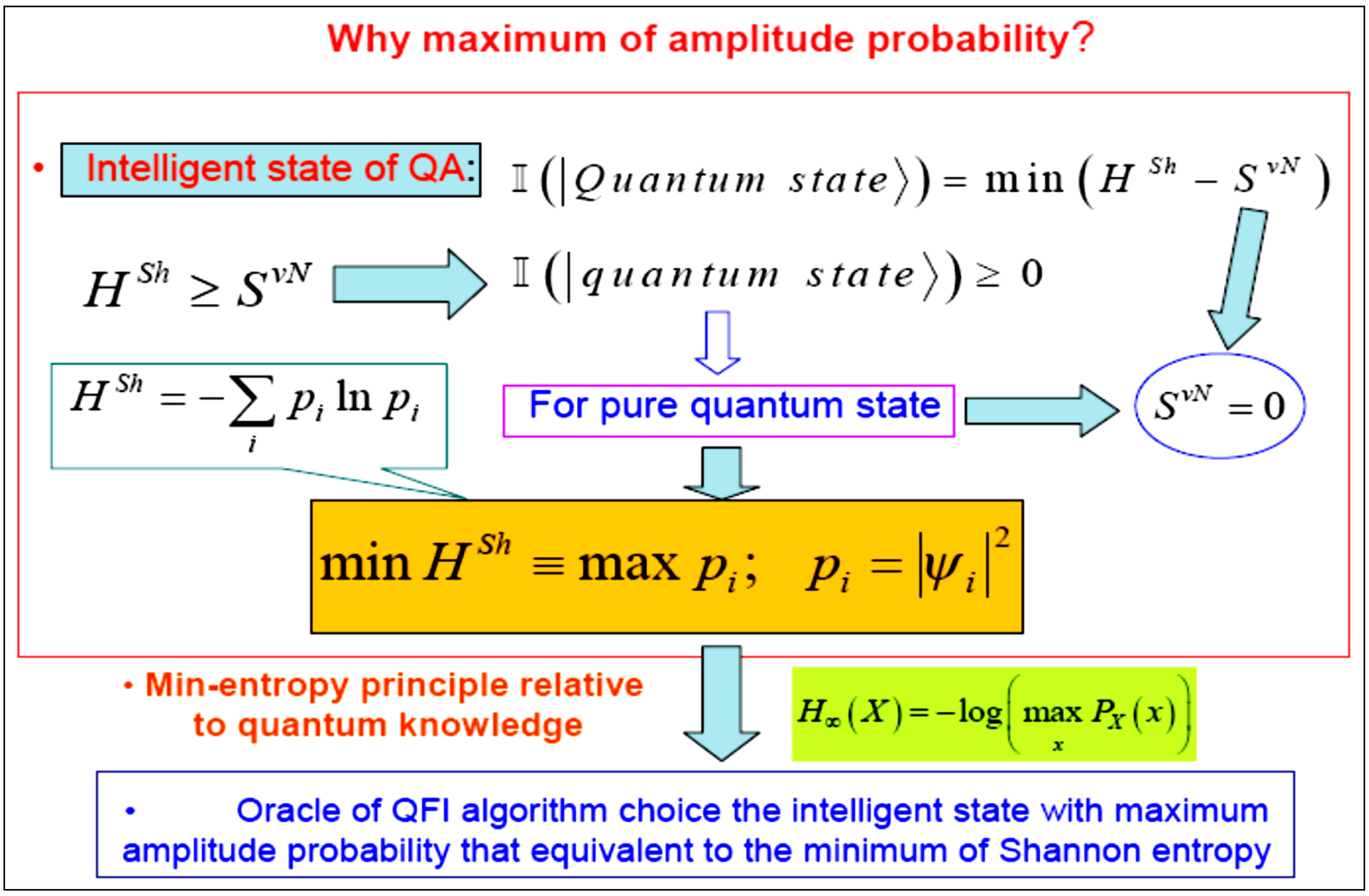

*Figure 2: The selection of the intelligent state by maximum probability amplitude.*

This Box plays the role of the interference. Therefore, *quantum oracle* model is realized with calculation of amplitude probabilities in superposition state with mixed types of quantum correlations and selection state with maximum amplitude probability. According to definition, quantum oracle has full necessary information about solution. Extraction of this information (as analog of interference) is realized together of decoding operation. In our case decoding operation is the standard procedure (inner product in Hilbert space).

***3.2.6 Intelligent output states of Deutsch-Jozsa algorithm.*** The von Neumann entropy is interpreted as the degree of information in a vector (describing the property of function $f$ ), namely as a measure of the information stored in quantum correlation about the function $f$ . The Shannon entropy must be interpreted as the measure of the degree of inaccessibility to this information through the measurement. In this context, the QAG $G$ of Eq. (3) transfers information from $f$ into the output vector minimizing the quantity of unnecessary noise producible by the measurement, or, more technically, minimizing the non-negative quantity:

$$N_{\{j\}}\left(|output\rangle\right) = E_{\{j\}}^{Sh}\left(|output\rangle\right) - E_{\{j\}}^{vN}\left(|output\rangle\right)$$

for the first $n$ qubits. The measure of intelligence of an output state according to the definition in Eq. (1) is

$$\mathcal{J}\left(|output\rangle\right) = 1 - \left\langle N\left(|output\rangle\right)\right\rangle \qquad \textbf{(30)}$$

where

$$\left\langle N\left(|output\rangle\right)\right\rangle = \frac{1}{n}\sum_{j\in\{1,\ldots,n\}} N_{\{j\}}\left(|output\rangle\right) \qquad \textbf{(31)}$$

is the mean unnecessary noise. According to this definition, the action of the Walsh-Hadamard transform in Deutsch-Jozsa algorithm is to associate to every possible function $f$ a maximally intelligent output state, namely a state $|output\rangle$ such that $\mathcal{J}\left(|output\rangle\right) = 1$.

Physically, the measure of intelligent QA, described by Eq. (30), characterizes the amount of value information necessary for decision making regarding successful solution of the QA.

*Figures 3 (a)* and *3 (b)* show the measure of intelligent QA for two cases in Eqs (5) and (6) according to Eq. (30).

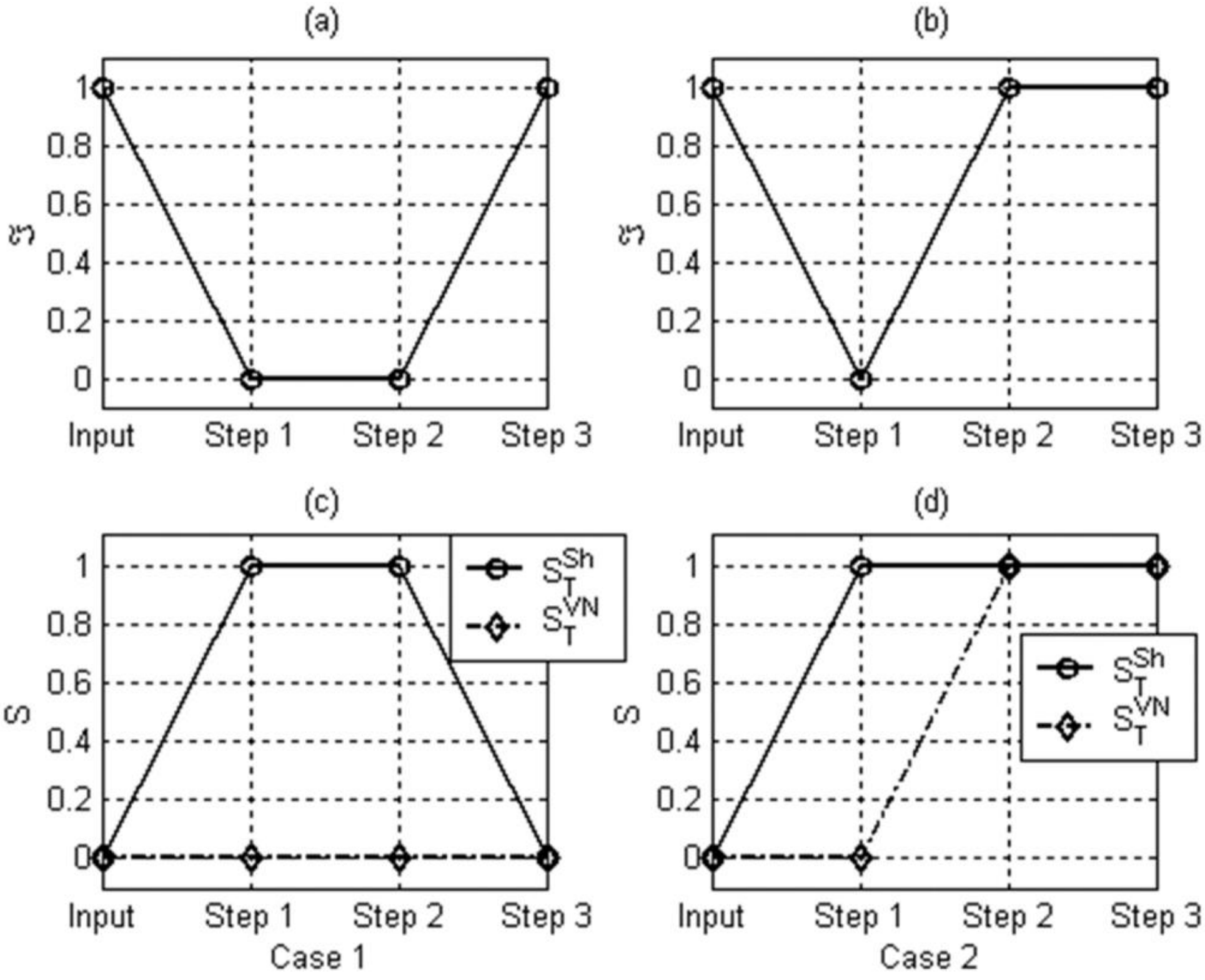

*Figure 3: Measure of intelligence for Deutsch-Jozsa's QA for two cases from Table 5.*

In both cases this measure has the maximal value, i.e., equal to 1. It means that Deutsch-Jozsa QA

as decision-making algorithm is intelligent QA with maximum degree 1. From Eq. (29), although optimality is fulfilled for the intelligence of the output states, the von Neumann entropy is still a lower bound for the Shannon entropy (see *Figs 3 (c, d)*). If the von Neumann entropy is too high, even a maximally intelligent output state may be «too random» when the measurement is performed. Therefore, in Deutsch-Jozsa algorithm the information transfer from $f$ into the output state is done by paying attention that the quantity of entanglement does not exceed. The role of «entanglement controller» is played by the pair «superposition-entanglement» operators.

In order to illustrate this concept, consider the following matrix:

$$U_F = \begin{pmatrix} C & 0 & 0 & 0 \\ 0 & I & 0 & 0 \\ 0 & 0 & I & 0 \\ 0 & 0 & 0 & C \end{pmatrix} \quad \textbf{(32)}$$

which encodes a balanced function for $n = 2$. The von Neumann entropy of the first $n$ qubits, when $U_F$ is applied, is from Eq. (23)

$$E_{\{1\}}^{vN}\left(|\psi_2\rangle\right) = E_{\{2\}}^{vN}\left(|\psi_2\rangle\right) = \boxed{0} \quad \textbf{(33)}$$

If the role of the superposition was played by the operator ${}^{n}H \otimes I$ instead of ${}^{n+1}H$ and the input vector of dimension $n+1$ was

$$|input\rangle = |0\rangle \otimes |0\rangle \otimes \ldots \otimes |0\rangle \otimes |0\rangle \quad \textbf{(34)}$$

then the von Neumann entropy for the first $n$ qubits after the same step would be

$$E_{\{1\}}^{vN}\left(|\psi_2\rangle\right) = E_{\{2\}}^{vN}\left(|\psi_2\rangle\right) = \boxed{1} \quad \textbf{(35)}$$

And, therefore, the Shannon entropy could not be reduced by the interference operator. The output would be completely random and the algorithm would not work.

Now consider similar examples with $n = 3$ and $f_1, f_2$ be defined as in Table 3 for different entanglement operators that illustrate additional properties of information flows from the Deutsch-Jozsa QA. Let $n = 3$ and $f_1, f_2$ be defined as in Table 3 and consider two additional cases of quantum entanglement operators as

$$U_{F_3} = \begin{pmatrix} I & 0 & 0 & 0 & 0 & 0 & 0 & 0 \\ 0 & C & 0 & 0 & 0 & 0 & 0 & 0 \\ 0 & 0 & I & 0 & 0 & 0 & 0 & 0 \\ 0 & 0 & 0 & I & 0 & 0 & 0 & 0 \\ 0 & 0 & 0 & 0 & I & 0 & 0 & 0 \\ 0 & 0 & 0 & 0 & 0 & C & 0 & 0 \\ 0 & 0 & 0 & 0 & 0 & 0 & C & 0 \\ 0 & 0 & 0 & 0 & 0 & 0 & 0 & C \end{pmatrix} (Case\ 3) \quad \textbf{(36)}$$

and $U_{F_2}$ is written as a block matrix as follows:

$$U_{F_4} = I_2 \otimes \begin{pmatrix} I & 0 \\ 0 & C \end{pmatrix} (Case\ 4) \quad \textbf{(37)}$$

*Tables 6* and *7* show the dynamics of the Deutsch-Jozsa algorithm for the Cases 3 and 4, correspondingly.

| Step | Classical Entropy $H_{Sh}(\|\psi\rangle)$ | Quantum Entropy $S_{\|\psi\rangle}(\{j\})$ | States and Gate Operations | Dynamics of Solution Probabilities |
|---|---|---|---|---|
| Input | 0 | 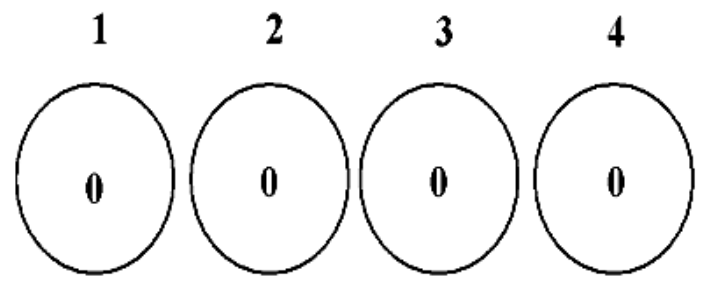 | $\|000\rangle \otimes \|1\rangle$ <br> ${}^4H$ | 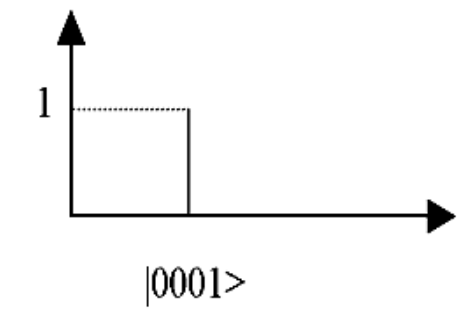 |
| Super-position | (1,1,1,1)<br>4 | 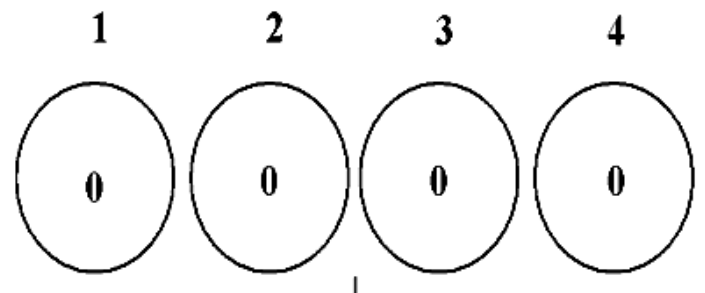 | $\frac{\|0\rangle+\|1\rangle}{\sqrt{2}} \otimes \frac{\|0\rangle+\|1\rangle}{\sqrt{2}} \otimes \frac{\|0\rangle+\|1\rangle}{\sqrt{2}} \otimes \frac{\|0\rangle-\|1\rangle}{\sqrt{2}}$ <br> $U_F$ | 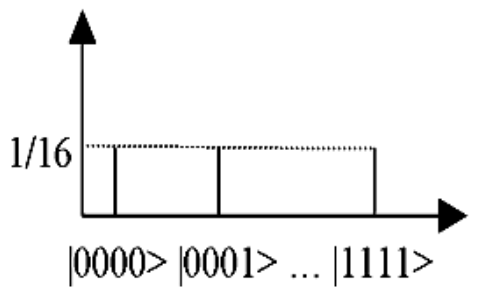 |
| Entangle-ment | (1,1,1,1)<br>4 | 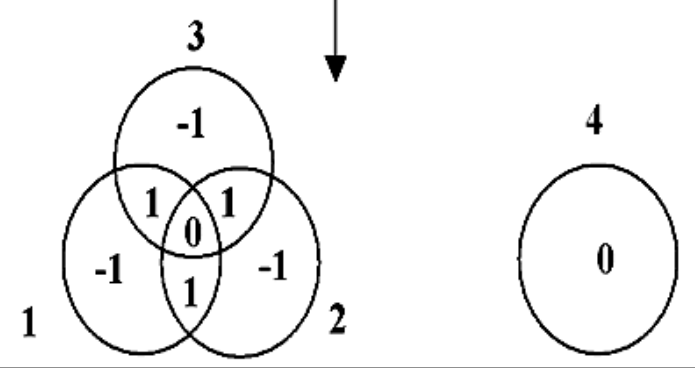 | $\frac{1}{2\sqrt{2}}\begin{pmatrix}\|000\rangle-\|001\rangle+\|010\rangle+ \\ \|011\rangle+\|100\rangle-\|101\rangle- \\ -\|110\rangle-\|111\rangle\end{pmatrix} \otimes \frac{\|0\rangle-\|1\rangle}{\sqrt{2}}$ <br> ${}^3H \otimes I$ | 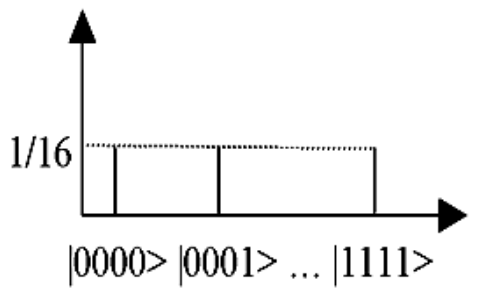 |
| Inter-ference | (1,1,1,0)<br>3 | 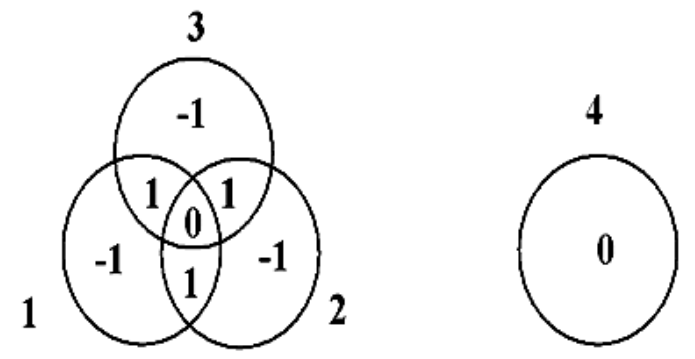 | $\frac{1}{2}\begin{pmatrix}\|001\rangle+\|011\rangle+ \\ +\|100\rangle-\|110\rangle\end{pmatrix} \otimes \frac{\|0\rangle-\|1\rangle}{\sqrt{2}}$ | 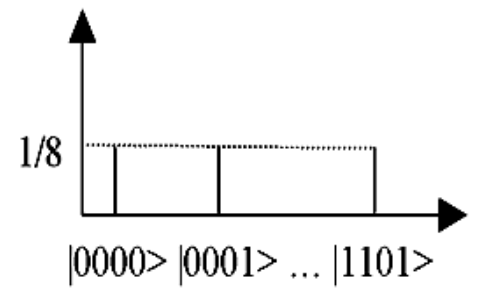 |

*Table 6: Information Analysis of Deutsch-Jozsa's Algorithm for case 3.*

| Step | Classical Entropy $H_{Sh}(\|\psi\rangle)$ | Quantum Entropy $S_{\|\psi\rangle}(\{j\})$ | States and Gate Operations | Dynamics of Solution Probabilities |
|---|---|---|---|---|
| Input | 0 | 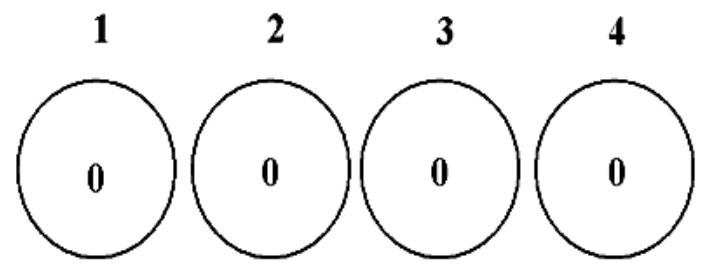 | $\|000\rangle \otimes \|1\rangle$ <br> ${}^4H$ | 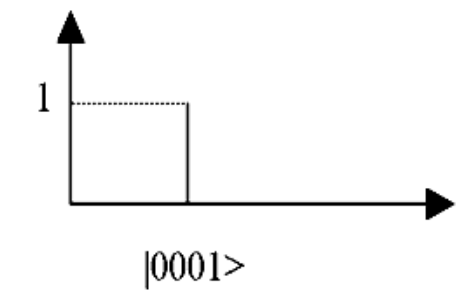 |
| Super-position | (1,1,1,1)<br>4 | 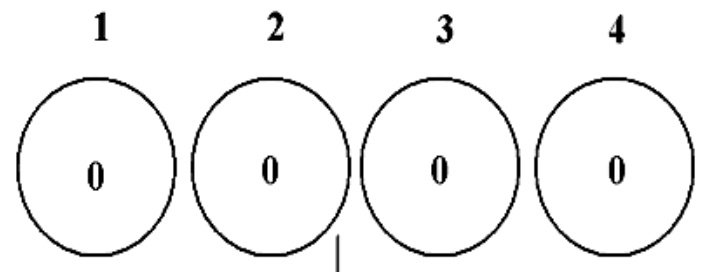 | $\frac{\|0\rangle+\|1\rangle}{\sqrt{2}} \otimes \frac{\|0\rangle+\|1\rangle}{\sqrt{2}} \otimes \frac{\|0\rangle+\|1\rangle}{\sqrt{2}} \otimes \frac{\|0\rangle-\|1\rangle}{\sqrt{2}}$ <br> $U_F$ | 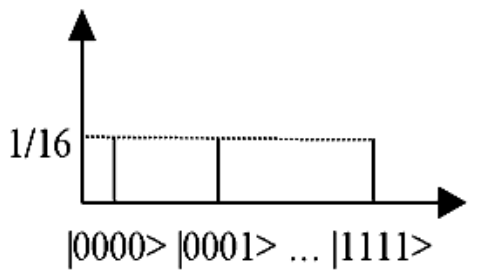 |
| Entan-glement | (1,1,1,1)<br>4 | 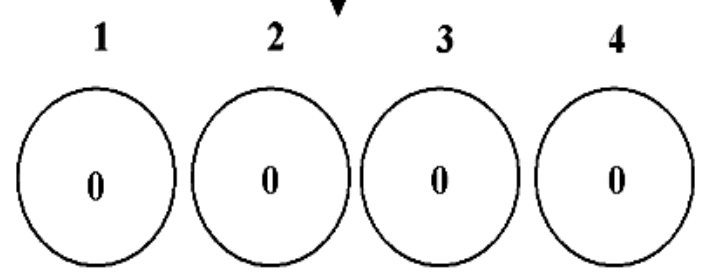 | $\frac{\|0\rangle+\|1\rangle}{\sqrt{2}} \otimes \frac{\|0\rangle+\|1\rangle}{\sqrt{2}} \otimes \frac{\|0\rangle-\|1\rangle}{\sqrt{2}} \otimes \frac{\|0\rangle-\|1\rangle}{\sqrt{2}}$ <br> ${}^3H \otimes I$ | 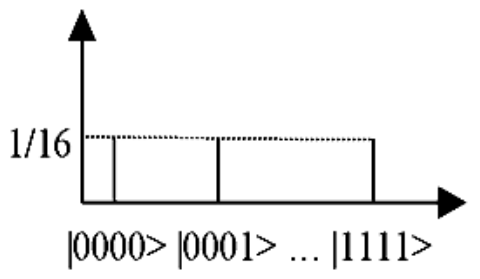 |
| Inter-ference | (0,0,1,0)<br>1 | 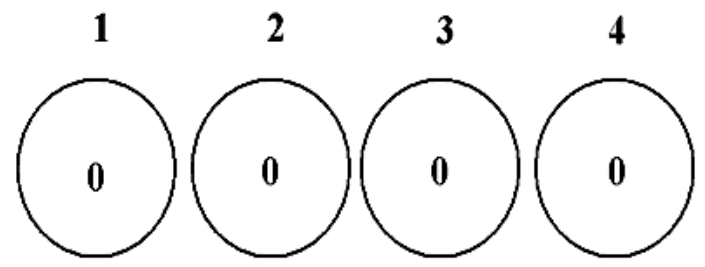 | $\|001\rangle \otimes \frac{\|0\rangle-\|1\rangle}{\sqrt{2}}$ | 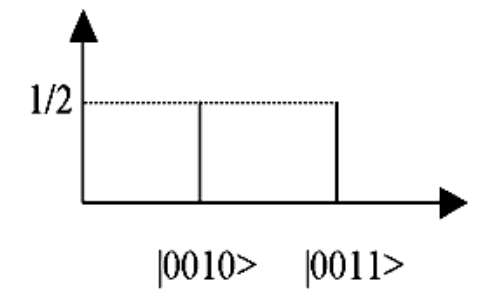 |

*Table 7: Information Analysis of Deutsch-Jozsa's Algorithm for case 4.*

For comparison of results *Table 8* shows the results of calculation for the Case 1.

| Step | Classical Entropy $H_{Sh}(\|\psi\rangle)$ | Quantum Entropy $S_{\|\psi\rangle}(\{j\})$ | States and Gate Operations | Dynamics of Solution Probabilities |
|---|---|---|---|---|
| Input | 0 | 1: 0, 2: 0, 3: 0, 4: 0 | $\|000\rangle\otimes\|1\rangle$ → ${}^4H$ | 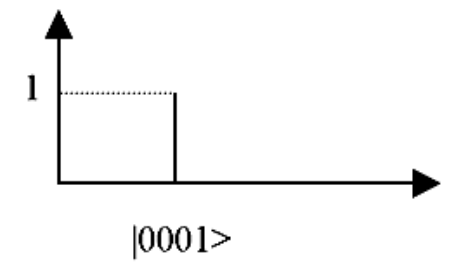 |
| Super-position | (1,1,1,1)<br>4 | 1: 0, 2: 0, 3: 0, 4: 0 | $\frac{\|0\rangle+\|1\rangle}{\sqrt{2}}\otimes\frac{\|0\rangle+\|1\rangle}{\sqrt{2}}\otimes\frac{\|0\rangle+\|1\rangle}{\sqrt{2}}\otimes\frac{\|0\rangle-\|1\rangle}{\sqrt{2}}$ → $U_F$ | 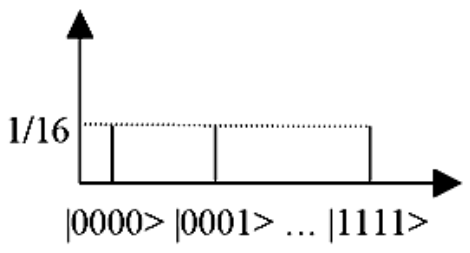 |
| Entan-glement | (1,1,1,1)<br>4 | 1: 0, 2: 0, 3: 0, 4: 0 | $\frac{\|0\rangle+\|1\rangle}{\sqrt{2}}\otimes\frac{\|0\rangle+\|1\rangle}{\sqrt{2}}\otimes\frac{\|0\rangle+\|1\rangle}{\sqrt{2}}\otimes\frac{\|0\rangle-\|1\rangle}{\sqrt{2}}$ → ${}^3H\otimes I$ | 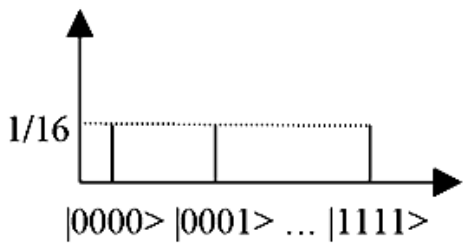 |
| Inter-ference | (0,0,0,0)<br>0 | 1: 0, 2: 0, 3: 0, 4: 0 | $\|000\rangle\otimes\frac{\|0\rangle-\|1\rangle}{\sqrt{2}}$ | 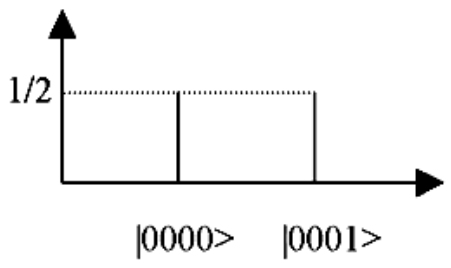 |

*Table 8: Information Analysis of Deutsch-Jozsa's Algorithm for case 1.*

By applying the above-mentioned information analysis to these three tables, one can draw the following physical interpretation of results and conclusions about classical and quantum entropy, changing after superposition, entanglement and interference have taken place:

| | | |
|---|---|---|
| 1 | | The input vector is a basis vector, therefore, the classical information of this state is 0; it is the tensor product of $n$ basis vectors of dimension 2, so the von Neumann entropy of every qubits composing it is also 0. |
| 2 | | The superposition operator $H_4$ increase the classical Shannon entropy its minimal value 0 to its maximal value 4, but leaves the situation unchanged from the quantum von Neumann entropy point of view. |
| 3 | | The entanglement operator is a classical unitary operator therefore it maps different basis vectors into different basis vectors leaving the classical information on the system unchanged. On the contrary it may create correlation among the different binary vectors in the tensor product describing the system state; this correlation is described by the von Neumann entropy of the different subparts of the system: |
| | 3.$a$ | The quantum information of the whole system is always 0, even when entanglement operator creates correlation, since the vector describing it is a pure state, whereas inner values for mutual information and conditional entropy may be positive or negative: they encode the quantum information necessary to decode the property being investigated for operator of entanglement $U_F$ . |
| | 3.$b$ | The states of the system before and after action of the entanglement operator takes place cannot be distinguished from a classical information point of view, since the Shannon entropy did not change. Only with a quantum information approach is the difference between these two states can be revealed. |
| 4 | | The interference operator leaves the quantum information picture unchanged maintaining encoded the information necessary to identify $U_F$ as a constant or balanced operator. On the contrary, it decreases the classical entropy making quantum information accessible; through the action of interference the vector acquires the minimum of classical entropy: such a vector according to the definition is an intelligent state because it represents a coherent output of QA computation with minimum entropy uncertainty relation (EUR) as successful result. |

A comparison of *Tables 6* and 7 reveals that:

- The entanglement operator in Case 3 effectively creates quantum correlation among different subparts of the system, whereas in Case 4 the general state is written as the tensor product of binary basis vectors and so no quantum correlation is involved;
- The interference operator in Case 3 reduces the classical Shannon entropy of 1 bit, whereas in Case 4 it reduces it of 3 bits.

This explains why it is important to choose carefully both the superposition operator and the input vector in order to store in quantum correlation the information actually needed to solve the problem.

The following general properties are observed for the Deutsch-Jozsa algorithm:

| | |
|---|---|
| 1 | The presence of quantum correlation appears as the degree of resistance (immunity) of the system to change its classical entropy, as the measure of state chaos and defines the internal degree of intelligent possibility of QA |
| 2 | The action of interference undergoes this property mapping $U_F$ into an intelligent state revealing it |

Now consider from *Table 8* the results of simulation for Case 1. In this situation, the entanglement operator creates no correlation. This is a common characteristic to all linear operators $U_F$ implementing a function $f:\{0,1\}^n \to \{0,1\}^m$ such that $f(x) = k \cdot x$ or $f(x) = \neg(k \cdot x)$ for some binary constant $k$, as it showed in *Table 9*.

| Step | Classical Entropy $H_{Sh}(\lvert\psi\rangle)$ | Quantum Entropy $S_{\lvert\psi\rangle}(\{j\})$ | States and Gate Operations | Dynamics of Solution Probabilities |
|---|---|---|---|---|
| Input | 0 | 1 … $2^n$-1 $2^n$: 0 … 0 0 | $\lvert 0 \ldots 0\rangle \otimes \lvert 1\rangle$; ${}^{n+1}H$ | 1; $\lvert 0\ldots 01\rangle$ |
| Super-position | $n+1$ | 1 … $2^n$-1 $2^n$: 0 … 0 0 | $\frac{\lvert 0\rangle+\lvert 1\rangle}{\sqrt{2}} \otimes \ldots \otimes \frac{\lvert 0\rangle+\lvert 1\rangle}{\sqrt{2}} \otimes \frac{\lvert 0\rangle-\lvert 1\rangle}{\sqrt{2}}$; $U_F$ | $1/2^n$; $\lvert 0\ldots 0\rangle$ $\lvert 0\ldots 1\rangle$ … $\lvert 1\ldots 1\rangle$ |
| Entan-glement | $n+1$ | 1 … $2^n$-1 $2^n$: 0 … 0 0 | $\frac{\lvert 0\rangle\pm\lvert 1\rangle}{\sqrt{2}} \otimes \ldots \otimes \frac{\lvert 0\rangle\pm\lvert 1\rangle}{\sqrt{2}} \otimes \frac{\lvert 0\rangle-\lvert 1\rangle}{\sqrt{2}}$; ${}^{n}H \otimes I$ | $1/2^n$; $\lvert 0\ldots 0\rangle$ $\lvert 0\ldots 1\rangle$ … $\lvert 1\ldots 1\rangle$ |
| Inter-ference | 1 | 1 … $2^n$-1 $2^n$: 0 … 0 0 | $\lvert k\rangle \otimes \frac{\lvert 0\rangle-\lvert 1\rangle}{\sqrt{2}}$ | 1/2; $\lvert k\,0\rangle$ $\lvert k\,1\rangle$ |

*Table 9: Information Analyses of Deutsch - Jozsa's Algorithm (Linear Functions).*

These functions among the input set of balanced and constant minimize to 0 the «gap» between the highest and lowest information values appearing in the Wenn - diagram of shown in the tables. Other balanced functions are mapped into less intelligent states that are higher classical entropy vectors.

This means that it is a non-success result as it is shown in *Table 10*.

| Step | Classical Entropy $H_{Sh}(\|\psi\rangle)$ | Quantum Entropy $S_{\|\psi\rangle}(\{j\})$ | States and Gate Operations | Dynamics of Solution Probabilities |
|---|---|---|---|---|
| Input | 0 | 1: 0 … $2^n$-1: 0, $2^n$: 0 | $\|0...0\rangle\otimes\|1\rangle$ ; $^{n+1}H$ | 1 ; \|0…01> |
| Super-position | $n+1$ | 1: 0 … $2^n$-1: 0, $2^n$: 0 | $\frac{\|0\rangle+\|1\rangle}{\sqrt{2}}\otimes...\otimes\frac{\|0\rangle+\|1\rangle}{\sqrt{2}}\otimes\frac{\|0\rangle-\|1\rangle}{\sqrt{2}}$ ; $U_F$ | $1/2^n$ ; \|0…0> \|0…1> … \|1…1> |
| Entan-glement | $n+1$ | 1: 1 … $2^n$-1: 1, $2^n$: 1 | $\|\psi\rangle\otimes\frac{\|0\rangle-\|1\rangle}{\sqrt{2}}$ ; $^nH\otimes I$ | $1/2^n$ ; \|0…0> \|0…1> … \|1…1> |
| Inter-ference | $S^{Sh}>1$ | 1: 0 … $2^n$-1: 0, $2^n$: 0 | $\|\mu\rangle\otimes\frac{\|0\rangle-\|1\rangle}{\sqrt{2}}$ | r ; \|0…0> \|0…1> … \|1…1> |

*Table 10: Information Analysis of Deutsch-Jozsa's Algorithm (Non-Linear Balanced Functions).*

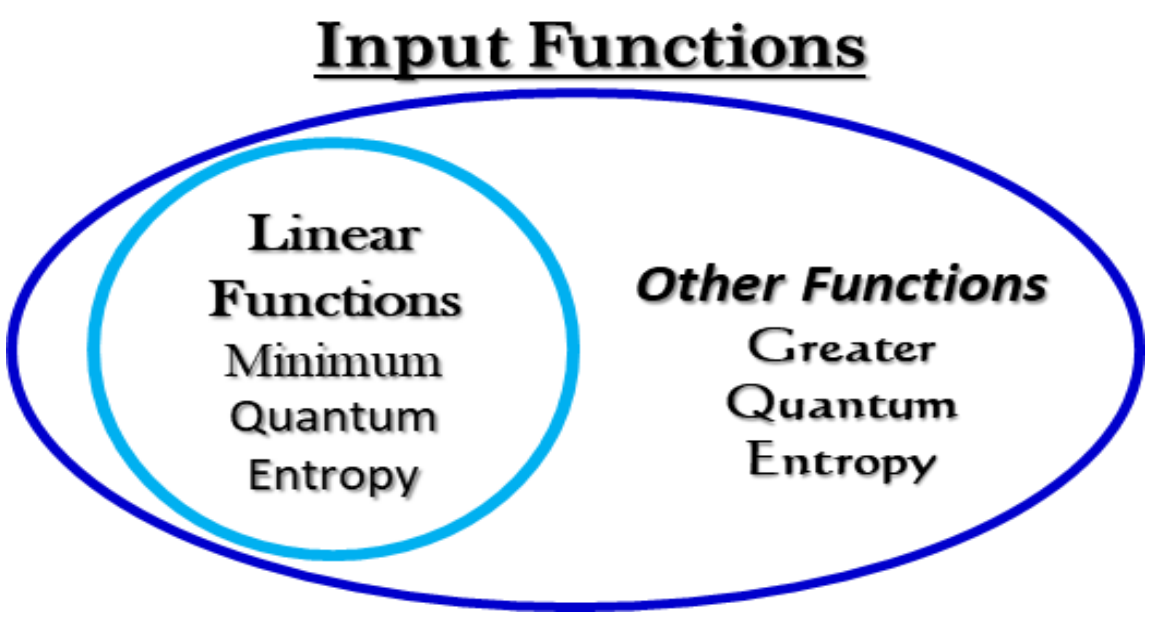

*Figure 4: Quantum Information Structure of Deutsch-Jozsa's Input Space.*

On the analysis level of the Deutsch-Jozsa algorithm, the next step is to investigate the relationship between the von Neumann entropy introduced by the entanglement operator and the problem the algorithm solves. It is possible to quantify the information about a given function that is sufficient to store in the system in order to decide if it is constant or balanced.

More in general, the same information analysis carried on for Deutsch-Jozsa algorithm should be done for the other QA's benchmarks in order to have a global picture of the known quantum information processing techniques.

## 4. Simulation results of QA-termination problem solution based on principle of Shannon/von Neumann minimum entropy.

From the step-by-step majorization principle of QA's it follows that for efficient termination of QA's that give the highest probability of successful result, the Shannon entropy was minimal for the step $m+1$. This is the principle of minimum Shannon entropy for termination of a QA with the successful result. This result also follows from the principle of QA maximum intelligent state as in Eq. (1). For this case according to Eq. (1)

$$\max \mathcal{J}_T(|\psi\rangle) = 1 - \min \frac{H_T^{Sh}(|\psi\rangle)}{|T|}, \qquad . \qquad \textbf{(38)}$$

$$S_T^{vN}(|\psi\rangle) = 0 \qquad \text{(for pure quantum state)}$$

Thus, the principle of maximal intelligence of QA's include as particular case the principle of minimum Shannon entropy for QA-termination problem solution.

### 4.1. QA-termination problem solving based on minimum Shannon/von Neumann dynamic simulation entropy.

Consider the complete basis vector $|A\rangle = a_1|0\ldots0\rangle + a_2|0\ldots1\rangle + \ldots + a_{2^n}|1\ldots1\rangle$. For this case, the Shannon entropy is:

$$S^{Sh}(|A\rangle) = -\sum_{i=1}^{2^n} a_i^2 \log a_i^2, \quad S^{vN}(|A\rangle) = -\sum_{i=1}^{2^n} \lambda_i\left(a_i^T a_i\right)\log \lambda_i\left(a_i^T a_i\right) \qquad \textbf{(39)}$$

where $\lambda_i\left(a_i^T a_i\right)$ are eigenvalues of quantum state $|A\rangle$.

Using the decomposition of quantum state vector $|A\rangle$ with the selection of measurement and calculation:

$$\begin{aligned}|A\rangle &= a_1|0\ldots0\rangle|0\rangle + a_2|0\ldots0\rangle|1\rangle + \ldots a_{2^{n-1}}|1\ldots1\rangle|0\rangle + a_{2^n}|1\ldots1\rangle|1\rangle \\ &= \left[a_1|0\ldots0\rangle + a_3|0\ldots1\rangle + \ldots + a_{2^{n-1}}|1\ldots1\rangle\right]|0\rangle \\ &+ \left[a_2|0\ldots0\rangle + a_4|0\ldots1\rangle + \ldots + a_{2^n}|1\ldots1\rangle\right]|1\rangle\end{aligned}$$

or

$$|A\rangle = |A\rangle_{|0\rangle} + |A\rangle_{|1\rangle}. \qquad \textbf{(40)}$$

Then the partial entropy can be calculated as:

| | | | | | |
|---|---|---|---|---|---|
| $S^{Sh}_{\vert 0\rangle}$ | = | $-\sum_i a_i^2 \log a_i^2$ | $S^{vN}_{\vert 0\rangle}$ | = | $-\sum_i \lambda_i\left(a_i^T a_i\right)\log \lambda_i\left(a_i^T a_i\right)$ |
| $S^{Sh}_{\vert 1\rangle}$ | = | $-\sum_{j\neq i} a_j^2 \log a_j^2$ | $S^{vN}_{\vert 1\rangle}$ | = | $-\sum_i \lambda_i\left(a_i^T a_i\right)\log \lambda_i\left(a_i^T a_i\right)$ |

(41)

where $i = 1,3,\ldots,2^{n-1}, \quad j = 2,4,\ldots,2^n$.

In the more general case of Shor's QA, $i = \left[i_0, i_0 + 2^n, \ldots, 2^{2n} - i_0 + 1\right]$, where $i = 1$ for the state vector $|A\rangle_{|0\ldots0\rangle}$ and $i = 2^n$ for the state vector $|A\rangle_{|1\ldots1\rangle}$.

*Figures 5 (a) — (d)* show simulation results of dynamic behavior for Shannon and von Neumann entropies for termination of Deutsch and Deutsch-Jozsa, QAs respectively.

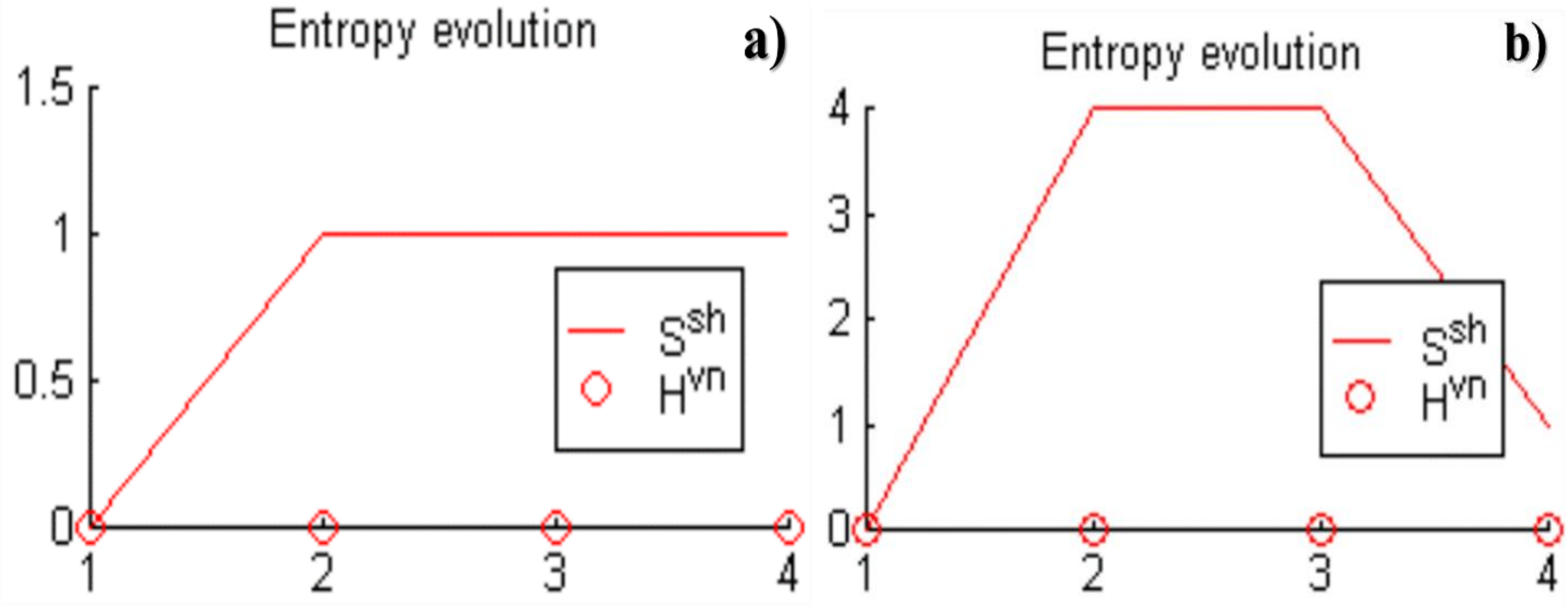

*Figure 5 (a, b): Simulation results of dynamics behavior for Shannon and von Neumann entropies of Deutsch's and Deutsch - Jozsa's QA.*

The complex vector entering the quantum algorithmic gate - QAG is considered as an information source both from the classical and the quantum level.

# 5. Quantum Search Algorithms: Information Entropy Analysis of Dynamic Flows and Min - Entropy Termination Problem Solution

The technique from quantum information theory can be used in the domain of quantum algorithm (QA) synthesis and simulation. For this purpose, the classical and quantum information flow in the Shor's and Grover's QA is analyzed. The quantum algorithmic gate (QAG) $G$, which is based on superposition, quantum super correlation (entanglement) and interference, when acting on the input vector, stores information into the system state, and in this case (similar to Deutsch-Jozsa QA case) minimizing the gap between the classical Shannon entropy and the quantum von Neumann entropy. This principle provides both a methodology to design a QG and a technique to efficiently simulate its behavior on a classical computer. Methodologically, the principle of maximal intelligence on output states for synthesizing QAGs is used. In general, the joint action of the superposition operator and of entanglement operator is to introduce the information necessary to solve the problem in the system quantum correlation. the principle of maximal intelligence of QA's include as particular case the principle of minimum Shannon entropy for QA-termination problem solution. From the application standpoint, the existence of a measure for the intelligence degree of a state allows the combination of QA techniques for encoding functions with some other computational methods, such as genetic algorithms. In this context, the measure of unnecessary noise becomes a fitness function in order to measure the desirability of a result. Thus, quantum computing provides a way of processing information that can be used in the classical problem-solving domain. One of the most radical changes that is being pursued in the last 20 years, is a change in the way that calculations are implemented physically. Rather than trying to avoid the quantum mechanical effects that manufacturers of processors are faced with, researchers in the field of quantum computing try to harness these effects. Computers that are based on these quantum mechanical laws are collectively referred to as quantum computers, as opposed to classical computers that rely on the classical laws of physics. One of the most important problem is sophisticated information dynamic flow analysis of quantum algorithm models. For example, quantum search algorithms as Shor's and Grover's algorithms that from information theory point view in this article are considered.

## 5.1. Information analysis of Shor's quantum search algorithm: The information role of entanglement and interference in Shor's QAG.

In Shor's algorithm, the dynamics of quantum computation states are analyzed from the classical and quantum information points of view. The Shannon entropy is interpreted (as mentioned above) as the degree of information accessibility through measurement, while the von Neumann entropy is employed to measure the quantum information of entanglement. The intelligence of a state with respect to a subset of qubits is defined. Similar to abovementioned results with Deutsch-Jozsa QA, the intelligence of a state is maximal if the gap between the Shannon and the von Neumann entropy for the chosen result qubits is minimal. The quantum Fourier transform (QFT) creates maximally intelligent states with respect to the first $n$ qubits for Shor's problem, since it annihilates the gap between the classical and quantum entropies for the first $n$ qubits of every output states.

***5.1.1. Computational dynamics of Shor's QA-gate (QAG).*** In the Shor's algorithm an integer number $n > 0$ and a function $f:\{0,1\}^n \to \{0,1\}^n$ are given such that $f$ have period *r*, namely $f$ is such that $\forall x \in \{0,1\}^n : f(x) \equiv f(x+r) \bmod n$, and $f$ is injective with respect to its period. The problem is to find *r*. The function $f$ is first encoded into the injective function

$$F:\{0,1\}^n \times \{0,1\}^n \to \{0,1\}^n \times \{0,1\}^n$$

such that:

$$F\{x,y\} = (x, y \oplus f(x)) \quad \textbf{(42)}$$

where $\oplus$ is the bitwise XOR operator and $F$ is then encoded into a unitary operator $U_F$. This purpose is fulfilled by mapping every binary input string of length $2n$ into a vector in a Hilbert space of dimension $2^{2n}$ according to the following recursive encoding scheme $\tau$:

$$\tau(0) = |0\rangle \quad \tau(0) = |0\rangle \quad \tau(z_1 \dots z_{2n}) = |z_1 \dots z_{2n}\rangle \quad \textbf{(43)}$$

where $|z_1 \dots z_{2n}\rangle$ stands for the tensor product $|z_1\rangle \otimes \dots \otimes |z_{2n}\rangle$, being $z_1, \dots, z_{2n} \in \{0,1\}$. Each bit value in a string is mapped into a vector of a Hilbert sub-space of dimension 2. This subspace is called a qubit. Similarly, a sequence *f* successive bit values of length $l$ is mapped into a vector of dimension $2^l$. This subspace is a quantum register of length $l$. Using this scheme, the operator $U_F$ is defined as a squared matrix of order $2^{2n}$ such that:

$$\forall z \in \{0,1\}^{2n} : U_F |z\rangle = |F(z)\rangle . \quad \textbf{(44)}$$

Practically, the operator $U_F$ is as follows:

$$[U_F]_{i,j} = \delta_{i, 1+\left[F\left([j-1]_{(2)}\right)\right]_{(10)}} \quad \textbf{(45)}$$

where $[q]_{(b)}$ is the basis $b$ representation of number $q$ and $\delta_{i,j}$ is the Kronecker delta. The idea of encoding a function $f$ into unitary operator is not a peculiarity of the Shor's algorithm, but it is typical of all known QA's. In general, $U_F$ contains the information about the function $f$ needed to solve the problem.

In the Shor's case, one could calculate the period $r$ of $f$ by testing the operator $U_F$ on the input vectors $\underbrace{|0 \dots 0\rangle}_{n} \otimes \underbrace{|0 \dots 0\rangle}_{n}$ obtaining $\underbrace{|0 \dots 0\rangle}_{n} \otimes \underbrace{|f(0 \dots 0)\rangle}_{n}$, $\underbrace{|0 \dots 1\rangle}_{n} \otimes \underbrace{|0 \dots 0\rangle}_{n}$ obtaining $\underbrace{|0 \dots 1\rangle}_{n} \otimes \underbrace{|f(0 \dots 1)\rangle}_{n}$ and so on until a vector $\underbrace{|x_1 \dots x_n\rangle}_{n}$ for the first register of length $n$ is found such that the corresponding vector $\underbrace{|f(x_1 \dots x_n)\rangle}_{n}$ in the second register of length $n$ coincides with $\underbrace{|f(0 \dots 0)\rangle}_{n}$. The period $r$ of $f$ coincides with the binary number $x_1 \dots x_n$. The number of $U_F$ tests required by this algorithm is $r$ [2].

Since the period $r$ of the function varies from 1 to $2^n$, the temporal complexity of this algorithm is exponential for the worst case. In order to extract the information stored in $U_F$ more effectively, a different perspective is used. The operator $U_F$ must in fact be used in order to transfer as much information as possible from operator to the input vector each time $U_F$ works. To this purpose, it is embedded into a unitary operator $G$, called the quantum algorithmic gate (*QAG*), having the following general form [10]:

$$G = (IF \otimes I_m)\cdot U_F \cdot (IF \otimes I_m) \quad \textbf{(46)}$$

where $IF$ stands for a unitary squared matrix of order $2^n$ and $I_m$ for identity matrix of order $2^n$. In the case of the Shor's algorithm, $U_F$ is embedded into the unitary QAG

$$G = (QFT_n \otimes I_n)\cdot U_F \cdot (QFT_n \otimes I_n) \quad \textbf{(47)}$$

The symbol $QFT_n$ denotes the unitary quantum Fourier transform of order $n$:

$$[QFT_n]_{ij} = \frac{1}{2^{n/2}} \exp\left\{2\pi J \left[\frac{(i-1)(j-1)}{2^n}\right]\right\}, \quad \textbf{(48)}$$

where $J$ is the imaginary unit. Subsequently, the gate does not act on many different basis input vectors. On the contrary, it always gets as input the starting vector $\underbrace{|0\ldots0\rangle}_{n} \otimes \underbrace{|0\ldots0\rangle}_{n}$. The corresponding computation evolves according to the following steps [10]:

| Step | Computational algorithm | Equation |
|---|---|---|
| *Step 0* | $\|input\rangle = \underbrace{\|0\ldots0\rangle}_{n} \otimes \underbrace{\|0\ldots0\rangle}_{n}$ | **(49)** |
| *Step 1* | $\|\psi_1\rangle = (QFT_n \otimes I_n)\|input\rangle = \frac{1}{2^{n/2}} \sum_{i_1,\ldots,i_n} \|i_1\ldots i_n\rangle \otimes \underbrace{\|0\ldots0\rangle}_{n}$ | **(50)** |
| *Step 2* | $\|\psi_2\rangle = U_F\|\psi_1\rangle = \frac{1}{2^{n/2}} \sum_{i_1,\ldots,i_n} \|i_1\ldots i_n\rangle \otimes \underbrace{\|f(i_1\ldots i_n)\rangle}_{n}$ | **(51)** |
| *Step 3* | $\|output\rangle = (QFT_n \otimes I_n)\|\psi_2\rangle = \frac{1}{2^{n/2}} \sum_{\substack{j_1,\ldots,j_n \\ i_1,\ldots,i_n}} a_{i_1\ldots i_n}^{j_1\ldots j_n} \|j_1\ldots j_n\rangle \otimes \underbrace{\|f(i_1\ldots i_n)\rangle}_{n}$ | **(52)** |

Where:

$$a_{i_1\ldots i_n}^{j_1\ldots j_n} = \frac{1}{2^{n/2}} \exp\left\{2\pi J \left[\frac{(i_1,\ldots i_n)_{(10)}(j_1,\ldots j_n)}{2^n}\right]\right\}.$$

If $k = \frac{2^n}{r}$ is an integer number, the output state can be written as

$$|output\rangle = \frac{1}{r}\sum_{p=0}^{r-1}\sum_{s=0}^{r-1} \exp\left\{2\pi Js[i_1\ldots i_n]_{(10)} \frac{l_p}{r}\right\} \left|\left[\frac{s2^n}{r}\right]_{(2)}\right\rangle \otimes \left|[p]_{(2)}\right\rangle, \quad \textbf{(53)}$$

where $l_p$ is an integer positive number and binary representation are obtained using $n$ bits. Therefore, the first quantum register of length $n$ of the output state generates a periodical probability distribution with period $k$ for every possible vector of the second register. By repeating the algorithm a number of times polynomial in $n$ and by performing a measurement each time, one can reconstruct the value of $r$.

***5.1.2. Physical interpretation of Shor's algorithm steps.*** In Step 1 the operator

$\left(QFT_n \otimes {}^nI\right)$ acts on a basis vector. It transforms the vector source $\underbrace{|0\ldots0\rangle}_{n} \otimes \underbrace{|0\ldots0\rangle}_{n}$ into a linear combination of equally weighted basis vectors of the form $|i_1\ldots i_n\rangle \otimes \underbrace{|0\ldots0\rangle}_{n}$. Since every basis vector is interpreted as a possible observable state of the system, the $QFT_n$ plays the role of the superposition operator for the first $n$ qubits. In Step 3, the operator $\left(QFT_n \otimes {}^nI\right)$ acts on every basis vector belonging to the linear combination $|\psi_2\rangle$. This means that every vector of such a combination generates a superposition of basis vectors, whose complex weights (i.e., amplitudes of probability) are equal in modulus, but different in phase. Every basis vector is now weighted by the summation of the probability amplitudes coming from the different source vectors. This summation can increase or decrease the resulting amplitude of probability. Since this phenomenon is very similar to classical wave interference, in Step 3, the operator $QFT_n$ plays the role of the interference operator.

From the mathematical point of view, when the matrix $QFT_n$ acts as a superposition operator (Step 1), the first matrix column only is involved in the calculation of the resulting vector. On the contrary, when it acts for the second time (Step 3), all matrix columns are involved and the interference among the weights coming from the different source vectors takes place. The operator $U_F$ (Step 2) acts between the first and the second application of $QFT_n$. Its effect is to map every basis vector of $|\psi_1\rangle$ into another basis vector *injectively*. In this way it may create nonlocal correlation among qubits. Therefore, $U_F$ plays the role of the entanglement operator [10].

The QAGs of the best known algorithms can all be described as the composition of a superposition, an entanglement, and interference operators, where the superposition and the interference operators coincide, but play different roles, as it is in the case for $QFT_n$ in the above Step 1 and Step 3. From a qualitative point of view, the action of the superposition operator is to exploit the potential parallelism of the system by preparing the system itself in a superposition of all its possible states. When the entanglement operator acts on this superposed state the whole information about $f$ contained in $U_F$ is transferred to the resulting vector. Finally, the interference operator makes this information accessible by measurement in order to solve the problem. To illustrate this, consider, for example, Shor's algorithm with $n=3$ and $f_1$, $f_2$ defined as in *Table 11*.

| $x$ | $f_1(x)$ | $f_2(x)$ |
|---|---|---|
| 000 | 001 | 000 |
| 001 | 111 | 010 |
| 010 | 001 | 100 |
| 011 | 111 | 110 |
| 100 | 001 | 000 |
| 101 | 111 | 010 |
| 110 | 001 | 100 |
| 111 | 111 | 110 |

*Table 11: Example of periodical functions.*

Then

$$U_{F_1} = I_2 \otimes \begin{pmatrix} I_2 & 0 \\ 0 & C_2 \end{pmatrix} \otimes C \ (Case\ 1) \quad \textbf{(54)}$$

And

$$U_{F_2} = I \otimes \begin{pmatrix} I_2 & 0 & 0 & 0 \\ 0 & I \otimes C & 0 & 0 \\ 0 & 0 & C \otimes I & 0 \\ 0 & 0 & 0 & C_2 \end{pmatrix} \otimes I \, , C = \begin{pmatrix} 0 & 1 \\ 1 & 0 \end{pmatrix} \ \text{(Case2)} \quad \textbf{(55)}$$

The computation involved with these two cases of operators is shown in *Table 12* and *Table 13*.

| *Step* | *State* |
|---|---|
| *Input* | $\lvert 000\rangle \otimes \lvert 000\rangle$ |
| *Step* 1 | $\frac{(\lvert 0\rangle + \lvert 1\rangle)}{\sqrt{2}} \otimes \frac{(\lvert 0\rangle + \lvert 1\rangle)}{\sqrt{2}} \otimes \frac{(\lvert 0\rangle + \lvert 1\rangle)}{\sqrt{2}} \otimes \lvert 000\rangle$ |
| *Step* 2 | $\frac{(\lvert 0\rangle + \lvert 1\rangle)}{\sqrt{2}} \otimes \frac{(\lvert 0\rangle + \lvert 1\rangle)}{\sqrt{2}} \otimes \frac{(\lvert 000\rangle + \lvert 111\rangle)}{\sqrt{2}} \otimes \lvert 1\rangle$ |
| *Step* 3 | $\frac{1}{\sqrt{2}} \left( \frac{(\lvert 0\rangle + \lvert 1\rangle)}{\sqrt{2}} \otimes \lvert 0000\rangle + \frac{(\lvert 0\rangle - \lvert 1\rangle)}{\sqrt{2}} \otimes \lvert 0011\rangle \right) \otimes \lvert 1\rangle$ |

*Table 12: Shor's QG information flow with* $f = f_1$.

| **Step** | **State** |
|---|---|
| *Input* | $\lvert 000\rangle \otimes \lvert 000\rangle$ |
| *Step* 1 | $\frac{(\lvert 0\rangle + \lvert 1\rangle)}{\sqrt{2}} \otimes \frac{(\lvert 0\rangle + \lvert 1\rangle)}{\sqrt{2}} \otimes \frac{(\lvert 0\rangle + \lvert 1\rangle)}{\sqrt{2}} \otimes \lvert 000\rangle$ |
| *Step* 2 | $\frac{(\lvert 0\rangle + \lvert 1\rangle)}{\sqrt{2}} \otimes \frac{1}{2}(\lvert 0000\rangle + \lvert 0101\rangle + \lvert 1010\rangle + \lvert 1111\rangle) \otimes \lvert 0\rangle$ |
| *Step* 3 | $\frac{1}{\sqrt{2}} \begin{pmatrix} \lvert 000\rangle \otimes \frac{(\lvert 0\rangle + \lvert 1\rangle)}{\sqrt{2}} \otimes \frac{(\lvert 0\rangle + \lvert 1\rangle)}{\sqrt{2}} + \lvert 010\rangle \otimes \frac{(\lvert 0\rangle - \lvert 1\rangle)}{\sqrt{2}} \otimes \frac{(\lvert 0\rangle - J\lvert 1\rangle)}{\sqrt{2}} \\ + \lvert 100\rangle \otimes \frac{(\lvert 0\rangle + \lvert 1\rangle)}{\sqrt{2}} \otimes \frac{(\lvert 0\rangle - \lvert 1\rangle)}{\sqrt{2}} + \lvert 110\rangle \otimes \frac{(\lvert 0\rangle - \lvert 1\rangle)}{\sqrt{2}} \otimes \frac{(\lvert 0\rangle - J\lvert 1\rangle)}{\sqrt{2}} \end{pmatrix} \otimes \lvert 0\rangle$ |

*Table 13: Shor's QG information flow with* $f = f_2$.

### 5.2. Information analysis of the Shor's QAG.

To understand how the intelligence of $\lvert\psi\rangle$ changes while the Shor's algorithm runs the set of the first $n$ qubits, namely $T = \{1, \ldots, n\}$ is considered for the case where $2^n$ is multiple of $r$. The input vector defined in Eq. (49) is such that

$$E_T^{Sh}(\lvert input\rangle) = E_T^{vN}(\lvert input\rangle) = 0 \quad \textbf{(56)}$$

The intelligence of the state is:

$$\mathcal{J}\left(|input\rangle\right)=1 \quad \textbf{(57)}$$

Eq. (56) is easily proved by observing that

$$|input\rangle\langle input|_T = |0\rangle\langle 0|_n \quad \textbf{(58)}$$

($|0\rangle\langle 0|_n$ is the $n-th$ tensor power of $|0\rangle\langle 0|$). Since $\log_2 1=0$, $\log_2 |0\rangle\langle 0|_n$ corresponds to the null squared matrix of order $2^n$. Then it follows from Eq. (50) that the value of $S_T^{Sh}\left(|input\rangle\right)$ and $S_T^{vN}\left(|input\rangle\right)$ are both 0. In other words, the input state belongs to the measurement basis $\mathcal{B}$, therefore, both its Shannon and von Neumann entropy with respect to $T$ are zero.

When $\left(QFT_n \otimes I_n\right)$ is applied [Step 1, Eq. (50)], the first $n$ qubits undergo a unitary change of basis. This means their von Neumann entropy is left unchanged. On the contrary, the Shannon entropy increases. From Eq. (50) the Shannon entropy value is obtained from the main diagonal values. This means that after Step 1 it is given by

$$E_T^{Sh}\left(|\psi_1\rangle\right)=n, \quad E_T^{vN}\left(|\psi_1\rangle\right)=0 \quad \textbf{(59)}$$

The intelligence of the state with respect to the first $n$ qubits is at this point $\mathcal{J}_T\left(|\psi_1\rangle\right)=0$. The application of $U_F$ (Step 2) entangles the first $n$ qubits with the last $n$. In fact, being $f$ periodical with period $r$ and being $k=\frac{2^n}{r}$ an integer number, the state $|\psi_2\rangle$ can be written

$$|\psi_2\rangle=\sum_{l=0}^{r-1}\left(\left|[l]_{(2)}\right\rangle+\left|[l+r]_{(2)}\right\rangle+\left|\left[l+\left(\frac{2^n}{r}-1\right)r\right]_{(2)}\right\rangle\right)\otimes|f(l)\rangle. \quad \textbf{(60)}$$

From Eq. (60), the density matrix $|\psi_2\rangle\langle\psi_2|_T$ is written as a $k\times k$ block matrix

$$|\psi_2\rangle\langle\psi_2|_T=\frac{1}{2^n}\begin{pmatrix} I(r) & I(r) & \dots & I(r)\\ I(r) & I(r) & \dots & I(r)\\ \dots & \dots & \dots & \dots\\ I(r) & I(r) & \dots & I(r)\end{pmatrix}, \quad \textbf{(61)}$$

where $I(r)$ denotes the identity matrix of order $r$. The matrix $|\psi_2\rangle\langle\psi_2|_T$ can be decomposed into the tensor product of $1+\log_2 k$ smaller density matrices:

$$|\psi_2\rangle\langle\psi_2|_T=\frac{1}{2^{\log_2 k}}\begin{pmatrix}1 & 1\\ 1 & 1\end{pmatrix}_{\log_2 k}\otimes\frac{1}{r}I(r). \quad \textbf{(62)}$$

The von Neumann entropy of a tensor product can be written as the summation of the von Neumann entropies of its factors.

Therefore:

$$S_T^{vN}\left(|\psi_2\rangle\right)=-\left(\log_2 k\right)Tr\left(\frac{1}{2}A\log_2\left(\frac{1}{2}A\right)\right)-Tr\left(\frac{1}{r}I(r)\log_2\left(\frac{1}{r}I(r)\right)\right), \quad \textbf{(63)}$$

where $A=\begin{pmatrix}1 & 1\\ 1 & 1\end{pmatrix}$. Since $A/2$ is similar to $|1\rangle\langle 1|$ through a unitary change of basis, then Eq. (63) is written as

$$S_T^{vN}\left(|\psi_2\rangle\right) = -(\log_2 r)Tr\left(|1\rangle\langle 1|\log_2|1\rangle\langle 1|\right) - Tr\left(\frac{1}{r}I_r\log_2\left(\frac{1}{r}I_r\right)\right) = \boxed{\log_2 r}. \qquad \textbf{(64)}$$

The first equality in Eq. (64) is obtained by noting that $Tr\left(|1\rangle\langle 1|\log_2|1\rangle\langle 1|\right) = 0$. From the structure of the matrix in Eq. (61) it follows that the Shannon entropy did not change. Then for the set $T$ of the first $n$ qubits

$$S_T^{Sh}\left(|\psi_2\rangle\right) = n, \quad S_T^{vN}\left(|\psi_2\rangle\right) = \log_2 r. \qquad \textbf{(65)}$$

This means that

$$\mathcal{J}_T\left(|\psi_2\rangle\right) = \frac{1}{n}\log_2 r \qquad \textbf{(66)}$$

Finally, when $\left(QFT_n \otimes {}^n I\right)$ is applied [Step 3, Eq. (63)], the last $n$ qubits are left unchanged, whereas the first $n$ qubits undergo a unitary change of basis through the QFT. This implies that the von Neumann entropy is reduced. Indeed, from Eq. (64), the input superposition of the first $n$ qubits is periodic with period $k = \frac{2^n}{r}$ and only $r$ different basis vectors can be measured, everyone with probability $\frac{1}{r}$. This means

$$S_T^{Sh}\left(|output\rangle\right) = \log_2 r, \quad S_T^{vN}\left(|output\rangle\right) = \log_2 r \qquad \textbf{(67)}$$

The intelligence of the output state with respect to $T$ is

$$\mathcal{J}_T\left(|output\rangle\right) = 1 \qquad \textbf{(68)}$$

From Eq. (68) it follows that the QFT preserves the von Neumann entropy and the Shannon entropy of the first $n$ qubits as much as possible.

The two operators represented in Table 11 produce the information flow reported in *Table 14* and *Table 15*. It is worth observing how the intelligence of the state increases and decreases while the algorithm evolves.

| *Step* | $E_T^{Sh}\left(\lvert\psi\rangle\right)$ | $E_T^{vN}\left(\lvert\psi\rangle\right)$ | $\mathcal{J}_T\left(\lvert\psi\rangle\right)$ |
|---|---|---|---|
| *Input* | 0 | 0 | 1 |
| *Step* 1 | 3 | 0 | 0 |
| *Step* 2 | 3 | 1 | $\frac{1}{3}$ |
| *Step* 3 | 1 | 1 | 1 |

*Table 14: Shor's QAG information flow with* $f = f_1$.

| *Step* | $E_T^{Sh}\left(\lvert\psi\rangle\right)$ | $E_T^{vN}\left(\lvert\psi\rangle\right)$ | $\mathcal{J}_T\left(\lvert\psi\rangle\right)$ |
|---|---|---|---|
| *Input* | 0 | 0 | 1 |
| *Step* 1 | 3 | 0 | 0 |
| *Step* 2 | 3 | 2 | $\frac{2}{3}$ |
| *Step* 3 | 2 | 2 | 1 |

*Table 15: Shor's QAG information flow with* $f = f_2$.

Now, for comparison, consider the case $n = 2$ for the function $f_2$ with the period 4 and the entanglement operator as the particular case of Eq. (55)

$$U_{F_2} = \begin{pmatrix} I\otimes I & 0 & 0 & 0 \\ 0 & I\otimes C & 0 & 0 \\ 0 & 0 & C\otimes I & 0 \\ 0 & 0 & 0 & C\otimes I \end{pmatrix} (Case\ 3) \qquad \textbf{(69)}$$

Information analysis of Shor's algorithm is presented in *Table 16*.

| Step | Classical Entropy $H_{Sh}(\lvert\psi\rangle)$ | Quantum Entropy $S_{\lvert\psi\rangle}(\{j\})$ | States and Gate Operations | Dynamics of Solution Probabilities |
|---|---|---|---|---|
| Input | 0 | 1, 2, 3, 4, 5, 6: 0, 0, 0, 0, 0, 0 | $\lvert 000\rangle \otimes \lvert 000\rangle$ <br> ${}^3H\otimes{}^3I$ <br> ${}^4H$ | 1 <br> \|000000> |
| Super-position | 3 | 1, 2, 3, 4, 5, 6: 0, 0, 0, 0, 0, 0 | $\frac{\lvert 0\rangle+\lvert 1\rangle}{\sqrt{2}} \otimes \frac{\lvert 0\rangle+\lvert 1\rangle}{\sqrt{2}} \otimes \frac{\lvert 0\rangle+\lvert 1\rangle}{\sqrt{2}} \otimes \lvert 000\rangle$ <br> $U_F$ | 1/8 <br> \|000000> \|00100> … \|111000> |
| Entan-glement | 3 | (1), 2, 6: 0, 0, 0; 3, 4, 5: -1, -1, -1, 1, 1, 1, 0 | $\frac{\lvert 0\rangle+\lvert 1\rangle}{\sqrt{2}} \otimes \frac{\lvert 0\rangle+\lvert 1\rangle}{\sqrt{2}} \otimes \frac{\lvert 000\rangle+\lvert 111\rangle}{\sqrt{2}} \otimes \lvert 1\rangle$ <br> $QFT_3 \otimes I$ | 1/8 <br> \|000001> \|001111> … \|111111> |
| Inter-ference | 1 | 2, (3), 6: 0, 0, 0; 1, 4, 5: -1, -1, -1, 1, 1, 1, 0 | $\left[\frac{\lvert 0\rangle+\lvert 1\rangle}{2} \otimes \lvert 0000\rangle + \frac{\lvert 0\rangle-\lvert 1\rangle}{2} \otimes \lvert 0011\rangle\right] \otimes \lvert 1\rangle$ | 1/4 <br> \|000001> \|000111> \|100001> \|100111> |

*Table 16: Information Analysis of Shor's Algorithm (case1, $f = f_1$) in Eq. (54).*

*Table 17* shows the evolution of the algorithm when applied with this operator.

| Step | Classical Entropy $H_{Sh}(\lvert\psi\rangle)$ | Quantum Entropy $S_{\lvert\psi\rangle}(\{j\})$ | States and Gate Operations | Dynamics of Solution Probabilities |
|---|---|---|---|---|
| Input | 0 | 1, 2, 3, 4: 0, 0, 0, 0 | $\lvert 00\rangle \otimes \lvert 00\rangle$ <br> ${}^2H\otimes{}^2I$ <br> ${}^4H$ | 1 <br> \|0000> |
| Super-position | (1,1,1,1) <br> 4 | 1, 2, 3, 4: 0, 0, 0, 0 | $\frac{\lvert 0\rangle+\lvert 1\rangle}{\sqrt{2}} \otimes \frac{\lvert 0\rangle+\lvert 1\rangle}{\sqrt{2}} \otimes \lvert 00\rangle$ <br> $U_F$ | 1/4 <br> \|0000> \|0100> … \|1100> |
| Entan-glement | (1,1,1,1) <br> 4 | 1, 2; 3, 4: -1, 2, -1 | $\frac{1}{2}\begin{pmatrix} \lvert 00\rangle\otimes\lvert 00\rangle + \lvert 01\rangle\otimes\lvert 01\rangle + \\ + \lvert 10\rangle\otimes\lvert 10\rangle + \lvert 11\rangle\otimes\lvert 11\rangle \end{pmatrix}$ <br> $QFT_3 \otimes I$ | 1/4 <br> \|0000> \|0101> … \|1111> |
| Inter-ference | (1,0,1,0) <br> 2 | 1, 2; 3, 4: -1, 2, -1 | $\frac{1}{4}\left(\lvert 0000\rangle + \lvert 0001\rangle + \ldots - \lvert 0001\rangle + \lvert 1111\rangle\right)$ | 1/16 <br> \|0000> \|0001> … \|1111> |

*Table 17: Information Analysis of Shor's Algorithm for case 3.*

From *Table 17* the following is observed:

- The entanglement operator creates very strong quantum correlation among vectors 1, 2, 3 and 4. This correlation identifies the input function that has the maximal period (and so maximal entanglement).
- The entanglement operator creates the quantum correlation with negative conditional entropies between qubits 1 and 2, and between qubits 3 and 4. This is the nonclassical super correlation effect from entanglement operator.
- The interference operator preserves quantum correlation, but does not decrease the classical entropy because entanglement is too great (degree of resistance is very high).

In *Fig. 6* this structure is pictured.

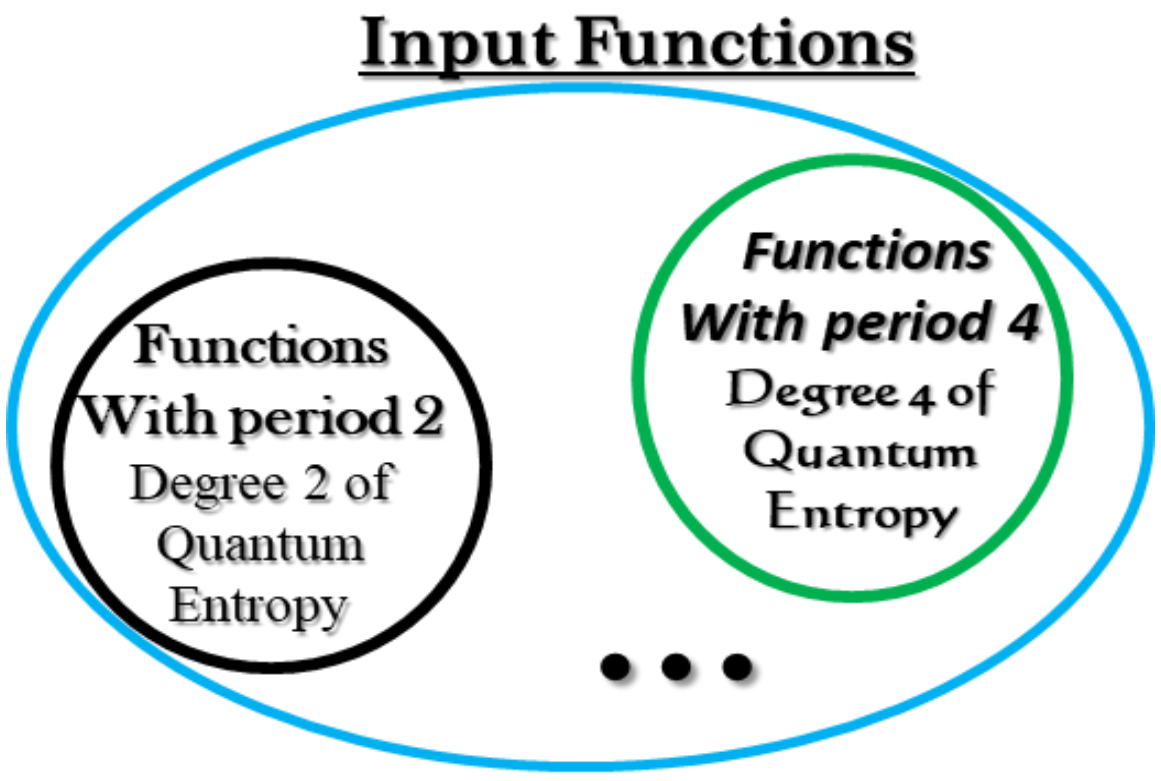

*Figure 6: Quantum Information Structure of Shor's Input Space.*

Shor's algorithm undergoes the special structure of its input space: periodical functions. Every function is characterized by its capacity to create quantum entanglement, which depends on its period.

### 5.3. Physical interpretation of the analysis results.

| | |
|---|---|
| **1** | When the QA computation begins, the Shannon entropy coincides with the von Neumann entropy, but they are both zero. The intelligence is then maximal since the system is in a basis state. |
| **2** | The superposition operator increases the Shannon entropy of the first $n$ qubits to its maximum, but leaves the von Neumann entropy unchanged. The intelligence is minimal since the degree of disorder is at its maximum but has not been stored into the system yet. |
| **3** | The entanglement operator increases the von Neumann entropy of the first $n$ qubits according to $f$, but leaves the Shannon entropy unchanged. The intelligence increase at this step since some information is stored into quantum correlation using super-correlation that created by the present of a new effect in evolution of QA's as the negative conditional entropy between the partial qubits |
| **4** | ***4.1.*** The interference operator preserves quantum correlation, but transpose it between basis vectors; this transposing maintains the period of the input function encoded, but it has as side effect to reduce the classical entropy, letting possible to access the period information generating an intelligent state, namely a state containing all required quantum information but with minimum classical entropy as a qualitative measure of free energy. |
| | ***4.2.*** The interference operator does not change the value of the von Neumann entropy introduced by the entanglement operator, but decreases the value of the Shannon entropy to its minimum, that is to the value of the von Neumann entropy itself. The intelligence of the state reaches its maximum again, but now with a non-zero quantity of information in quantum correlation. |

As described above, the von Neumann entropy can be interpreted as the degree of information in a vector, namely, as a measure of the information stored in quantum correlation about the function $f$. The Shannon entropy is interpreted as the measure of the degree of inaccessibility of this information through the measurement. In this context, the QAG $G$ of Eq. (47) transfers information from $f$ into the output vector minimizing the quantity of unnecessary noise producible by the measurement, or, more technically, minimizing the non-negative, according to:

$$S_T^{Sh}\left(|output\rangle\right) - S_T^{vN}\left(|output\rangle\right) = \Delta S\left(|output\rangle\right)$$

(the defect exchange of measurement entropy) with $T = \{1,\dots,n\}$ .The intelligence of the output state increases while the average unnecessary noise decreases. According to this definition, the action of the QFT in the Shor's algorithm is to associate with every possible function $f$ a maximally intelligent output state, namely a state $|output\rangle$ such that $\mathcal{J}_T\left(|output\rangle\right) = 1$. This is clear from the graphical representation of the information flow and the intelligence relative to the two functions considered in Example in *Fig. 7*.

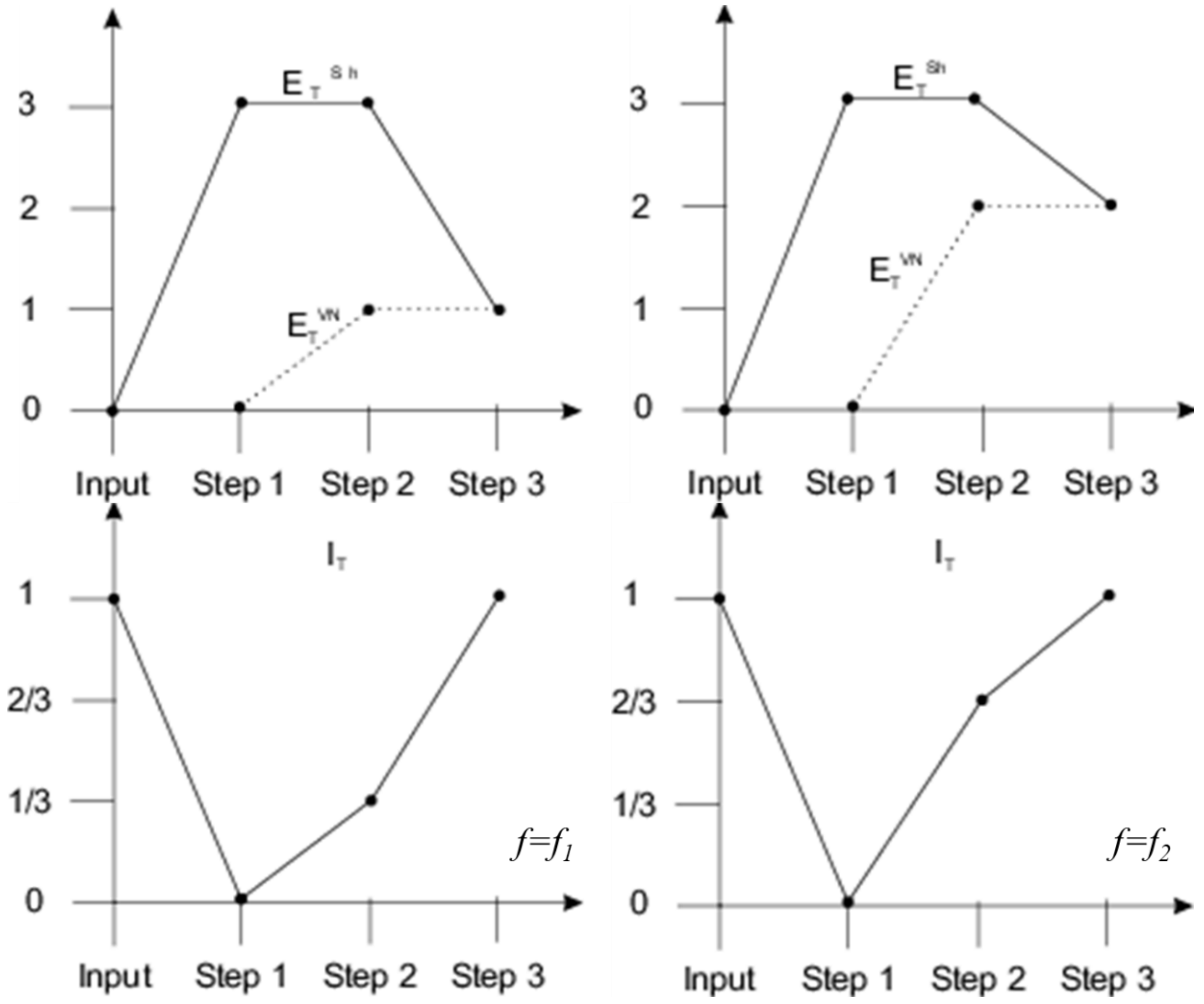

*Figure 7: Information flow and intelligence of the Shor's QA.*

If the period $r$ does not divide $2^n$ exactly, then the QFT is not optimal. In fact, the final superposition for the first $n$ qubits are not a periodical superposition. Nevertheless, by increasing the number $n$ of qubits used for encoding input strings, it is possible to approximate this periodical superposition as well as desired.

The way the function $f$ is encoded into the operator $U_F$ and the set $T$ used for the calculation of the QA-intelligence $\mathcal{J}_T\left(|output\rangle\right)$ are problem dependent.

Consider, for instance, the Deutsch-Jozsa decision making QA problem: one must decide if a Boolean function $f:\{0,1\}^n \to \{0,1\}^n$ is constant or balanced.

The same encoding scheme of the Shor's algorithm can be used to solve this problem. In this case, after the entanglement operator has acted, the von Neumann entropy of every proper subset of the first $n$ qubits is always 0, whereas the von Neumann entropy of the first $n$ qubits is 1 for some balanced functions. This means that for these functions no interference operator in the form $Int \otimes I_m$ can increase the intelligence of the state with respect to the first $n$ qubits, as it happens in the Shor's algorithm.

In other words, the state of the system after entanglement has been applied is already maximally intelligent with respect to first $n$ qubits and the information accessibility cannot be increased through the application of any interference operator. One solution to this problem is to encode $f$ into the unitary operator $U_F \cdot \left[ I_n \otimes (H \cdot C) \right]$ where $U_F$ is obtained as in the Shor's algorithm.

With this encoding scheme, the von Neumann entropy of the first $n$ qubits after Step 2 is always 0. On the contrary, the von Neumann entropy of any subset of the first $n$ qubits can be positive, implying entanglement between this subset and the rest of the system. In particular, every singleton constituted by one qubit may be characterized by a positive value of the von Neumann entropy. The Deutsch-Jozsa QA interference operator is chosen in order to reduce as much as possible the gap between the Shannon and the von Neumann entropies of every one of these singletons.

This operator is the Walsh-Hadamard transform of order $n$, defined as the tensor power $H_n$. Indeed, it is easy to verify that for every state $|\psi\rangle$ and every qubit $i$, the matrix $\left( H^{-1} \cdot |\psi\rangle\langle\psi|_{\{i\}} \cdot H \right)$ is diagonal. This means the action of $H_n$ is to maximize the intelligence of every one of the first $n$ qubits by annihilating the gap between its Shannon and von Neumann entropies [10].

# 6. Information-theoretical analysis of Grover's quantum search algorithm

The searching problem can be stated in terms of a list $\mathcal{L}[0,1,...,N-1]$ with a number $N$ of unsorted elements. Denote by $x_0$ the marked element in $\mathcal{L}$ that are sought. The quantum mechanical solution of this searching problem goes through the preparation of a quantum register in a quantum computer to store the $N$ items of the list. This will allow exploiting quantum parallelism. Thus, assume that the quantum registers are made of $n$ source qubits so that $N = 2^n$.

A target qubit is used to store the output of function evaluations or calls.

## 6.1. Quantum search algorithm.

To implement the quantum search, construct a unitary operation that discriminates between the marked item $x_0$ and the rest. The following function

$$f_{x_0}(x) = \begin{cases} 0, & \text{if } x \neq x_0 \\ 1, & \text{if } x = x_0 \end{cases},$$

and its corresponding unitary operation $U_{f_{x_0}} |x\rangle|y\rangle = |x\rangle\left|y \oplus f_{x_0}(x)\right\rangle$. Count how many applications of this operation or oracle calls are needed to find the item.

The rationale behind the Grover algorithm is: 1) to start with a quantum register in a state where all the computational basis states are equally present; 2) to apply several unitary transformations to produce an outcome state in which the probability of catching the marked state $|x_0\rangle$ is large enough. The steps in Grover's algorithm are shown as in [21] in tabular form below (the quantum circuit shown in [2,19].)

According to Steps 2 - 4 above, the QAG of Grover's quantum search algorithm (QSA) is [19]

$$G = (D_n \otimes I) \cdot U_F \cdot \left( {}^n H \otimes H \right)$$

that acts on the initial state of both registers in the QSA. Computational analysis of Grover's QSA is similar to analysis of the Deutsch-Jozsa QA. The basic component of the algorithm is the quantum operation encoded in Steps 3 and 4, which is repeatedly applied to the uniform state $|\psi_2\rangle$ in order to find the marked element. Steps 5 and 6 in Grover's algorithm are also applied in Shor's QSA. Although this procedure resembles the classical strategy, Grover's operation enhances by constructive interference of quantum amplitudes the presence of the marked state.

| **Step** | **Computational algorithm** | **F.** |
|---|---|---|
| *Step 1* | Initialize the quantum registers to the state: $|\psi_1 = input\rangle := |00...0\rangle|1\rangle$. | **(60)** |
| *Step 2* | Apply bit-wise the Hadamard one-qubit gate to the source register, so as to produce a uniform superposition of basis states in the source register, and also to the target register: $|\psi_2\rangle := U_H^{\otimes(n+1)}|\psi_1\rangle = \frac{1}{2^{(n+1)/2}}\sum_{x=0}^{2^n-1}|x\rangle\boxed{\sum_{y=0,1}(-1)^y|y\rangle}$. | **(61)** |
| *Step 3* | Apply the operator $U_{f_{x_0}}$: $|\psi_3\rangle := U_{f_{x_0}}|\psi_2\rangle = \frac{1}{2^{(n+1)/2}}\sum_{x=0}^{2^n-1}(-1)^{f_{x_0}(x)}|x\rangle\boxed{\sum_{y=0,1}(-1)^y|y\rangle}$. Let $U_{x_0}$ be the operator by $U_{x_0}|x\rangle := (1-2|x_0\rangle\langle x_0|)|x\rangle = \begin{cases} -|x_0\rangle & \text{if } x = x_0 \\ |x\rangle & \text{if } x \neq x_0 \end{cases}$, that is, it flips the amplitude of the marked state leaving the remaining source basis states unchanged. The state in the source register of Step 3 equals precisely the result of the action of $U_{x_0}$, i.e., $|\psi_3\rangle = ([1-2|x_0\rangle\langle x_0|]\otimes 1)|\psi_2\rangle$. | **(62)** |
| *Step 4* | Apply next the operation $D$ known as inversion about the average. This operator is defined as follows $D := -\left(U_H^{\otimes n}\otimes I\right)U_{f_0}\left(U_H^{\otimes n}\otimes I\right)$, and $|output\rangle = D|\psi_3\rangle$, where $U_{f_0}$ is the operator in Step 3 for $x_0 = 0$. The effect of this operator on the source is to transform $\sum_x \alpha_x|x\rangle \mapsto \sum_x(-\alpha_x + \langle\alpha\rangle)|x\rangle$, where $\langle\alpha\rangle := 2^{-n}\sum_x \alpha_x$ is the mean of the amplitudes, so its net effect is to amplify the amplitude of $|x_0\rangle$ over the rest. | **(63)** |
| *Step 5* | Iterate Steps 3 and 4 a number of times *m*. | |
| *Step 6* | Measure the source qubits (in the computational basis). The number *m* is determined such that the probability of finding the searched item $x_0$ is maximal. | |

The operator encoding the input function is

$$U_F = \begin{pmatrix} I & 0 & 0 & 0 & 0 & 0 & 0 & 0 \\ 0 & C & 0 & 0 & 0 & 0 & 0 & 0 \\ 0 & 0 & I & 0 & 0 & 0 & 0 & 0 \\ 0 & 0 & 0 & I & 0 & 0 & 0 & 0 \\ 0 & 0 & 0 & 0 & I & 0 & 0 & 0 \\ 0 & 0 & 0 & 0 & 0 & I & 0 & 0 \\ 0 & 0 & 0 & 0 & 0 & 0 & I & 0 \\ 0 & 0 & 0 & 0 & 0 & 0 & 0 & I \end{pmatrix}. \quad (64)$$

## 6.2. Information analysis of Grover's QA.

*Table 18* represents a general iteration algorithm for information analysis of Grover's QSA according to the above presented algorithm.

| Step | Classical and Quantum Entropy $S_{\|\psi\rangle}(\{j\}) = -\lambda_1 \log \lambda_1 - \lambda_2 \log \lambda_2$ | States and Gate Operations |
|---|---|---|
| Input | $\lambda_{1/2} = \frac{\alpha^2}{2}\left(-1 + 2^n + \frac{\beta^2}{\alpha^2} \pm \sqrt{5 - 2^{n+2} + 2^{2n} + \left(2^{n+2} - 8\right)\frac{\beta}{\alpha} + 2\frac{\beta^2}{\alpha^2} + \frac{\beta^4}{\alpha^4}}\right)$ <br> $H_{Sh}(\|\psi\rangle) = -\left(2^n - 1\right)\alpha^2 \log \alpha^2 - \beta^2 \log \beta^2$ | $\alpha\left(\|0...0\rangle + ... + \frac{\beta}{\alpha}\|x\rangle + ... + \|1...1\rangle\right) \otimes$ <br> $\otimes \frac{\|0\rangle - \|1\rangle}{\sqrt{2}}$ |
| Entangle-ment | $\lambda_{1/2} = \frac{\alpha^2}{2}\left(-1 + 2^n + \frac{\beta^2}{\alpha^2} \pm \sqrt{5 - 2^{n+2} + 2^{2n} - \left(2^{n+2} - 8\right)\frac{\beta}{\alpha} + 2\frac{\beta^2}{\alpha^2} + \frac{\beta^4}{\alpha^4}}\right)$ <br> $H_{Sh}(\|\psi\rangle) = -\left(2^n - 1\right)\alpha^2 \log \alpha^2 - \beta^2 \log \beta^2$ | $\alpha\left(\|0...0\rangle + ... - \frac{\beta}{\alpha}\|x\rangle + ... + \|1...1\rangle\right) \otimes \frac{\|0\rangle - \|1\rangle}{\sqrt{2}}$ |
| Interference | $\lambda_{1/2} = \frac{(\alpha - m)^2}{2}\left(-1 + 2^n + \frac{(\beta + m)^2}{(\alpha - m)^2} \pm \sqrt{5 - 2^{n+2} + 2^{2n} - \left(2^{n+2} - 8\right)\frac{\beta + m}{\alpha - m} + 2\left(\frac{\beta + m}{\alpha - m}\right)^2 + \frac{(\beta + m)^4}{(\alpha - m)^4}}\right)$ <br> $H_{Sh}(\|\psi\rangle) = -\left(2^n - 1\right)(\alpha - m)^2 \log(\alpha - m)^2 - (\beta + m)^2 \log(\beta + m)^2$ | $(\alpha - m)\left(\|0...0\rangle + ... + \frac{\beta + m}{\alpha - m}\|x\rangle + ... + \|1...1\rangle\right)$ <br> $\otimes \frac{\|0\rangle - \|1\rangle}{\sqrt{2}}$ |

*Table 18: Information Analysis of Grover's Algorithm (General Iteration).*

*Tables 19* and *20* two iterations of this algorithm with the operator $U_F$ from Eq. (33) are reported.

| Step | Classical Entropy $H_{Sh}(\|\psi\rangle)$ | Quantum Entropy $S_{\|\psi\rangle}(\{j\})$ | States and Gate Operations | Dynamics of Solution Probabilities |
|---|---|---|---|---|
| Input | 0 | 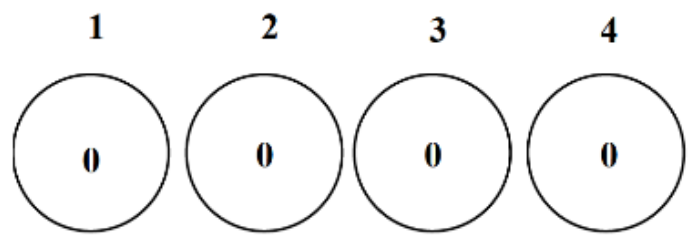  | $\|000\rangle \otimes \|1\rangle$ <br> $^4H$ | 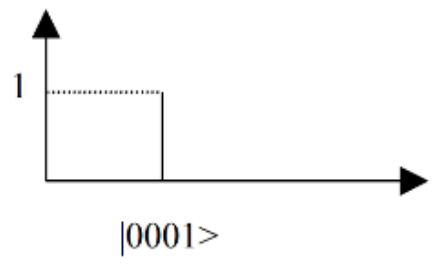  |
| Super-position | 4 | 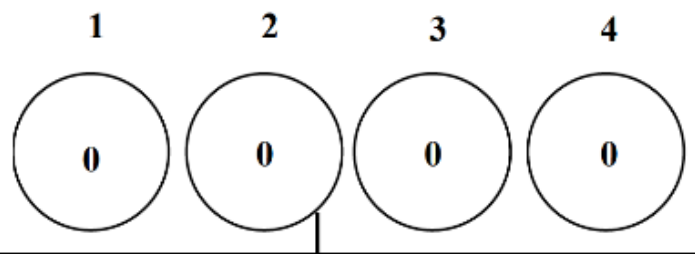  | $\frac{\|0\rangle + \|1\rangle}{\sqrt{2}} \otimes \frac{\|0\rangle + \|1\rangle}{\sqrt{2}} \otimes \frac{\|0\rangle + \|1\rangle}{\sqrt{2}} \otimes \frac{\|0\rangle - \|1\rangle}{\sqrt{2}}$ <br> $U_F$ | 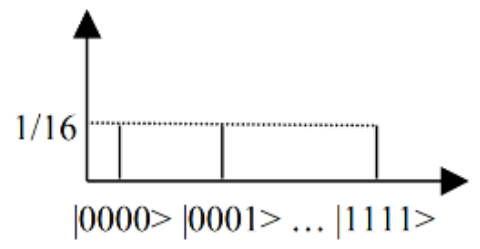  |
| Entan-glement | 4 | $L = 2 - \frac{3}{4}\log 3$ 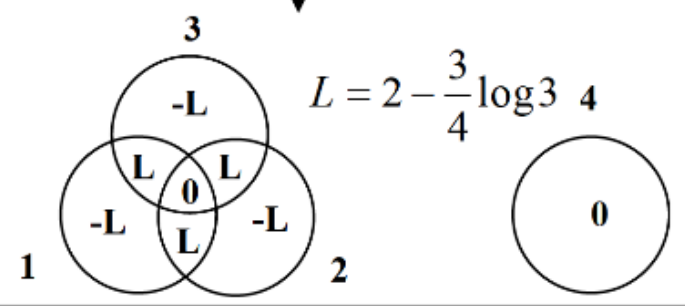  | $\frac{1}{2\sqrt{2}}\left(\begin{array}{l}\|000\rangle - \|001\rangle + \|010\rangle + \\ \|011\rangle + \|100\rangle + \|101\rangle + \\ + \|110\rangle + \|111\rangle\end{array}\right) \otimes \frac{\|0\rangle - \|1\rangle}{\sqrt{2}}$ <br> $D_3 \otimes I$ | 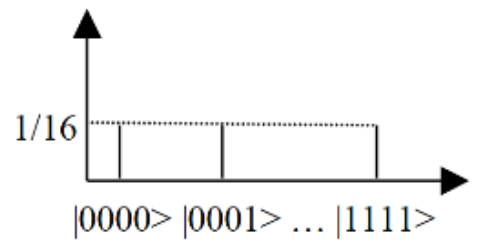  |
| Inter-ference | $5 - \frac{25}{16}\log 5$ | $L = -\frac{1}{8}\left(4 - \sqrt{13}\right)\log\left[\frac{1}{8}\left(4 - \sqrt{13}\right)\right] - \frac{1}{8}\left(4 + \sqrt{13}\right)\log\left[\frac{1}{8}\left(4 + \sqrt{13}\right)\right]$ 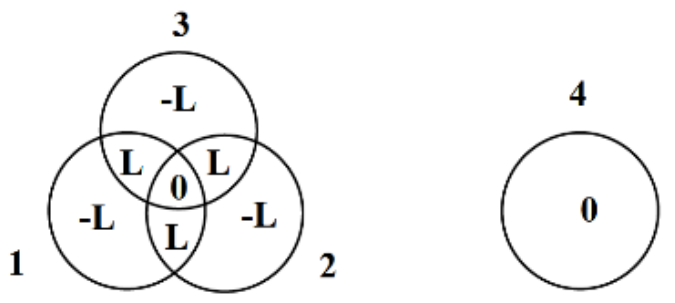  | $\frac{1}{4\sqrt{2}}\left(\begin{array}{l}\|000\rangle + 5\|001\rangle + \|010\rangle + \\ \|011\rangle + \|100\rangle + \|101\rangle + \\ + \|110\rangle + \|111\rangle\end{array}\right) \otimes \frac{\|0\rangle - \|1\rangle}{\sqrt{2}}$ | 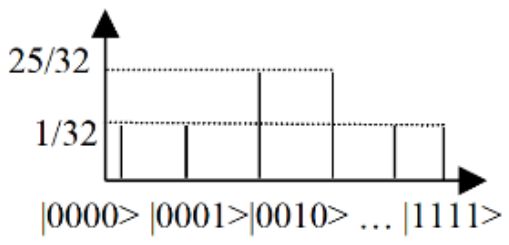  |

*Table 19: Information Analysis of Grover's Algorithm (First Iteration).*

| Step | Classical Entropy $H_{Sh}(\|\psi\rangle)$ | Quantum Entropy $S_{\|\psi\rangle}(\{j\})$ | States and Gate Operations | Dynamics of Solution Probabilities |
|---|---|---|---|---|
| Entan-glement | $5-\frac{25}{16}\log 5$ | $L=-\frac{1}{16}(8-\sqrt{37})\log\left[\frac{1}{16}(8-\sqrt{37})\right]-\frac{1}{16}(8+\sqrt{37})\log\left[\frac{1}{16}(8+\sqrt{37})\right]$ 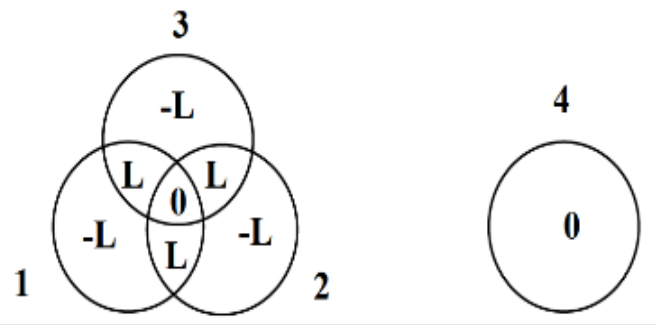  | $U_F$ <br> $\frac{1}{4\sqrt{2}}\begin{pmatrix}\|000\rangle-5\|001\rangle+\|010\rangle+\\ \|011\rangle+\|100\rangle+\|101\rangle+\\ +\|110\rangle+\|111\rangle\end{pmatrix}\otimes\frac{\|0\rangle-\|1\rangle}{\sqrt{2}}$ | 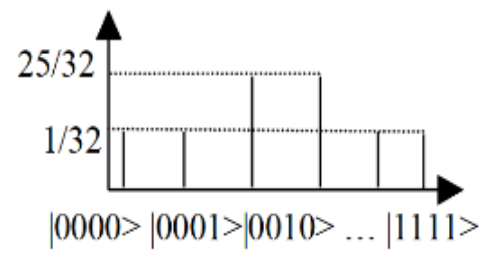  |
| Inter-ference | $6-\frac{121}{64}\log 11$ | $L=-\frac{1}{256}(128-\sqrt{14656})\log\left[\frac{1}{256}(128-\sqrt{14656})\right]-$ $-\frac{1}{256}(128+\sqrt{14656})\log\left[\frac{1}{256}(128+\sqrt{14656})\right]$ 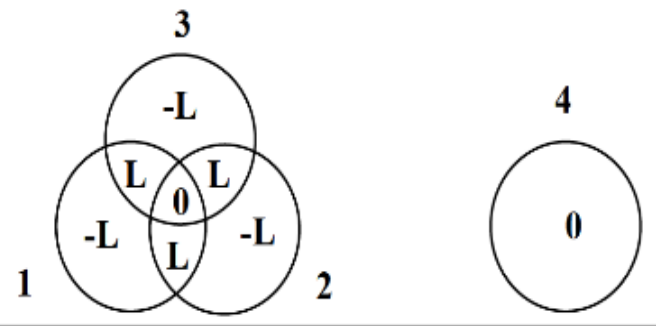  | $D_3\otimes I$ <br> $-\frac{1}{8\sqrt{2}}\begin{pmatrix}\|000\rangle-11\|001\rangle+\|010\rangle+\\ \|011\rangle+\|100\rangle+\|101\rangle+\\ +\|110\rangle+\|111\rangle\end{pmatrix}\otimes\frac{\|0\rangle-\|1\rangle}{\sqrt{2}}$ | 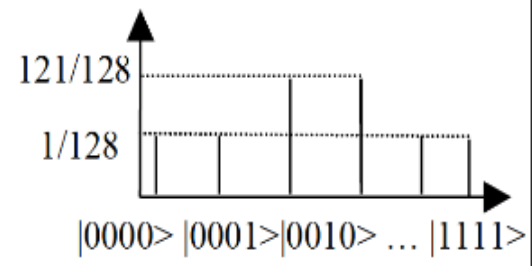  |

*Table 20: Information Analysis of Grover's Algorithm (Second Iteration).*

From these tables, the following are observed:

| | |
|---|---|
| 1 | The entanglement operator in each iteration increase correlation among the different qubits. |
| 2 | Super-correlation from entanglement operator is created according to negative conditional entropies between different qubits. |
| 3 | Interference operator reduces the classical Shannon entropy but it destroys part of the quantum super-correlation measure by conditional and von Neumann entropy. |

Grover's QSA builds in several iterations any intelligent states. Since each iteration encodes the searched function by entanglement, but then partly destroys the encoded information by the interference operator, several iterations are needed in order to conceal both the need to have encoded information and the need to access it. Methodologically, the principle of maximal intelligence on output states is used for synthesizing QAGs. In general, the joint action of the superposition operator and of entanglement operator is to introduce the information necessary to solve the problem in the system quantum correlation. This information, measured through the von Neumann entropy of every qubit, cannot explode, since this would mean too much randomness in the final outcome. The interference operator must reduce the randomness of the output state as much as possible. This means the interference operator is chosen in such that it preserves the von Neumann entropy, but makes the Shannon entropy collapse on its lower bound.

From the application standpoint, the existence of a measure for the intelligence degree of a state allows the combination of QA techniques for encoding functions with some other computational methods, such as genetic algorithms. In this context, the measure of unnecessary noise becomes a fitness function in order to measure the desirability of a result. Thus, quantum computing provides a way of processing information that can be used in the classical problem-solving domain.

### 6.3. Simulation results of QA-termination problem solution based on principle of Shannon/von Neumann minimum entropy

From the step-by-step majorization principle of QA's it follows that for efficient termination of QA's that give the highest probability of successful result, the Shannon entropy was minimal for the step $m+1$. This is the principle of minimum Shannon entropy for termination of a QA with the successful result [18]. This result also follows from the principle of QA maximum intelligent state as in Eq. (1). For this case according to Eq. (1)

$$\max \mathcal{J}_T\left(|\psi\rangle\right)=1-\min\frac{H_T^{Sh}\left(|\psi\rangle\right)}{|T|}, \quad S_T^{vN}\left(|\psi\rangle\right)=0 \quad \text{(for pure quantum state)} \qquad \textbf{(65)}$$

Thus, the principle of maximal intelligence of QA's include as particular case the principle of minimum Shannon entropy for QA-termination problem solution.

***6.3.1. QA-termination problem solving based on minimum Shannon/von Neumann dynamic simulation entropy.*** Consider the complete basis vector:

$$|A\rangle = a_1|0\ldots0\rangle + a_2|0\ldots1\rangle+\ldots+a_{2^n}|1\ldots1\rangle .$$

For this case, the Shannon entropy is

$$S^{Sh}\left(|A\rangle\right)=-\sum_{i=1}^{2^n}a_i^2\log a_i^2, \quad S^{vN}\left(|A\rangle\right)=-\sum_{i=1}^{2^n}\lambda_i\left(a_i^Ta_i\right)\log\lambda_i\left(a_i^Ta_i\right) \qquad \textbf{(66)}$$

where $\lambda_i\left(a_i^Ta_i\right)$ are eigenvalues of quantum state $|A\rangle$. Using the decomposition of quantum state vector $|A\rangle$ with the selection of measurement and calculation:

$$\begin{aligned}|A\rangle &= a_1|0\ldots0\rangle|0\rangle+a_2|0\ldots0\rangle|1\rangle+\ldots a_{2^{n-1}}|1\ldots1\rangle|0\rangle+a_{2^n}|1\ldots1\rangle|1\rangle\\ &=\left[a_1|0\ldots0\rangle+a_3|0\ldots1\rangle+\ldots+a_{2^{n-1}}|1\ldots1\rangle\right]|0\rangle\\ &+\left[a_2|0\ldots0\rangle+a_4|0\ldots1\rangle+\ldots+a_{2^n}|1\ldots1\rangle\right]|1\rangle\end{aligned}$$

or

$$|A\rangle = |A\rangle_{|0\rangle}+|A\rangle_{|1\rangle} . \qquad \textbf{(67)}$$

Then the partial entropy can be calculated as

| | | | | | |
|---|---|---|---|---|---|
| $S_{\lvert 0\rangle}^{Sh}$ | $=$ | $-\sum_i a_i^2\log a_i^2$ | $S_{\lvert 0\rangle}^{vN}$ | $=$ | $-\sum_i \lambda_i\left(a_i^Ta_i\right)\log\lambda_i\left(a_i^Ta_i\right)$ |
| $S_{\lvert 1\rangle}^{Sh}$ | $=$ | $-\sum_{j\neq i} a_j^2\log a_j^2$ | $S_{\lvert 1\rangle}^{vN}$ | $=$ | $-\sum_i \lambda_i\left(a_i^Ta_i\right)\log\lambda_i\left(a_i^Ta_i\right)$ |

**(68)**

where $i=1,3,\ldots,2^{n-1}$, $j=2,4,\ldots,2^n$. In the more general case of Shor's QA, $i=\left[i_0,i_0+2^n,\ldots,2^{2n}-i_0+1\right]$, where $i=1$ for the state vector $|A\rangle_{|0\ldots0\rangle}$ and $i=2^n$ for the state vector $|A\rangle_{|1..1\rangle}$.

*Figure 8* shows the final simulation results of dynamic behavior for Shannon and von Neumann entropies according to Grover's operator (after 3 and 7 iterations) action (the case when intermediate results after superposition and entanglement applications are not shown).

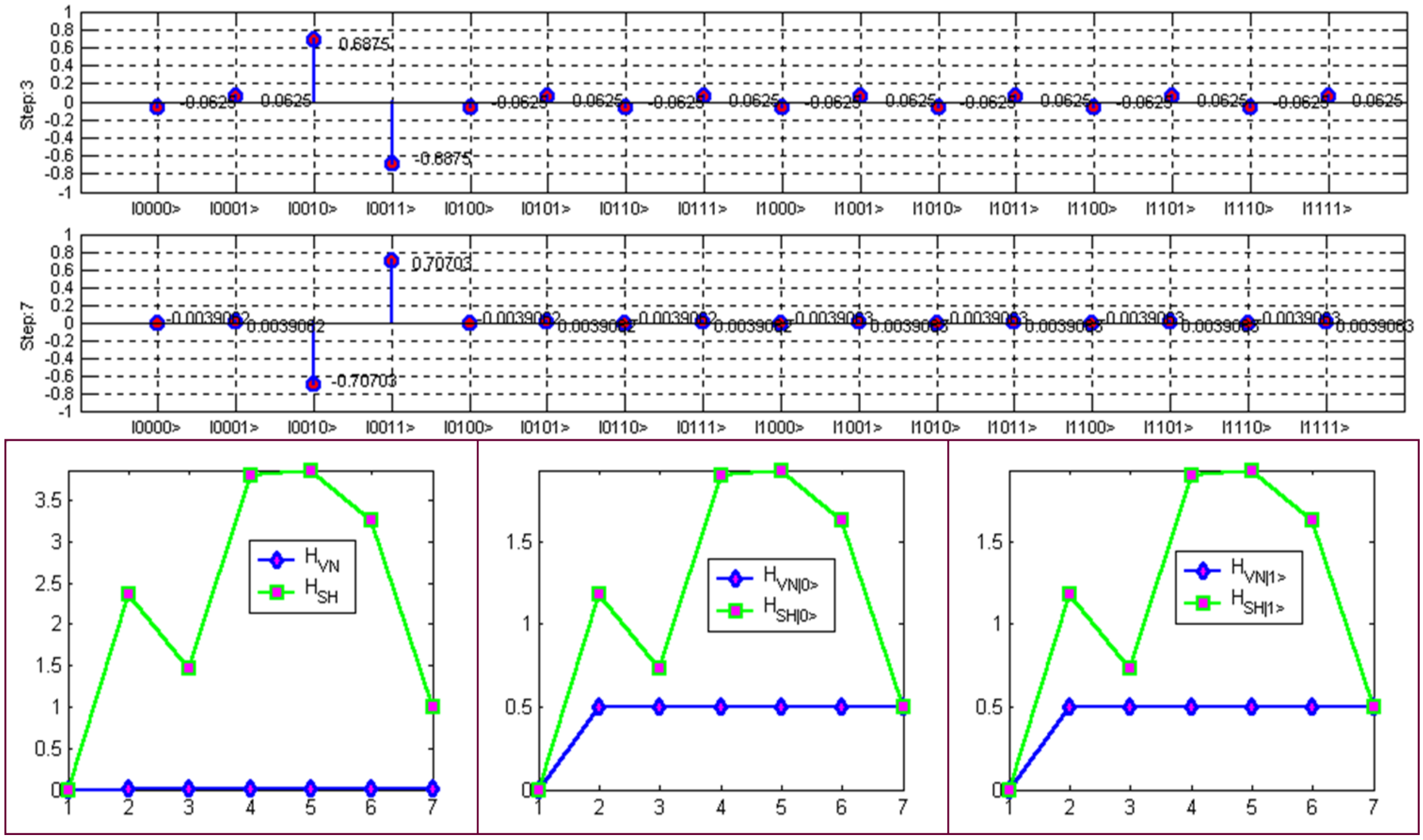

*Figure 8: Simulation results of Grover algorithm (intermediate results after superposition and entanglement are not shown).*

*Figure 9* shows the simulation results of dynamic behavior for Shannon and von Neumann entropies according to superposition and entanglement operator (after 3 and 7 iterations) actions (the case when intermediate results after superposition and entanglement applications are shown).

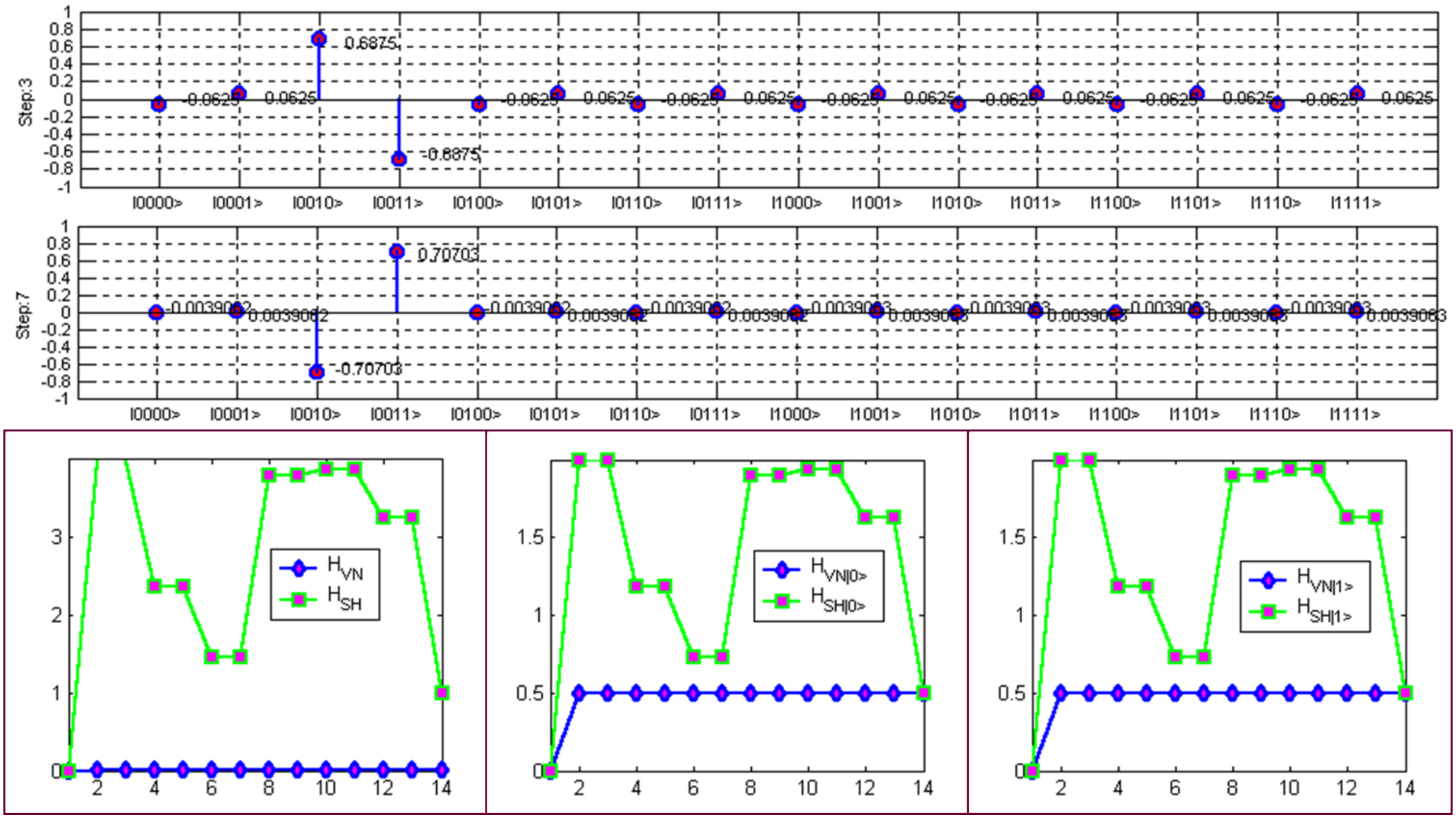

*Figure 9: Simulation results of Grover algorithm (intermediate results after superposition and entanglement are shown).*

*Figures 8* and *9* show the action of constructive interference that created the maximal value 1 of intelligent state of Grover's QSA after 7 iterations.

*Figure 10* shows the final simulation results of dynamic behavior for Shannon and von Neumann entropies for Shor's QA (the case of 2 bit function – period 2) without intermediate simulation results after actions of superposition and entanglement operators.

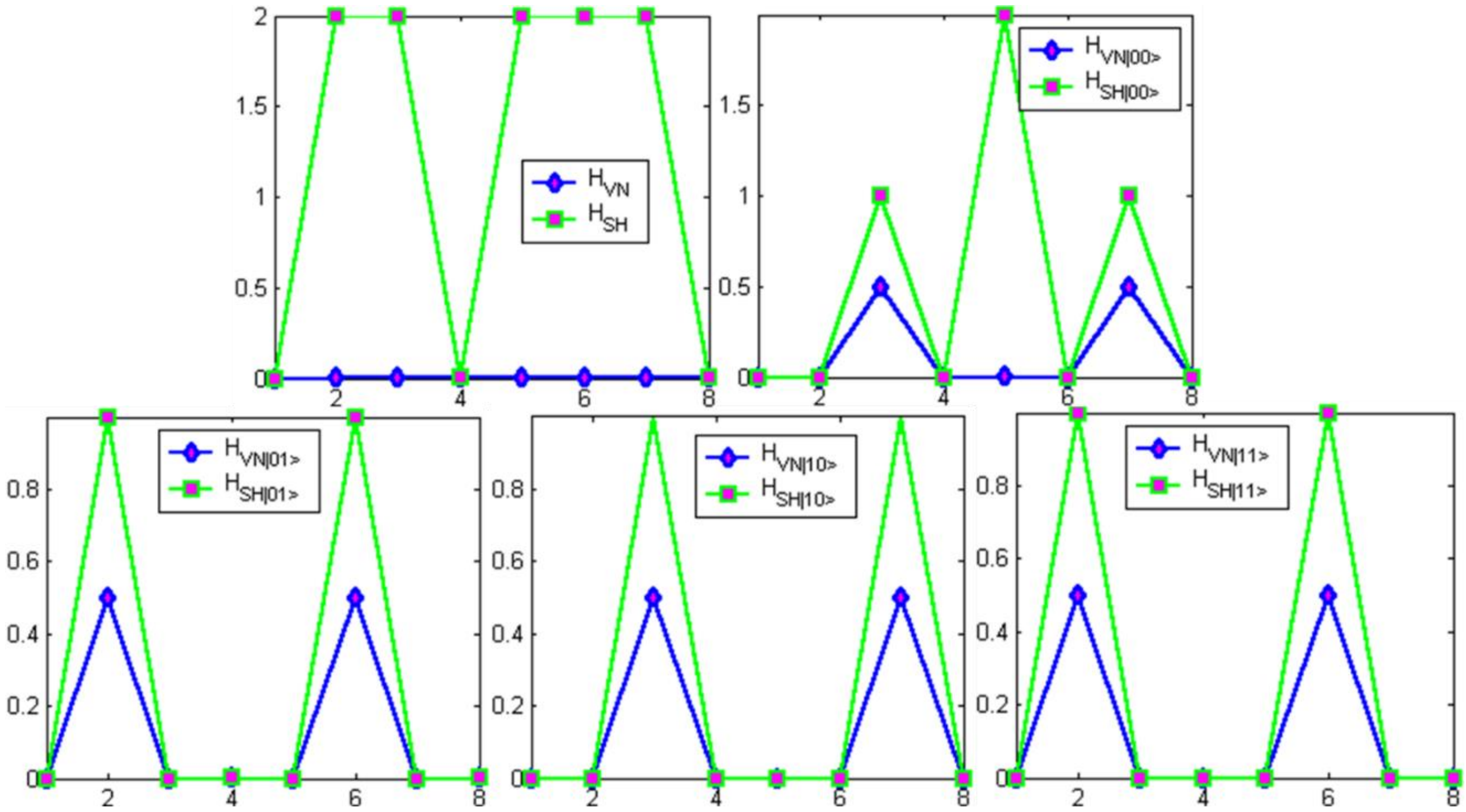

*Figure 10:* *Simulation results of Shor algorithm (intermediate results after superposition and entanglement are not shown, 2 bit function).*

*Figure 11* shows the intermediate simulation results of dynamic behavior for Shannon and von Neumann entropies for Shor's QA (the case of 2 bit function – period 2) after actions of superposition and entanglement operators.

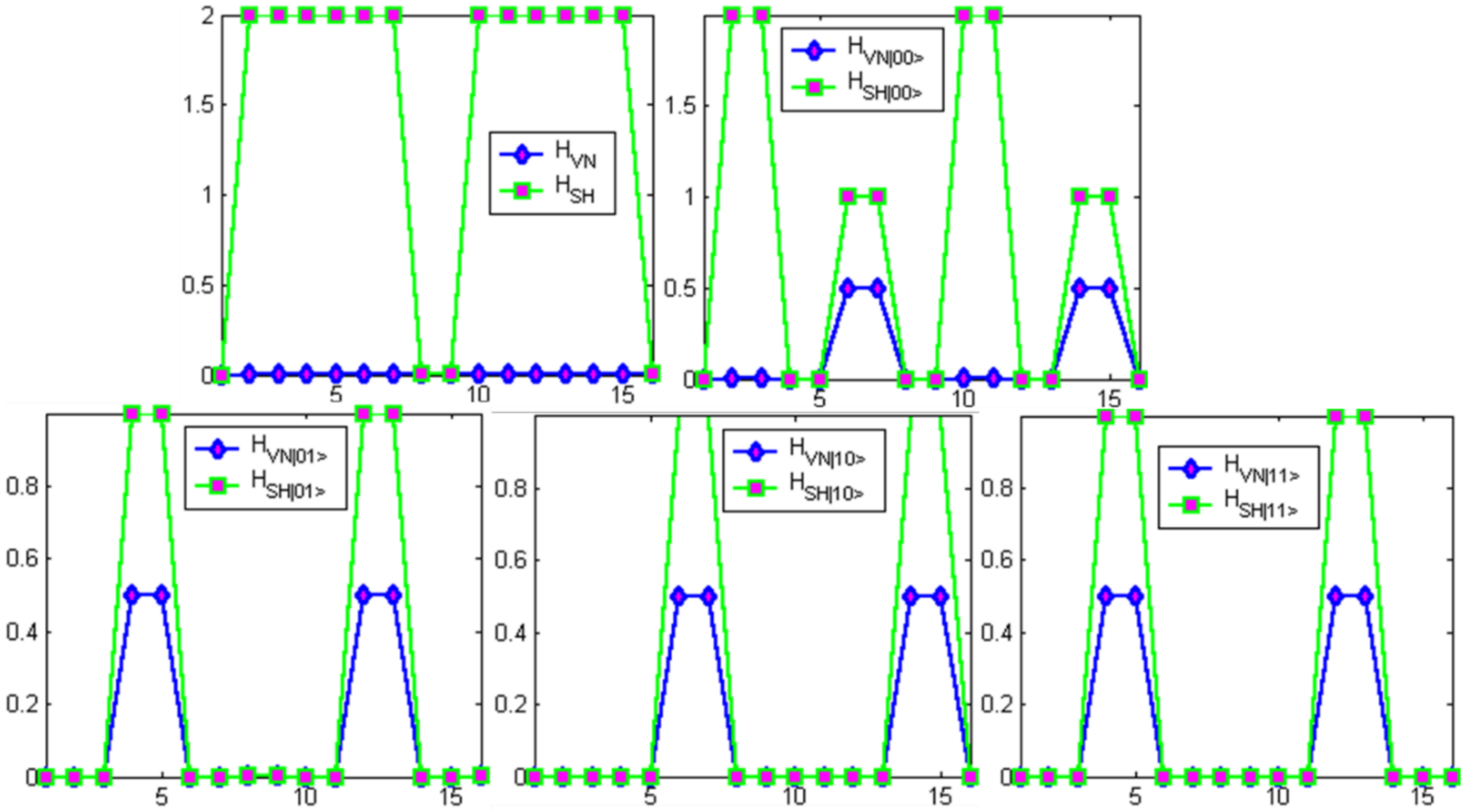

*Figure 11:* *Simulation results of Shor algorithm (intermediate results after superposition and entanglement are shown, 2 bit function (4 qubit)).*

*Figures 10* and *11* shows the action of constructive interference that created also as in Grover's QSA the maximal value 1 of intelligent state of Shor's QA after 14 iterations.

*Figures 12 (a) — (d)* show simulation results of dynamic behavior for Shannon and von Neumann entropies for termination of Deutsch, Deutsch-Jozsa, Grover's (with different search items number) and Shor's QAs respectively.

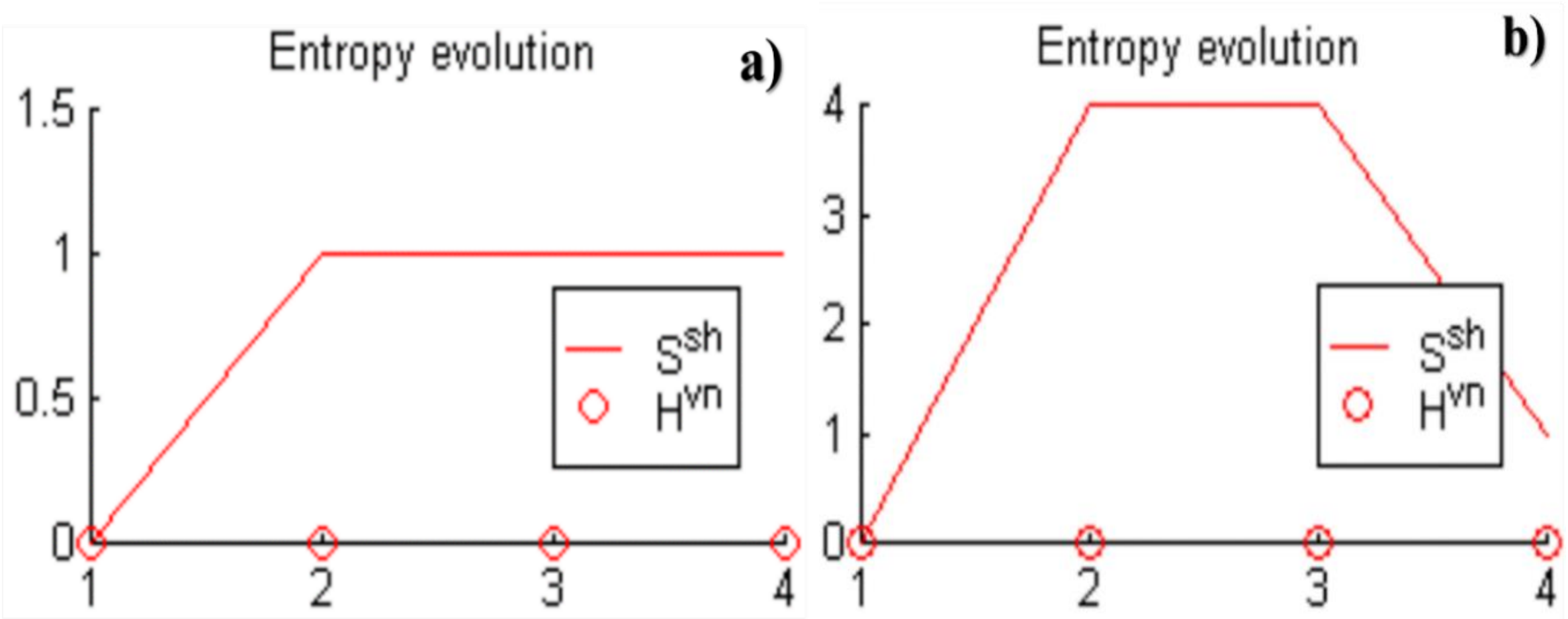

*Figure 12 (a, b): Simulation results of dynamics behavior for Shannon and von Neumann entropies of Deutsch's and Deutsch - Jozsa's QA.*

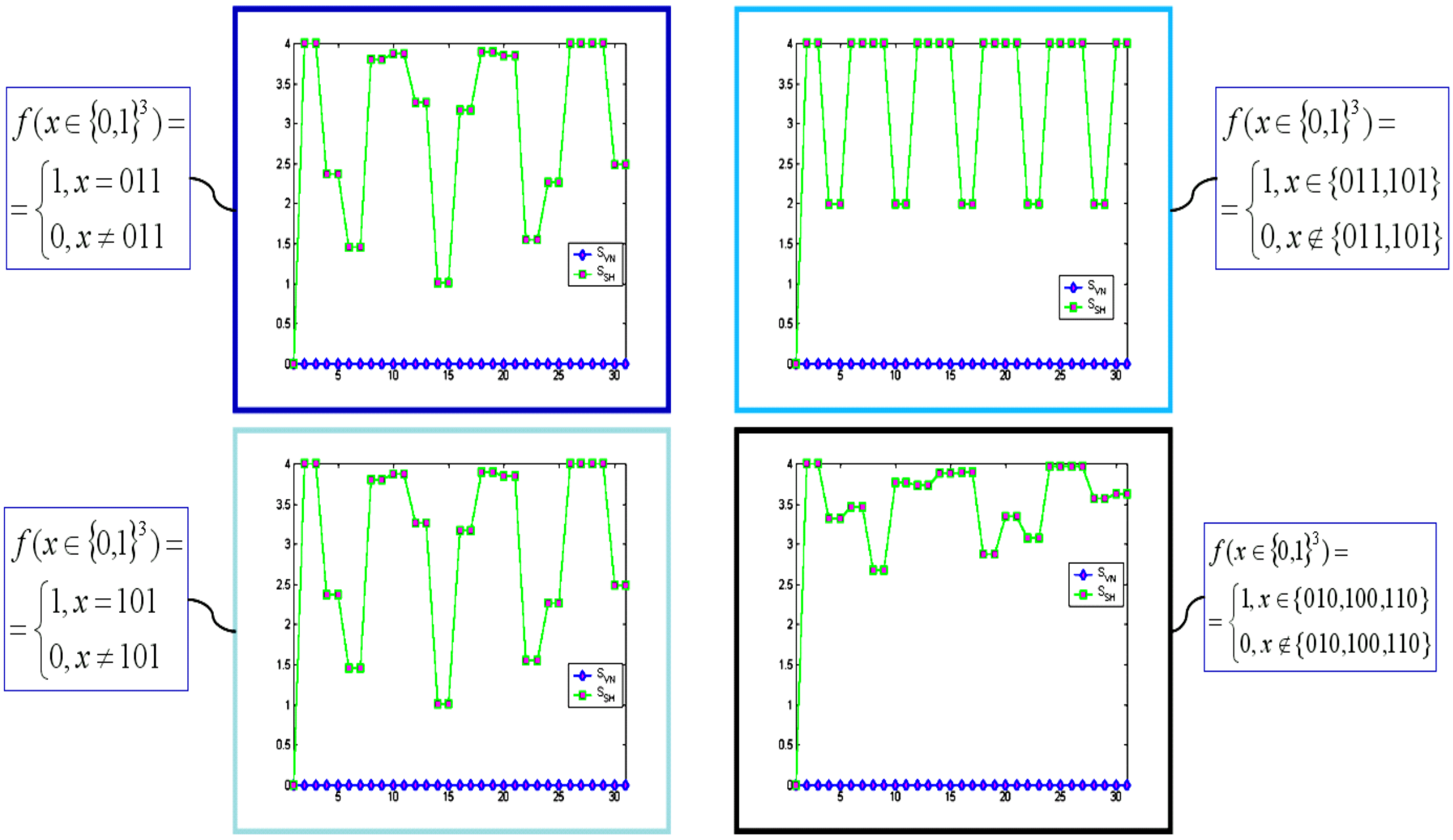

*Figure 12 (c): Simulation results of dynamics behavior for Shannon and von Neumann entropies of Grover's QA.*

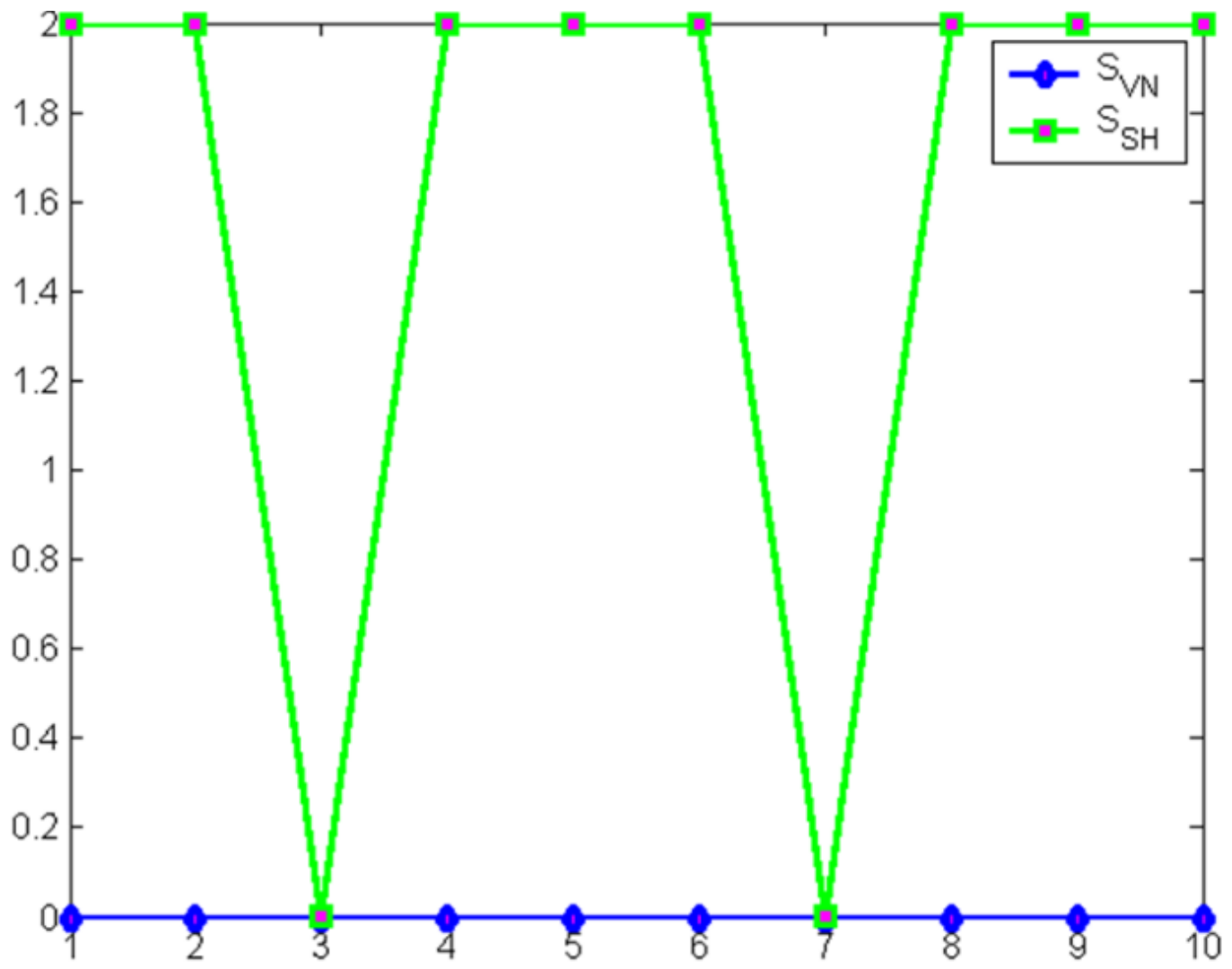

*Figure 12 (d): Simulation results of dynamics behavior for Shannon and von Neumann entropies of Shor's QA.*

***6.3.2. Analysis of QA-termination problem solving based on minimum Shannon/von Neumann dynamic simulation entropy.*** The diagonal matrix elements in Grover's QSA-operators are connected a database state to itself and the off-diagonal matrix elements are connected a database state to its neighbors in the database. The diagonal elements of the diffusion matrix have the opposite sign from the off-diagonal elements. The magnitudes of the off-diagonal elements are roughly equal, so we can write the action of the matrix on the initial state, as example:

$$\begin{pmatrix} -a & b & b & b & b & b \\ b & -a & b & b & b & b \\ b & b & -a & b & b & b \\ b & b & b & -a & b & b \\ b & b & b & b & -a & b \\ b & b & b & b & b & -a \end{pmatrix} \begin{pmatrix} 1 \\ 1 \\ -1 \\ 1 \\ 1 \\ 1 \end{pmatrix} \frac{1}{\sqrt{N}} = \begin{pmatrix} -a+(N-3)b \\ -a+(N-3)b \\ +a+(N-1)b \\ -a+(N-3)b \\ -a+(N-3)b \\ -a+(N-3)b \end{pmatrix} \frac{1}{\sqrt{N}}, \text{ where } a = 1-b.$$

If one of the states is marked, i.e. has its phase reserved with respect to those of the others, the multimode interference conditions are appropriate the constructive interference to the marked state, and destructive interference to the others. That is, the population in the marked bit is amplified. The form of this matrix is identical to that obtained through the inversion about the average procedure in Grover's QSA. This operator produce a contrast in the probability density of the final states of the database of $\frac{1}{N}\left[a+(N-1)b\right]^2$ for marked bit versus $\frac{1}{N}\left[a-(N-3)b\right]^2$ for the unmarked bits; $N$ is the number of bits in the data register.

Grover algorithm is a optimal and it is very efficient search algorithm. And Grover-based software is currently used for search routines in large database.

*Example:* A quantitative measure of success in the database search problem is the reduction of the information entropy of the system following the search algorithm. Entropy $S^{Sh}(P_i)$ in this example of a single marked state is defined as

$$S^{Sh}(P_i) = -\sum_{i=1}^{N} P_i \log P_i, \tag{69}$$

where $P_i$ is the probability that the marked bit resides in orbital $i$. In general, according to, the von Neumann entropy is not a good measure for the usefulness of Grover's algorithm.

For practically every value of entropy, there exit states are good initializers and states that are not. For example, $S\left(\rho_{(n-1)-mix}\right) = \log_2 N - 1 = S\left(\rho_{\left(\frac{1}{\log_2 N}\right)-pure}\right)$, but when initialized in $\rho_{(n-1)-mix}$, the Grover algorithm is as bad as guessing the market state. Another example may be given using pure states $H|0\rangle\langle 0|H$ and $H|1\rangle\langle 1|H$. With the first, Grover arrives to the marked state quadratic speed-up, while the second is practically unchanged by the algorithm.

We apply the Shannon information entropy for optimization of the termination problem of Grover's QSA. Information analysis of Grover's QSA based on using of Eq. (69), gives a lower bound on necessary amount of entanglement for searching of success result and of computational time: any QSA that uses the quantum oracle calls $\{O_s\}$ as $I-2|s\rangle\langle s|$ must call the oracle at least $T \geq \left(\frac{1-P_e}{2\pi} + \frac{1}{\pi \log N}\right)\sqrt{N}$ times to achieve a probability of error $P_e$ [22].

The information system consists of the $N$-state data register. Physically, when the data register is loaded, the information is encoded as the phase of each orbital. The orbital amplitudes carry no information [23]. While state-selective measurement gives as result only amplitudes, the information is completely hidden from view, and therefore the entropy of the system is maximum:

$$S_{init}^{Sh}\left(P_i\right) = -\log\left(1/N\right) = \log N \, .$$

The rules of quantum measurement ensure that only one state will be detected each time. If the algorithm works perfectly, the marked state orbital is revealed with unit efficiently, and the entropy drops to zero.

Otherwise, unmarked orbitals may occasionally be detected by mistake. The entropy reduction can be calculated from the probability distribution, using Eq. (69).

*Figure 13* show the result of entropy calculation for the simulation quantum search of one marked state in the case $N = 7$ .

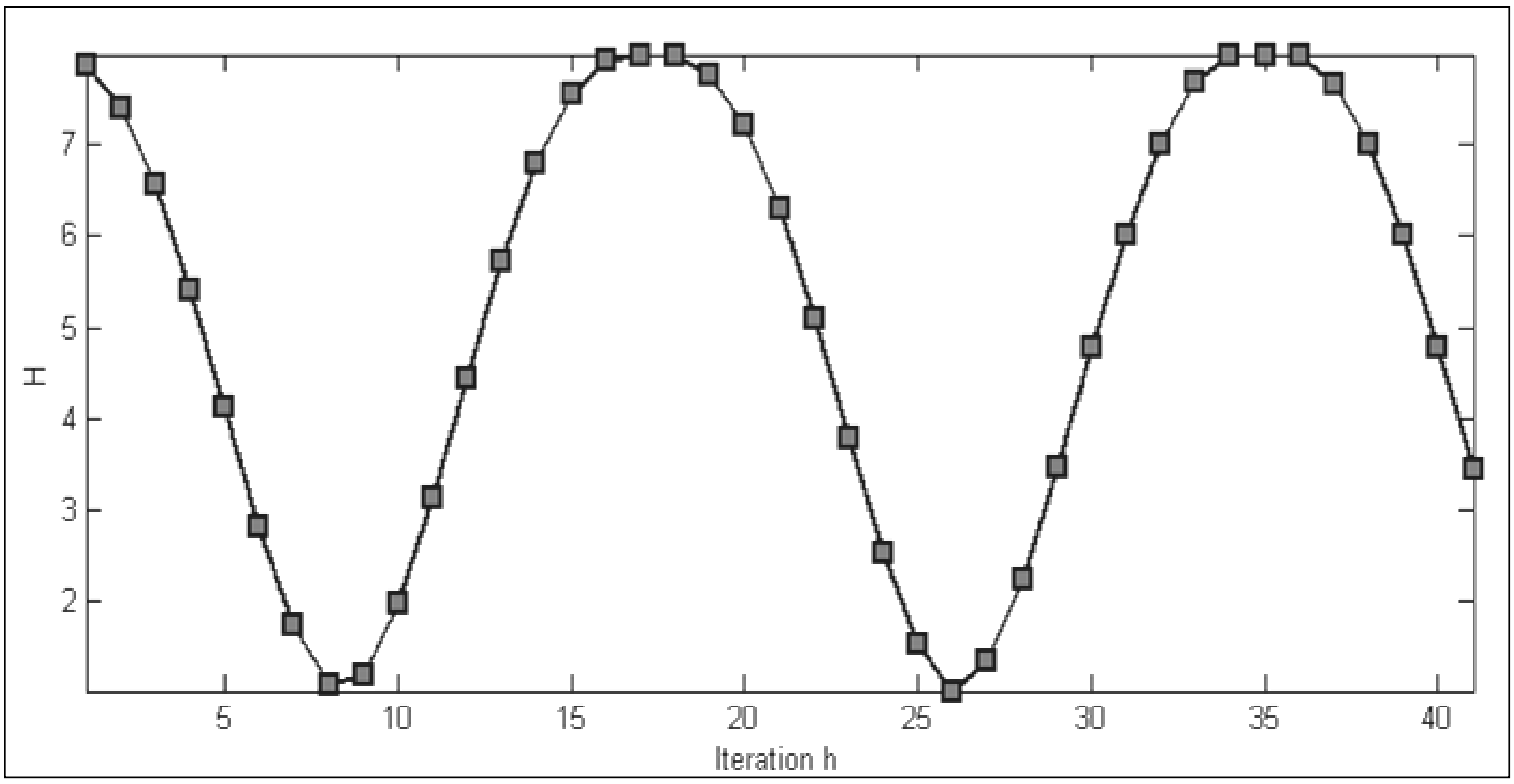

*Figure 13: Shannon entropy analysis of Grover's QSA dynamics with seven inputs.*

Additional results of information analysis of quantum algorithms can be found in [23-39].

## Conclusions

- The analysis of the classical and quantum information flow in Deutsch and Deutsch-Jozsa algorithm is used. It is shown that QAG G, based on superposition of states, quantum entanglement and interference, when acting on the input vector, stores information into the system state, minimizing the gap between classical Shannon entropy and quantum von Neumann entropy.
- Minimizing of the gap between Shannon and von Neumann entropies is considered as a termination criterion of QA computational intelligence measure.
- Methodologically, the principle of maximal intelligence on output states for synthesizing QAGs is applied.
- The technique from quantum information theory applied in the domain of quantum algorithm (QA) synthesis and simulation. For this purpose, the classical and quantum information flow in the Shor's and Grover's QA is analyzed.
- The quantum algorithmic gate, which is based on superposition, quantum super correlation (entanglement) and interference, when acting on the input vector, stores information into the system state, and in this case (similar to Deutsch-Jozsa QA case) minimizing the gap between the classical Shannon entropy and the quantum von Neumann entropy.
- This principle provides both a methodology to design a QAG and a technique to efficiently simulate its behavior on a classical computer.

## References

[1] Bose S., Rallan L., Vedral V. Communication capacity of quantum computation // Physical Review Letters. – 2000. –Vol. 85. – No 25. – pp. 5448 – 5451.

[2] Nielsen M.A., Chuang I.L. Quantum Computation and Quantum Information. – Cambridge Univ. Press, UK, 2000.

[3] Ulyanov S.V., Litvintseva L.V. Quantum information and quantum computational intelligence: Quantum decision making and search algorithms. – Note del Polo Ricerca, Università degli Studi di Milano (Polo Didattico e di Ricerca di Crema). – Vols. 84 & 85. – Milan. – 2005.

[4] Cerf N.J., Adami C. Negative entropy and information in quantum mechanics // Physical Review Letters. – 1997. – Vol. 79. – No 18. – pp. 5195 – 5197.

[5] Adami Ch. The physics of information // arXiv: quant-ph/0405005v1, 3 May 2004, 28 p.

[6] Adami Ch. Information theory in molecular biology // Physics of Life Reviews. –2004. – Vol. 1. – No 1. – pp. 3 – 22.

[7] Vedral V. The role of relative entropy in quantum information theory // Rev. Mod. Phys. 2002. – Vol. 74. – No 1. Pp. 197 – 234.

[8] Horodecki M., Oppenheim J., Winter A. Quantum information can be negative // arXiv: quant-ph/0505062v1, 9 May 2005, 8 p.

[9] Horodecki M., Oppenheim J., Winter A. Partial quantum information // Nature. –2005. – Vol. 436. – № 7051.

[10] Ghisi F., Ulyanov S.V. The information role of entanglement and interference in Shor quantum algorithm gate dynamics // J. of Modern Optics. – 2000. –Vol. 47. – № 12. – pp. 1012 – 1023.

[11] Benenti G., Casati G., Strini G. Principles of quantum computation and information. – Singapore: World Scientific, Vol. I. 2004; Vol. II. 2007.

[12] Janzing D. Computer science approach to quantum control. – Habilitation: Univ. Karlsruhe (TH) Publ. Germany, 2006.

[13] Ulyanov S.V., Litvintseva L.V. Quantum information and quantum computational intelligence: Applied quantum soft computing in AI, computer science, quantum games and self-organization, informatics and design of intelligent wise robust control. – Note del Polo Ricerca. Milano: Universita degli Studi di Milano Publ. – V. 86. – 2007.

[14] Ingarden R.S. Quantum information theory // Reports on Math. Physics. – 1976. Vol. 10. – № 1. – pp. 43 – 72.

[15] Ulyanov S.V. Generalized information measures and axiomatic information theory // Engineering Cybernetics (Science and Technics Investigations). – M.: VINITI Academy of Science USSR. – 1973. – Vol. 5. – pp. 352 – 385.

[16] Lomonaco S.L. Notes on quantum information theory. - Personal Lecture Notes. Vol. 3. (available: http://www.cs.umbc.edu/~lomonaco/lecturenotes/index.html)

[17] Arikan E. An information-theoretic analysis of Grover's algorithm // arXiv: quant-ph/0210068, 10 Oct 2002, 9p.

[18] Ulyanov S.V., Kurawaki I., Yazenin A.V. et al. Information analysis of quantum gates for simulation of quantum algorithms on classical computers // Proceedings of Intern. Conf. on Quantum Communication, Measurements and Computing (QCM&C'2000). – Capri. Italy, 2000. Kluwer Acad. /Plenum Publ. – 2001. – pp. 207 – 214.

[19] Ulyanov S.V. System and method for control using quantum soft computing – US patent No 6,578,018B1. – 2003.

[20] Ulyanov S.V., Litvintseva L.V. Quantum information and quantum computational intelligence: Quantum probability, Physics of quantum information and Information geometry, Quantum computational logic and Quantum complexity. – Note del Polo Ricerca. Milano: Universita degli Studi di Milano Publ. – Vol. 83. – 2007 (available: http://www.qcoptimizer.com/).

[21] Galindo A., Martın-Delgado M.A. Information and Computation: Classical and Quantum Aspects // arXiv:quant-ph/0112105; https://doi.org/10.48550/arXiv.quant-ph/0112105 (see, Rev. Mod. Phys. – 2002. – Vol. 74. – Pp. 347 - 423).

[22] Arikan, E. An Information-Theoretic Analysis of Grover's Algorithm. In: Shumovsky, A.S.,

Rupasov, V.I. (eds) Quantum Communication and Information Technologies. NATO Science Series, vol 113. Springer, Dordrecht. 2003 https://doi.org/10.1007/978-94-010-0171-7_1; IEEE International Symposium on Information Theory, 2003. Proceedings. DOI: 10.1109/ISIT.2003.1228418

[23] Ivancova O.V., Korenkov V.V., Ulyanov S.V. QUANTUM SOFTWARE ENGINEERING - QUANTUM SUPREMACY MODELLING. Part I: Design IT and information analysis of quantum algorithms. – M.: Kurs. – 2020.

[24] Serrano M.A., Perez-Castillo R., Piattini M. Quantum Software Engineering. Springer Verlag Publ., 2022, 330 p.

[25] Korenkov V.V. et al. Quantum software engineering. Vol. 2. M.: Kurs, 2022, 452 p.

[26] Ivancova O.V. et al. Quantum software engineering. Pt. 1. M.: Kurs, 2022, 448p.

[27] Reshetnikov A.G. et al. Intelligent cognitive robotics. Vol. 3: Quantum computational toolkit. M.: Kurs, 2023, 432 p.

[28] Fox M.F.J. et al. Preparing for the quantum revolution: What is the role of higher education? // PHYS. REV. PHYSICS EDUCATION RESEARCH. – 2020. – Vol. 16. – Pp. 020131.

[29] Ulyanov S.V. et al. Quantum information and quantum computational intelligence: Design & classical simulation of quantum algorithm gates. – Universita degli Studi di Milano: Polo Didattico e di Ricerca di Crema Publ. – 2005. – Vol. 80.

[30] Abhijith J. et al. Quantum Algorithm Implementations for Beginners. ArXiv, 2022, art. 1804.03719v3. Available at: https://arxiv.org/abs/1804.03719v3 (accessed Jun 27, 2022). ACM Transactions on Quantum Computing, Vol. 3, No. 4, Article 18. Publication date: July 2022; https://doi.org/10.1145/3517340

[31] Childs A.M. et al. Quantum algorithms and the power of forgetting. ArXiv, 2022, art. 2211.12447v2. Available at: https://arxiv.org/abs/2211.12447v2 (accessed Nov 12, 2022). https:// doi.org/10.4230/LIPIcs.IYCS.2023.37.

[32] Ulyanov S.V. System and Method for Control Using Quantum Soft Computing. Patent US-6578018- B1. USA, 2003.

[33] Cumming R., Thomas T. Using a quantum computer to solve a real-world problem – what can be achieved today? ArXiv, 2022, art. 2211.13080v1. Available at: https://arxiv.org/abs/2211.13080v1 (accessed Nov 18, 2022).

[34] de Wolf R. Quantum Computing: Lecture Notes // arXiv:1907.09415v5 [quant-ph] 16 Jan 2023.

[35] Lin L. Lecture Notes on Quantum Algorithms for Scientific Computation. Department of Mathematics, University of California, Berkeley Challenge Institute of Quantum Computation, University of California, Berkeley Computational Research Division, Lawrence Berkeley National Laboratory. – 2022.

[36] Nyman P. Simulation of Quantum Algorithms with a Symbolic Programming Language. ArXiv, 2022, art. 0705.3333v2. Available at: https://arxiv.org/abs/0705.3333v2 (accessed May 24, 2007).

[37] Juliá-Díaz B. et al. QDENSITY - A Mathematica quantum computer simulation. Computer Physics Communications, 2009, vol. 180. pp. 474. doi:10.1016/j.cpc.2008.10.006.

[38] Johansson N., Larsson A-K. Efficient classical simulation of the Deutsch–Jozsa and Simon's algorithms. Quantum Inf. Process, 2017, vol. 16. pp. 233. DOI 10.1007/s11128-017-1679-7.

[39] Li H. et al. Distributed exact quantum algorithms for Deutsch-Jozsa problem // arXiv:2303.10663v1 [quant-ph] 19 Mar 2023.

[40] Bush P., Lahti P.J. Maximal uncertainty and maximal information for quantum position and momentum // J. Phys. A: Math. Gen. - 1987. - Vol. 20. - No 14. - Pp. 899-906.

[41] Pati A.K. Uncertainty relation of Anandan - Aharonov and intelligent states // Physics Letters. - 1999. - Vol. A262. - No 3. - Pp. 296–301.

[42] Ivancova O.V., Korenkov V.V., Ulyanov S.V. QUANTUM SOFTWARE ENGINEERING QUANTUM SUPREMACY MODELLING. Part I: Design IT and information analysis of quantum algorithms. – M.: Kurs. – 2020.

[43] El Kinani A.H., Daoud M. Generalized Intelligent States for an Arbitrary Quantum System // arXiv:quant-ph/0311029v1 6 Nov 2003.

## *Appendix.* Intelligent coherent states with minimum uncertainty and maximal information.

The minimum-uncertainty coherent states (as example, for the harmonic-oscillator potential) can be defined as those states that minimize the uncertainty relation of Heisenberg (leading to the equality in the uncertainty relations), subject to the added constraint that the ground state is a member of the set. They are considered to be as close as possible to the classical states. Beyond the harmonic-oscillator system, coherent states have also been developed for quantum (Schrodinger) systems with general potentials and for general Lie symmetries. These states are called (general) minimum-uncertainty coherent states and (general) displacement-operator coherent states. There is also a different generalization of the coherent states of the harmonic-oscillator system. This is the concept of "squeezed" states. (Squeezing is a reduction of quadrature fluctuations below the level associated with the vacuum) [40,41].

**A**. The *even* and *odd* coherent states for one-mode harmonic oscillator were introduced in 1970s. These states, which have been called Schrodinger cat states, were studied in detail. These states are representatives of non-classical states. Schrodinger cat states have properties similar to those of the squeezed states, i.e. the squeezed vacuum state and the even coherent state contain Fock states with an even number of photons.

***Definition***: *Intelligent states are quantum states, which satisfy the equality in the uncertainty relation for non-commuting observables*.

In quantum mechanics two non-commuting observables cannot be simultaneously measured with arbitrary precision. This fact, often called the Heisenberg uncertainty principle, is a fundamental restriction that is related neither to imperfection of the existing real-life measuring devices nor to the experimental errors of observation. It is rather the intrinsic property of the quantum states itself.

The uncertainty principle provides (paradoxically enough) the only way to avoid many interpretation problems. The uncertainty principle specified for given pairs of observables finds its mathematical manifestation as the uncertainty relations.

The first rigorous derivation of the uncertainty relation from the basic non-commuting observables (i.e., for the position and moment, $[\hat{x}, \hat{p}] = i\hbar$ ) is due to Kennard (1927). This derivation (repeated in most textbooks on quantum mechanics ever since) leads to the inequality: $\Delta\hat{x}\Delta\hat{p} \geq \frac{1}{2}\hbar$ . In fact, it can be considered as a simple consequence of the properties of the Fourier transform that connects the wave functions of the system in the position and momentum representation (more general form of uncertainty inequality with Wigner-Yanase-Dyson skew information in [42] are described).

**B**. It is possible to present quantum uncertainty relations (UR) in terms of entropy or information ("entropic UR" –EUR). The usual "standard UR" (for standard deviations)

$$(\Delta_\varphi A)^2 (\Delta_\varphi B)^2 \geq \frac{1}{4}\left|\langle [A,B]_- \rangle_\varphi\right|^2 + \frac{1}{4}\left|\langle \{A,B\}_+ \rangle_\varphi - 2\langle A\rangle_\varphi \langle B\rangle_\varphi\right|^2$$

(note that the second term in this inequality represents the covariance, or correlation,

$$\mathrm{cov}_\varphi(A,B) := \frac{1}{2}\langle\varphi|AB + BA|\varphi\rangle - \langle\varphi|A|\varphi\rangle\langle\varphi|B|\varphi\rangle$$

between the observables $A$ and $B$ in the state $|\varphi\rangle$) presented by an inequality of the entropic form $S^{(A)} + S^{(B)} \geq S_{AB}$ or in information form $I_\varphi(A) + I_\varphi(B) \leq I_\varphi(A,B)$ as more adequate expressions for the "uncertainty principle". It is known that given two non-commuting observables, we can derive an uncertainty relation for them and the class of states that satisfy the equality sign in the inequality are called intelligent states (see, Definition).

*Example*. If we have any continuous parameter $\lambda$ and any Hermitian observable $A(\lambda)$ which is the generator of the parametric evolution, then UR give us $\langle \Delta A(\lambda) \rangle \Delta\lambda \geq \frac{\hbar}{4}$ where $\langle \Delta A(\lambda) \rangle = \frac{1}{(\lambda_2 - \lambda_1)} \int_{\lambda_1}^{\lambda_2} \Delta A(x) dx$ is the parameter average of the observable uncertainty and $\Delta\lambda = \frac{\pi}{s_0}(\lambda_2 - \lambda_1)$ is the scaled displacement in the space of the conjugate variable of $A$. This generalized UR would hold for position-momentum, phase-number or any combinations. For the case when initial and final states are orthogonal we know that all states of the form

$$|\psi(\lambda)\rangle = \frac{1}{\sqrt{2}} \left( e^{-\frac{i}{\hbar} a_i \lambda} |\psi_i\rangle + e^{-\frac{i}{\hbar} a_j \lambda} |\psi_j\rangle \right), i \neq j$$

are the only <u>intelligent</u> states which satisfy the equality $\langle \Delta A(\lambda) \rangle \Delta\lambda = \frac{\hbar}{4}$.

However, these states do not satisfy the equality when the initial and final states are non-orthogonal. In this case, if the generator of the parametric evolution $A$ can be split into two parts $A_0 + A_1$ such that $A_0$ has a complex basis of normalised eigenvectors $\{|\psi_i\rangle\}_{i \in I}$ which degenerate spectrum $\{a_0\}$, with $I$ a set of quantum numbers and $A_1$ has matrix elements $(A_1)_{ii} = 0 = (A_1)_{jj}$, and $(A_1)_{ij} = (A_1)_{ji} = a_1$, then all states of the form

$$|\psi(\lambda)\rangle = e^{-\frac{i}{\hbar} a_0 \lambda} \left[ \cos\left( a_1 \frac{\lambda}{\hbar} \right) |\psi_i\rangle - i \sin\left( a_1 \frac{\lambda}{\hbar} \right) |\psi_j\rangle \right], i \neq j$$

are "<u>intelligent states"</u> for <u>non-orthogonal initial</u> and <u>final states</u>.

**C.** It has been shown that the "Everett (entropic) UR" implies the famous Heisenberg UR as $\Delta q \Delta p \geq \frac{\hbar}{2}$. We shall compare various characterisations of "maximal information" and point out their connection with "minimum uncertainty". In the following we restrict ourselves mainly to "simple" observables (defined on the smallest non-trivial Boolean algebra $\Sigma = \{0, a, \neg a, 1\}$): we are interested in information with respect to single effect $E : I_\varphi(E) = E_\varphi \ln(E_\varphi) + E'_\varphi \ln(E'_\varphi), E' = I - E$.

Non-commutativity or incompatibility of (unsharp) properties $E$ and $F$ will, in general, exclude the possibility of measuring or preparing both of them simultaneously. In particular, if $E = E^Q(X), F = F^P(Y)$ are position and momentum spectral projections associated with bounded measurable sets $X, Y$, then $E^Q(X) \wedge E^P(Y) = 0$ holds or, equivalently

$$\langle \varphi | E^Q(X) | \varphi \rangle = 1 \Rightarrow \langle \varphi | E^P(Y) | \varphi \rangle < 1$$

$$\langle \varphi | E^P(Y) | \varphi \rangle = 1 \Rightarrow \langle \varphi | E^Q(X) | \varphi \rangle < 1$$

Thus "certain" position and momentum determinations exclude each other, and the question arises as to what "degree of uncertainty" they can be "known" simulataneously. One may take any reasonable characterisation of <u>maximal</u> joint knowledge, or joint information. In this case above mentioned statement can be put into the following equivalent form

$$\left.\begin{array}{l}\langle\varphi|E^{Q}(X)|\varphi\rangle=1\Rightarrow\langle\varphi|E^{P}(Y)|\varphi\rangle<1\\ \langle\varphi|E^{P}(Y)|\varphi\rangle=1\Rightarrow\langle\varphi|E^{Q}(X)|\varphi\rangle<1\end{array}\right\}\Rightarrow\begin{cases}E_{\varphi}+F_{\varphi}<2\\ E_{\varphi}\cdot F_{\varphi}<1\end{cases}$$

The "state of maximal information" can be defined through three values. The first expression $E_{\varphi}+F_{\varphi}$ can be maximised and an explicit construction procedure for the corresponding "state of maximal information" has been given below. Here we shall study the question of maxima for this quantity as well as for $E_{\varphi}\cdot F_{\varphi}$ and for $I_{\varphi}(E)+I_{\varphi}(F)$ for an arbitrary pair of effects, $E$ and $F$. In particular, we shall show that each quantity can be maximal only if there exist states which lead to minimal uncertainty product in UR.

Furthermore, in the case of projections the maxima of $I_{\varphi}(E)+I_{\varphi}(F)$ (if they exist) coincide with those of one of the quantities $E_{\varphi}^{\nu}+F_{\varphi}^{\eta}$ and $E_{\varphi}^{\nu}\cdot F_{\varphi}^{\eta}$ $\left(E^{\nu}\in\{E,E'\},F^{\eta}\in\{F,F'\}\right)$.

For maximal $E_{\varphi}+F_{\varphi}$ the variation of $\langle\varphi|E|\varphi\rangle+\langle\varphi|F|\varphi\rangle-\lambda\langle\varphi|\varphi\rangle$ must vanish which implies the following equations: $(E+F)|\varphi\rangle=(E_{\varphi}+F_{\varphi})|\varphi\rangle$. Multiplying with $E$ or with $F$ and taking the expectations yields

$$(\Delta_{\varphi}E)^{2}=(\Delta_{\varphi}F)^{2}=-\left(\langle\varphi|EF|\varphi\rangle-E_{\varphi}\cdot F_{\varphi}\right)=-\mathrm{cov}_{\varphi}(E,F),$$

which leads to a minimal UR: $(\Delta_{\varphi}E)^{2}\cdot(\Delta_{\varphi}F)^{2}=\left[\mathrm{cov}_{\varphi}(E,F)\right]^{2}$.

Similarly, maximising the product $E_{\varphi}\cdot F_{\varphi}$ gives $(F_{\varphi}E+E_{\varphi}F)|\varphi\rangle=2E_{\varphi}\cdot F_{\varphi}|\varphi\rangle$ and

$$(\Delta_{\varphi}E)^{2}F_{\varphi}^{\ 2}=(\Delta_{\varphi}F)^{2}E_{\varphi}^{\ 2}=-E_{\varphi}\cdot F_{\varphi}\,\mathrm{cov}_{\varphi}(E,F)$$

which leads again to a minimal UR, $E_{\varphi}\neq 0\neq F_{\varphi}$.

Finally, maximal information sum $I_{\varphi}(E)+I_{\varphi}(F)$ will be realised in states satisfying $(\ln E_{\varphi}-\ln E'_{\varphi})(E-E'_{\varphi})|\varphi\rangle+(\ln F_{\varphi}-\ln F'_{\varphi})(F-F'_{\varphi})|\varphi\rangle=0$. Generally, this equation contains all stationary points, e.g. the minimum $E_{\varphi}=E'_{\varphi}=F_{\varphi}=F'_{\varphi}=\frac{1}{2}$, or the joint eigenstates. Since we are looking for states of maximal information with respect to positive outcomes for $E,F$ we shall assume $E_{\varphi}>\frac{1}{2}$ and $F_{\varphi}>\frac{1}{2}$. Then this equality implies [40]:

$$(\alpha E+F)|\varphi\rangle=(\alpha F_{\varphi}+F_{\varphi})|\varphi\rangle,\alpha=\frac{\ln\left(\frac{E_{\varphi}}{E'_{\varphi}}\right)}{\ln\left(\frac{F_{\varphi}}{F'_{\varphi}}\right)}\geq 0$$

and $\alpha(\Delta_{\varphi}E)^{2}=\frac{1}{\alpha}(\Delta_{\varphi}F)^{2}=-\mathrm{cov}(E,F)$ which again gives rise to the minimal uncertainty product in UR.

We have thus shown that all three notions of maximal information are consistent in so far as they imply minimal uncertainty product.

*Example.* Let $E,F$ denote position and momentum spectral projections, respectively: $E=E^{Q}(X),F=F^{P}(Y)$. The sum of probabilities $E_{\varphi}+F_{\varphi}$ has been shown to be maximal in the state $\varphi=\varphi_{\min}$ with $|\varphi_{\min}\rangle=\left(\frac{1+a_{0}}{2a_{0}^{\ 2}}\right)^{1/2}E|g_{0}\rangle+\left(\frac{1-a_{0}}{2(1-a_{0}^{\ 2})}\right)^{1/2}E'|g_{0}\rangle$ provided that $X,Y$ are bounded

measurable sets. Here $a_0^2$ is the maximal eigenvalue of the compact operator (FEF) and $g_0$ is the corresponding eigenvector satisfying $FEF|g_0\rangle = a_0^2|g_0\rangle, F|g_0\rangle = |g_0\rangle, \||g_0\rangle\|_2^2 = 1$. It is clear from above description that $\varphi_{\min}$ must be an eigenstate of $(E+F)$. This can also be seen directly in the following way. Introduce $|f_0\rangle = a_0^{-1}E|g_0\rangle, \|f_0\|_2^2 = a_0^{-2}\langle g_0|FEF|g_0\rangle = 1, E|f_0\rangle = |f_0\rangle$. Then we have

$$EFE|f_0\rangle = a_0^2|f_0\rangle, |g_0\rangle = a_0^{-1}F|f_0\rangle$$

and $\varphi_{\min}$ can be written in the symmetric form

$$|\varphi_{\min}\rangle = \frac{1}{\sqrt{2(1+a_0)}}\left[|f_0\rangle + |g_0\rangle\right]$$

We conclude that $\varphi_{\min}$ maximises all the three quantities $(E_\varphi \cdot F_\varphi)$, $(E_\varphi + F_\varphi)$ and $(I_\varphi(E) + I_\varphi(F))$, and it minimises the uncertainty product $\Delta_\varphi E \cdot \Delta_\varphi F$.

Thus, maximal information (minimal entropy) and minimal uncertainty can be achieved on "intelligent coherent states" and will again coincide.
Additional properties of "intelligent coherent states" in [40-43] demonstrated.